\documentclass[]{aa}
\usepackage		[utf8]					{inputenc}
\usepackage		[english]				{babel}
\usepackage								{amsmath}
\usepackage								{amsfonts}
\usepackage								{amssymb}
\usepackage 					  		{graphicx}
\usepackage								{natbib}
\usepackage     [T1]                    {fontenc}
\usepackage                             {color}
\usepackage                             {hyperref}
\usepackage								{gensymb}
\usepackage								{subfigure}
\usepackage                             {longtable}
\usepackage                             {lscape}

\definecolor{darkblue}{rgb}{0.0,0.0,0.4}
\definecolor{darkgreen}{rgb}{0.0,0.4,0.0}
\hypersetup{
    colorlinks,
    linkcolor=black,
    citecolor=darkgreen,
    urlcolor=darkblue
    }

\newcommand{\ccol}{$^{13}$CO(2\,--\,1) }
\newcommand{\cool}{C$^{18}$O(2\,--\,1) }
\newcommand{\cco}{$^{13}$CO }
\newcommand{\coo}{C$^{18}$O }

\begin{document}

\title{SEDIGISM: The kinematics of ATLASGAL filaments}

\author{M. Mattern\inst{1}
          \and
          J. Kauffmann\inst{1,2}
          \and 
          T. Csengeri\inst{1}
          \and
          J. S. Urquhart\inst{3}
          \and
          S. Leurini\inst{4}
          \and
          F. Wyrowski\inst{1}
          \and
          A. Giannetti\inst{5}
          \and
          P. J. Barnes\inst{6,7}
          \and 
          H. Beuther\inst{8}
          \and 
          L. Bronfman\inst{9}
          \and
          A. Duarte-Cabral\inst{10}
          \and
          T. Henning\inst{8}
          \and
          J. Kainulainen\inst{8,11}
          \and
          K. M. Menten\inst{1}
          \and
          E. Schisano\inst{12}
          \and
          F. Schuller\inst{13}
          }
\institute{Max-Planck-Institut f\"ur Radioastronomie, Auf dem H\"ugel 69, D-53121 Bonn
           \email{mmattern@mpifr-bonn.mpg.de}
           \and
           Haystack Observatory, Massachusetts Institute of Technology, 99 Millstone Road, Westford, MA 01886, USA
           \and
           School of Physical Sciences, University of Kent, Ingram Building, Canterbury, Kent CT2 7NH, UK
           \and
           INAF - Osservatorio Astronomico di Cagliari, via della Scienza 5, 09047 Selargius (CA), Italy
           \and
           INAF - Istituto di Radioastronomia, and Italian ALMA Regional Centre, via P. Gobetti 101, 40129 Bologna, Italy
           \and
           Astronomy Department, University of Florida, PO Box 112055, Gainesville, FL 32611, USA
           \and
School of Science and Technology, University of New England, NSW 2351 Armidale, Australia
           \and
           Max-Planck-Institut für Astronomie, K\"onigstuhl 17, 69117 Heidelberg, Germany
           \and
           Departamento de Astronomía, Universidad de Chile, Casilla 36-D, Santiago, Chile
           \and
           School of Physics and Astronomy, Cardiff University, Queens Buildings, The Parade, Cardiff CF24 3AA, UK
           \and
           Dept. of Space, Earth and Environment, Chalmers University of Technology, Onsala Space Observatory, 439 92 Onsala, Sweden
           \and
           Istituto di Astrofisica e Planetologia Spaziali, INAF, via Fosso del Cavaliere 100, I-00133 Roma, Italy
           \and
           AIM, CEA, CNRS, Université Paris-Saclay, Université Paris Diderot, Sorbonne Paris Cité, F-91191 Gif-sur-Yvette, France           
           }
\date{Received ...; accepted ...}

\abstract{Analysing the kinematics of filamentary molecular clouds is a crucial step towards understanding their role in the star formation process. Therefore, we study the kinematics of 283 filament candidates in the inner Galaxy, that were previously identified in the ATLASGAL dust continuum data. The \ccol and \cool data of the SEDIGISM survey (Structure, Excitation, and Dynamics of the Inner Galactic Inter Stellar Medium) allows us to analyse the kinematics of these targets and to determine their physical properties at a resolution of $30\arcsec$ and $\rm 0.25~km\,s^{-1}$. To do so, we developed an automated algorithm to identify all velocity components along the line-of-sight correlated with the ATLASGAL dust emission, and derive size, mass, and kinematic properties for all velocity components. We find two-third of the filament candidates are coherent structures in position-position-velocity space. The remaining candidates appear to be the result of a superposition of two or three filamentary structures along the line-of-sight. At the resolution of the data, on average the filaments are in agreement with Plummer-like radial density profiles with a power-law exponent of $p \approx 1.5 \pm 0.5$, indicating that they are typically embedded in a molecular cloud and do not have a well-defined outer radius. Also, we find a correlation between the observed mass per unit length and the velocity dispersion of the filament of $m \propto \sigma_\text{v}^2$. We show that this relation can be explained by a virial balance between self-gravity and pressure. Another possible explanation could be radial collapse of the filament, where we can exclude infall motions close to the free-fall velocity.}


\maketitle

\section{Introduction}
Filamentary structures play an important role in the process of star formation. Observations at different wavelengths based on various tracers have revealed that filaments are ubiquitous in the interstellar medium \citep[e.g., ][]{Schneider1979, Molinari2010, Andre2010, Schisano2014, Ragan2014, Li2016b}. Filaments are seen in quiescent and star-forming clouds, in which a significant fraction of pre-stellar cores are located \citep{Andre2010}. Filamentary structures have wide ranges of masses ($\rm \sim 1\text{ -- }10^5 \, M_\odot$) and lengths ($\rm \sim 0.1\text{ -- }100 \, pc$) \citep[e.g., ][]{Bally1987, Jackson2010, Arzoumanian2011, Hernandez2012, Hacar2013, Kirk2013, Palmeirim2013, Li2016b, Kainulainen2013a, Beuther2015, Kainulainen2016, Abreu-Vicente2016, Zucker2017}.

The processes of filament formation and filament fragmentation to star-forming cores are not well understood. Because of the wide range of filament size scales and masses these processes might also differ among filaments. High-resolution magnetohydrodynamical simulations of molecular cloud evolution and filament formation show subsonic motions in the inner dense regions of filaments, but the surrounding low density gas is supersonic \citep{Padoan2001,Federrath2016}. Additionally, accretion flows along and radially onto the filament have been seen in observations and simulations \citep{Schneider2010, Peretto2013,Peretto2014a,Henshaw2014,Smith2015}. Therefore, the formation and evolution of filaments is a highly dynamical process and to constrain it is essential to study their kinematics. 

Studies of filaments have targeted mainly sources in nearby star-forming regions, e.g. Orion, Musca and Taurus \citep{Bally1987,Takahashi2013,Hacar2015,Kainulainen2015,Kainulainen2017}, where high resolution data ($\rm \sim 0.01~pc$, $\rm 0.1~km\,s^{-1}$) reveals sub-structures like fibers \citep{Hacar2013, Hacar2018}, or prominent mid-infrared extinction structures, e.g. ``Nessie'' and infrared dark clouds like G11.11$-$0.12 \citep{Johnstone2003,Pillai2006,Schneider2010,Jackson2010,Kainulainen2013a,Henshaw2014,Mattern2018a}. Detailed studies of these filaments led us to recognize their important role in star formation, and their internal structure, but studies of small samples do not allow to draw general conclusions. In particular the filaments towards the more distant, typically high-mass star forming regions have not yet been systematically studied. Therefore, it is necessary to study a large unbiased sample of filaments. Such studies have recently become feasible because of modern multi-wavelength surveys, which cover the Galactic plane at high resolution and sensitivity.

Several catalogues of filamentary structures have been conducted in the last years, which can be divided in two groups. The filaments in the catalogues of \cite{Schisano2014,Koch2015,Li2016b} were identified from continuum data and therefore, miss the kinematic information, and might be affected by line-of-sight projection effects. The catalogues of \cite{Ragan2014,Zucker2015,Abreu-Vicente2016,Wang2015, Wang2016} concentrate on the longest filamentary structures in the Galaxy. While the identification methods and criteria vary in these studies, all filaments are tested for a velocity coherent behaviour.

In this study, we target the largest catalogue of filamentary structures published so far \citep{Li2016b}, which is based on the ATLASGAL survey at $\rm 870 \mu m$ \citep{Schuller2009}. As these structures were identified in continuum dust emission data, the scope of this work is to use the SEDIGISM data \citep{Schuller2017} to assess their velocity structure. Because of the large number of targets, it is necessary to perform the analysis in a fully-automated way, which will be also presented in this work. 

In this paper, we will refer to the structures identified by \cite{Li2016b} as filament candidates. After the analysis of their velocity structure we will refer to the velocity coherent structures in the filament candidates as filaments, where one filament candidate can consist of multiple filaments. Some of these filaments may not meet the definitions of a filament, as they seem to be composed of a chain of dense clumps, or a dense clump with an elongated low column density environment. However, since filaments fragment, these structures could represent a late phase of evolution and should not be ignored.

The structure of the paper is as follows: Section \ref{data} introduces the survey data used in this study and the targeted catalogue of filament candidates. The methods used to separate the velocity components of a given filament candidate and to derive its filament parameters are described in Section \ref{methods}. In Section \ref{results} we present the resulting statistics of the velocity separation and the interpretation of the kinematics. We then discuss in Section \ref{discussion} the dependency of the filament mass with increasing radius, and the origin of the correlation found between the line-mass (mass per unit length) and velocity dispersion of the filaments. Finally, we summarize our results in Section \ref{conclusion}.

\section{Data and filament sample}
\label{data}

\subsection{Survey data}
\label{surveys}
Within this paper, we will make use of three surveys: ATLASGAL \citep[APEX Telescope Large Area Survey of the Galaxy, ][]{Schuller2009}, ATLASGAL+PLANCK \citep[ATLASGAL combined with PLANCK, ][]{Csengeri2016} and SEDIGISM \citep[Structure, Excitation and Dynamics of the Inner Galactic InterStellar Medium, ][]{Schuller2017}. 

The ATLASGAL survey was conducted with the Large APEX Bolometer Camera (LABOCA) at $\rm 870~\mu m$ between 2007 and 2010 at the Atacama Pathfinder Experiment (APEX) telescope \citep{Guesten2006} located on the Chajnantor plateau in Chile. The resolution of the survey is $19.2\arcsec$ ($6.0\arcsec$ per pixel) with a $1\sigma$ RMS noise in the range of $40$--$70\rm~mJy/beam$. It covers the inner Galactic plane between $-80\degree \leq \ell \leq 60\degree$ and $|\textit{b}|\leq 1.5\degree$. It is sensitive to the cold dust, and it traces mainly the high molecular hydrogen column density regions ($N_{\text{H}_2} \rm \geq 1.0\times10^{22}~cm^{-2}$) of the ISM.

As the ATLASGAL data is missing the large scale low column density emission due to sky noise subtraction, \cite{Csengeri2016} combined the survey with the data observed by the HFI instrument at $\rm 353~GHz$ ($\rm 850~\mu m$) with a resolution of $4.8'$ on board the PLANCK satellite \citep{Lamarre2010, PlanckI2014}. The combined ATLASGAL+PLANCK survey is sensitive to a wide range of spatial scales at a resolution of $21\arcsec$ covering the same region as the original ATLASGAL data on the same pixel grid.

The SEDIGISM survey \citep{Schuller2017} covers the inner Galactic plane between $-60\degree \leq \ell \leq 18\degree$ and $|\textit{b}|\leq 0.5\degree$, which was observed from 2013 to 2016 with the SHeFI heterodyne receiver \citep{Vassilev2008} at the APEX telescope. The prime targets of the survey are the \ccol and \cool molecular lines. The average root-mean-square (RMS) noise of the survey is $\rm 0.9~K$ (T$_{\rm MB}$) at a velocity resolution of $\rm 0.25~km\,s^{-1}$, an FWHM beam size of $30\arcsec$, and a pixel-size of $9.5\arcsec$. For this analysis we use the first data release (DR1, Schuller et al. in prep.).

\subsection{The ATLASGAL sample of filaments}

Based on ATLASGAL, \cite{Li2016b} produced a catalogue of filament candidates, which is the base for this study. The filaments were identified in the ATLASGAL only maps, after they were smoothed to a spatial resolution of $\rm 42\arcsec$. The source extraction was performed with the DisPerSE \citep[Discrete Persistent Structures Extractor, ][]{Sousbie2011} algorithm, which is optimized for the identification of large spatially coherent structures, and has been successfully used to trace filaments in previous studies \citep[e.g., ][]{Hill2011, Arzoumanian2011}. Because of the limited sensitivity and resolution (minimal mean column density $N_{\text{H}_2}=1.6\times10^{21}~\rm cm^{-2}$), the resulting catalogue is unlikely to be complete, however, as it covers a large fraction of the Galactic plane it is likely to include the full range of sizes and masses of filamentary type structures.

Not all of the identified structures are filamentary, but they cover a range of morphologies and complexity from roundish clumps to large web-like structures. Therefore, the identified structures were categorized by \cite{Li2016b} through visual inspection into six groups: unresolved clumps, marginally resolved elongated structures, filaments, networks of filaments, complexes, and unclassified structures. Here a filament was defined as single elongated linear structure with relatively few branches, an intensity clearly above the surrounding medium and an aspect ratio of at least 3, that is clearly resolved across its length and width. The high column densities found in the Galactic centre region lead to a higher probability of identifying more complex structures. Therefore, the number of filament candidates in the catalogue is lower towards the Galactic centre. This classification resulted in a catalogue of 517 filament candidates, providing the starting point of this study. For more details about the filament identification see \cite{Li2016b}. In the following, we will refer to the structures of the catalogue as filament candidates, as they have been identified only in position-position space, which leaves the possibility of line-of-sight projection effects. One of the objectives of this study is to investigate their velocity coherence.

\section{The automated filament analysis}
\label{methods}
The SEDIGISM survey covers the Galactic plane between $-60\degree \leq \ell \leq 18\degree$ and $|\textit{b}|\leq 0.5\degree$, which is only a part of the ATLASGAL survey. Therefore, we analyse the 283 filament candidates in the area covered by all three surveys described in section \ref{surveys}. This corresponds to $\sim 55 \%$ of the total number of filaments, and can therefore be considered representative of such structures in the inner Galactic plane. 

Because of the large number of filament candidates, it is necessary to use an automated approach to analyse them. However, as the sample is distributed over a large range of Galactic longitudes, it is unlikely to find homogeneous conditions in their surrounding material. Therefore, we choose a robust and efficient method to analyse the data in a systematic way, which leads to the following decisions for the analysis. We use the calculation of moments instead of multi-Gaussian fitting to identify the kinematics of the filaments. Also, we do not truncate or alter the skeletons of the filament candidates to fit the identified filaments more accurately, but rather neglect parts where we do not detect molecular gas. Therefore, there are two sets of pixels for a filament candidate used in the analysis: One that describes only the skeleton for the calculation of the kinematics and one that includes also the surrounding area within a dilation box with diameter of three beams used for the structure correlation. This approach results in larger uncertainties in the derived properties, but the homogeneous method enables the finding of correlations in the large scale properties of the filaments, which is the aim of this work.

From \cite{Li2016b} we have a set of positions defining the skeleton of each filament, which trace the highest ATLASGAL intensities, that form the backbone of the structure. For each candidate we extract the data around the skeleton from the surveys using a rectangular box that is $5 \arcmin$ larger on each side than the extrema of the skeleton points. This showed to be sufficient for the analysis of the most nearby $\rm < 2~kpc$ filaments, where the angular extend of the structure is the largest. We now describe the analysis performed on every filament candidate.

\subsection{The filament skeleton}

\begin{figure}[tbh]
\centering
\includegraphics[width=0.5\textwidth, clip=true, trim= 0cm 2.5cm 1.5cm 4cm]{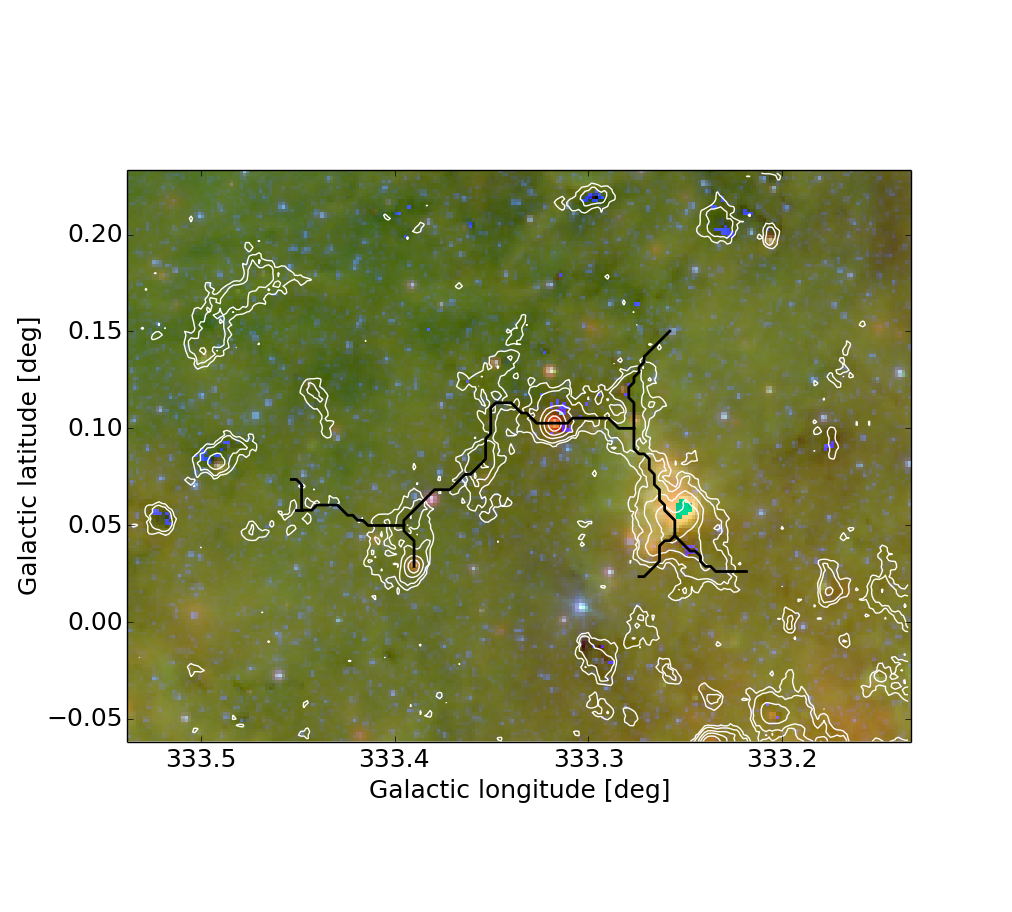} 
\caption{ATLASGAL contours (0.05, 0.1, 0.2, 0.5, 1.0, 2.0 Jy/beam) and skeleton derived by DisPerSE for the filament candidate G333.297+00.073, overlaid on an infrared three color image of the field (red: MIPSGAL 24 $\rm \mu m$; green: GLIMPSE 8.0 $\rm \mu m$; blue GLIMPSE 3.6 $\rm \mu m$.}
\label{skeleton_rgb}
\end{figure}

\begin{figure}[tbh]
\centering
\includegraphics[width=0.5\textwidth, clip=true, trim= 0.3cm 1.5cm 2.5cm 2cm]{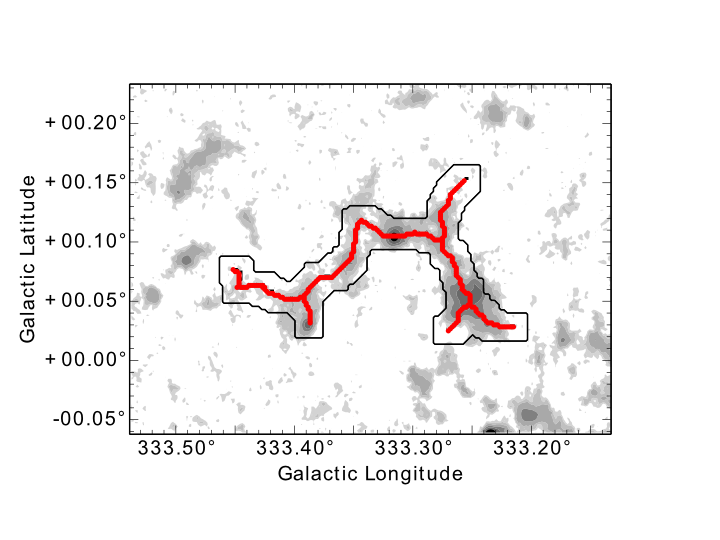} 
\caption{The skeleton of the filament candidate G333.297+00.073 derived by DisPerSE on top of the ATLASGAL grayscale contour map.  
The black contour indicates the dilation box used for the correlation in Section \ref{correlation}.}
\label{skeleton_final}
\end{figure}

For the analysis of each filament candidate we first have to transfer the skeleton coordinates (Fig. \ref{skeleton_rgb}) onto the SEDIGISM grid. To do so, we check whether all positions of the skeleton are covered by the SEDIGISM observations, and remove the positions if they are not covered. This allows us to continue with structures that are partially truncated by the data limits. Then we overlay the skeleton coordinates on the pixel grid of the molecular line data. We mark the pixels within a dilation box around each pixel that covers a position of the DisPerSE skeleton as part of the new pixel skeleton. The size of the dilation box is set to be larger than the maximum distance between two neighbouring skeleton points. Here a width of $1$ pixel ($9.5\arcsec$) is sufficient. As the resulting skeleton mask might have a width larger than one pixel, we use the thinning algorithm of \cite{Gonzalez1992} to truncate the pixel skeleton. The result is a ``chain'' of pixels which might have several branches (see Fig. \ref{skeleton_final}).

\subsection{Identification of velocity components}
\label{vel_identification}

\begin{figure}
\centering
\includegraphics[width=0.5\textwidth, clip=true, trim= 0cm 0cm 0cm 0cm]{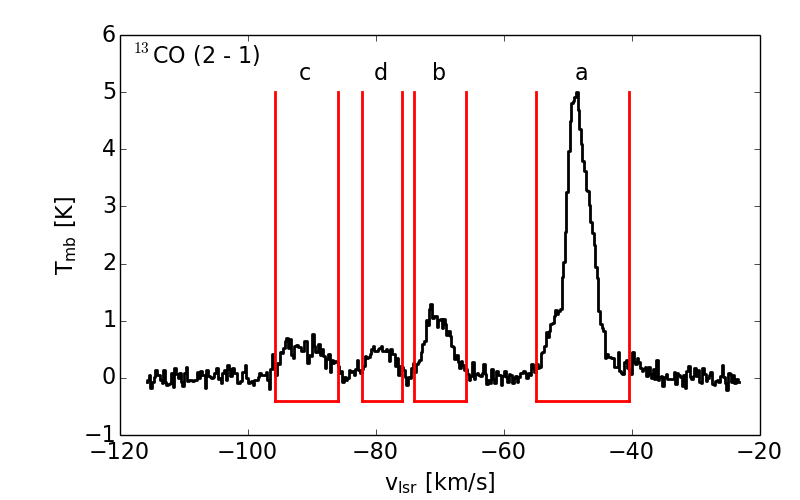}
\includegraphics[width=0.5\textwidth, clip=true, trim= 0cm 0cm 0cm 0cm]{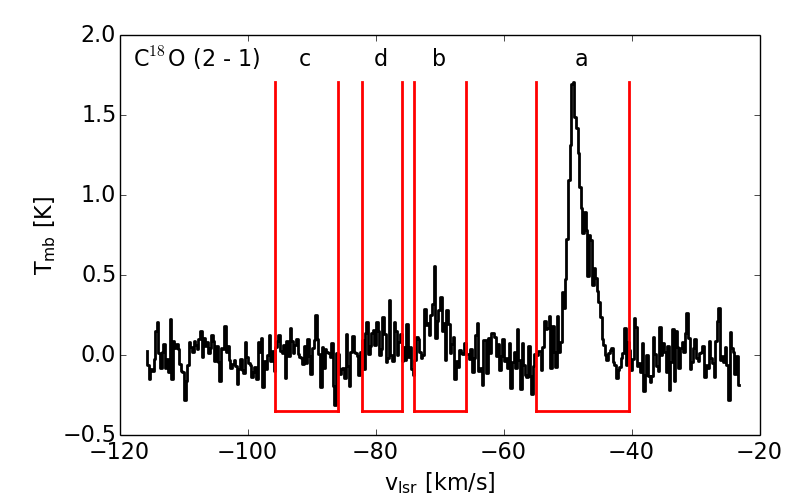}
\caption{Average \cco (top) and \coo (bottom) spectrum over the skeleton of filament candidate G333.297+00.073 (see Figs. \ref{skeleton_rgb} and \ref{skeleton_final}). The red lines mark the identified emission intervals named by letters.}
\label{spectrum}
\end{figure}

In order to investigate whether the filament candidates form a single structure in velocity, spectroscopic observations are indispensable. From a study of filaments in the SEDIGISM first science field \citep[7 filament candidates in $\rm 1.5~deg^2$,][]{Schuller2017}, we know that one can observe line emission at very different velocities towards one continuum structure due to projection effects through the Galactic plane. Therefore, we average all spectra located on the skeleton and identify the velocity ranges that show emission peaks in this spectrum. To identify the velocities we smooth the average spectrum with a Gaussian kernel with a dispersion of 4 channels ($= \rm 1\,km\,s^{-1}$) to reduce the noise. In case the signal-to-noise ratio (SNR) is low, peak intensity $\leq 5\sigma$, we double the kernel width. We then define the velocity range of each spectral component in the averaged spectrum as that between which the emission attains more than the $1\sigma$ noise level. This leaves us with a minimum separation limit of $\delta v_\text{min} = \rm 2.5~km\,s^{-1}$, which is described in detail later on. Furthermore, we only consider components with an SNR $\geq 5$ in their integrated intensity of the original data (for example, see Fig. \ref{spectrum}). 

We then define the ends of the velocity components where the emission peaks of the average spectrum exceeds the $1\sigma$ noise level, so the emission peak is not likely to be truncated, and accept only components with an integrated intensity $\rm SNR \geq 5$ (for example Fig. \ref{spectrum}). The above procedure is only applied to the \cco data because of their higher SNR. The \coo emission lines are narrower than the \cco data lines and so we use the same velocity ranges to calculate the moments in the \coo data. Thereafter, we calculate the zeroth, first, and second `order' moment of each velocity component, that indicate the integrated intensity, peak velocity, and velocity dispersion, respectively, for both molecules. This gives us a first impression of the kinematics of the filament. 

\begin{figure}[htb]
\centering
\includegraphics[width=0.5\textwidth]{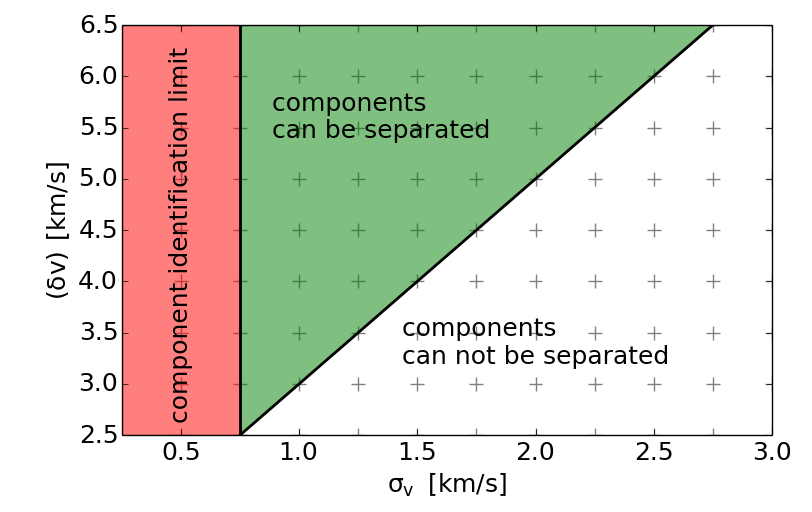}
\caption{Component separation limits dependent on the velocity dispersion of the two velocity components. The crosses indicate the modelled data points}
\label{vel_separation}
\end{figure}

Separating the velocity components of a filament along the line-of-sight is a crucial part of this work. Therefore, the technique of identifying the velocity range of emission needs to be tested in a systematic way. We created a simulated data cube with an RMS noise per position of $\rm 1~K$, typical for SEDIGISM, and include two filaments at the same 2D location, with the same velocity dispersion and intensity, but with different peak velocities. For the emission of the filaments, we assume a Gaussian line profile. We then vary the peak-to-peak velocity, the linewidth, and the intensity and analyse these cubes in exactly the same way as the observed data. From this modelling, we find for filaments with a signal-to-noise $\rm SNR > 4$, which is typical for our \cco data (channel width $\rm 0.25~km\,s^{-1}$), that the minimal separated peak-to-peak velocity $(\delta v)_{\min}$ depends linearly on the velocity dispersion $\rm \sigma_v$ like, 
\begin{equation}
(\delta v)_{\min} =\rm 2 \sigma_v + 1~km\,s^{-1}   $   ,$
\label{eq-separation}
\end{equation}
shown in Fig. \ref{vel_separation}. Emission lines with a velocity dispersion of $\rm \sigma_v < 0.75~km\,s^{-1}$ (3 channels) are not identified as an emission line. For filaments with low intensities ($\rm SNR \leq 4$) and for different intensities these limits must be degraded by $\rm \Delta \sigma_v = 0.25~km\,s^{-1}$ . Note, as the identification is done on the average spectrum over the filament skeleton, the velocity dispersion of an emission line can be larger than the intrinsic value because of velocity gradients along the skeleton. As a result, we are not resolving the kinematic substructure, like fibres \citep[$\delta \text{v} \approx \rm 1.0~km\, s^{-1}$, ][]{Hacar2013}, but can determine the large scale kinematics of the filament.

From previous studies \citep{Schuller2017} we know that \cco is likely to be optically thick towards the densest regions, which might lead to effects like self-absorption, and could affect the separation method. As the abundance of \coo is lower by $\text{\cco} / \text{\coo} = 8.3$ \citep{Miettinen2012}, it is likely to be optically thin over the whole filament. During visual comparison of the \cco and \coo spectra of all 604 ATLASGAL clumps within the filaments (position and velocity) optically thick \cco emission is seen in 76 clumps, corresponding to $13\%$. However, these clumps do not show a significant effect on the average spectrum over the whole skeleton. Hence, our method is unlikely to separate velocity components because of self-absorption features. Additionally, we conclude that the contribution of the dense core emission is small when compared to the emission integrated up over the whole filament.

\begin{figure}
\centering
\includegraphics[width=0.5\textwidth, clip=true, trim= 0cm 0cm 1cm 0cm]{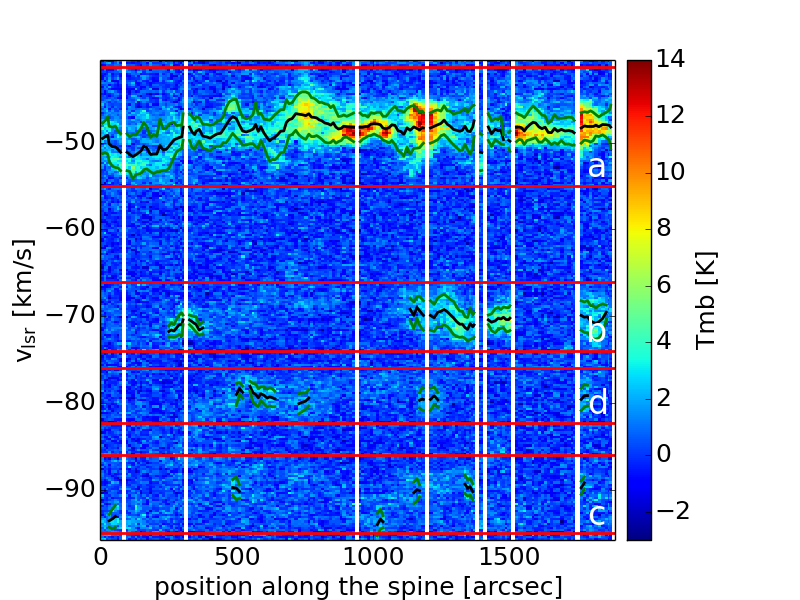}
\caption{Position-velocity plot of the intensity along the skeleton of the filament candidate G333.297+00.073. The white stripes indicate the beginning/end of a skeleton branch. The first five branches show the longest connection through the skeleton from higher to lower galactic longitude and the last four are the branches to the side in the same direction. The horizontal red lines mark the identified emission intervals shown in Fig. \ref{spectrum} with intervals a, b, d, c from top to bottom. The jagged black and green lines mark the per pixel measured peak velocity and the $1\sigma$ interval of the detected emission peak.}
\label{pv-plot}
\end{figure}

\begin{figure*}[tbh]
\centering
\begin{minipage}{0.48\textwidth}
\includegraphics[width=\textwidth, clip=true, trim= 1cm 4cm 3cm 2.0cm]{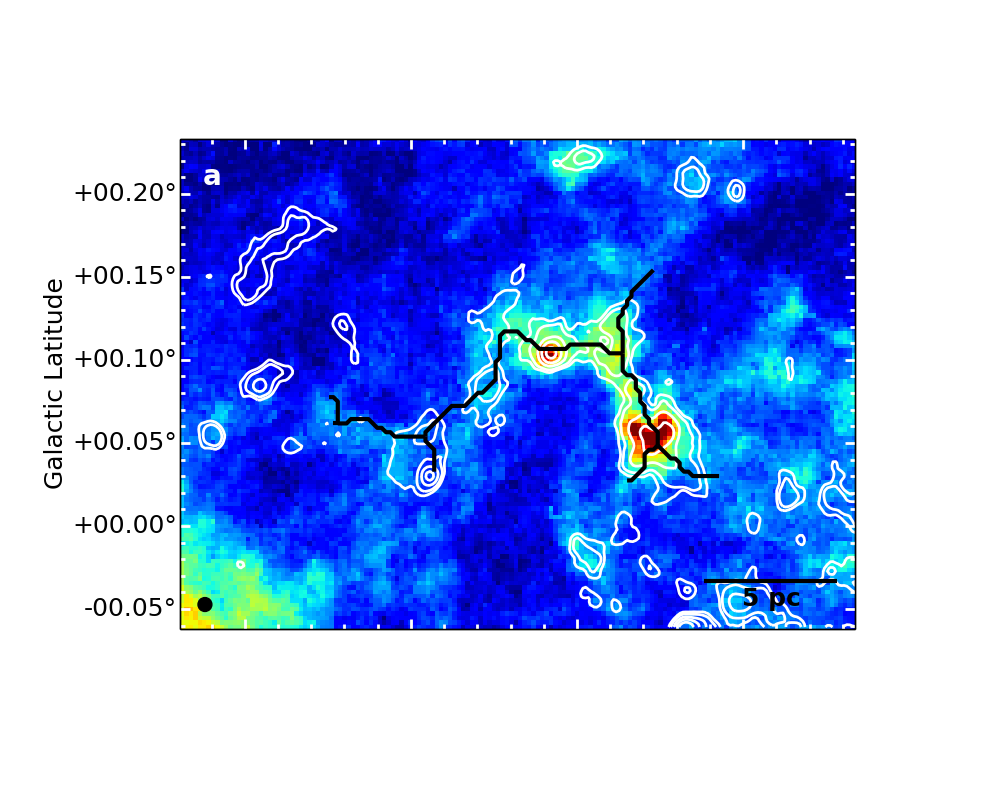}
\end{minipage}
\begin{minipage}{0.48\textwidth}
\includegraphics[width=\textwidth, clip=true, trim= 3cm 4cm 1cm 2.0cm]{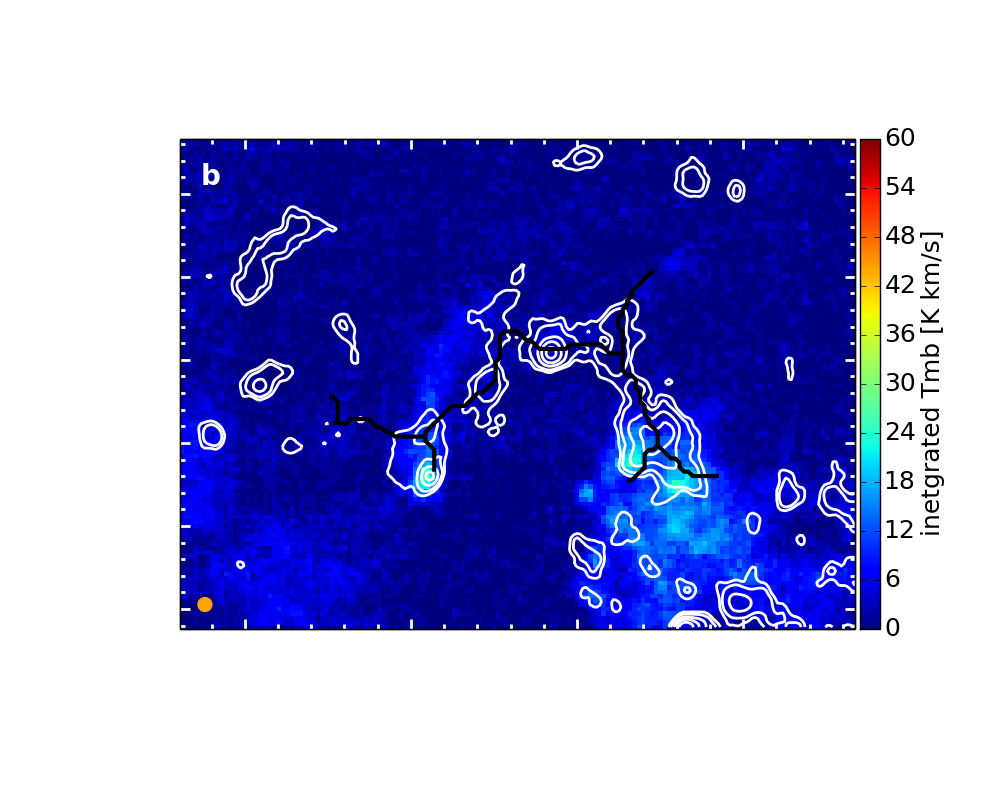}
\end{minipage}

\begin{minipage}{0.48\textwidth}
\includegraphics[width=\textwidth, clip=true, trim= 1cm 2cm 3cm 2.0cm]{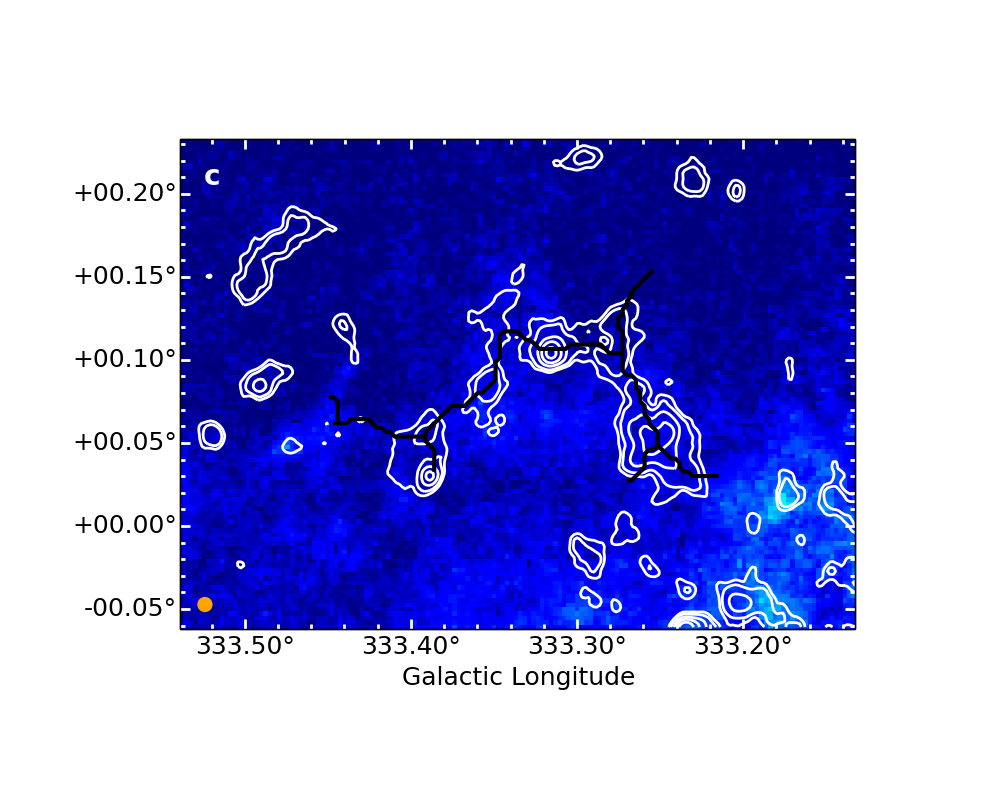}
\end{minipage}
\begin{minipage}{0.48\textwidth}
\includegraphics[width=\textwidth, clip=true, trim= 3cm 2cm 1cm 2.0cm]{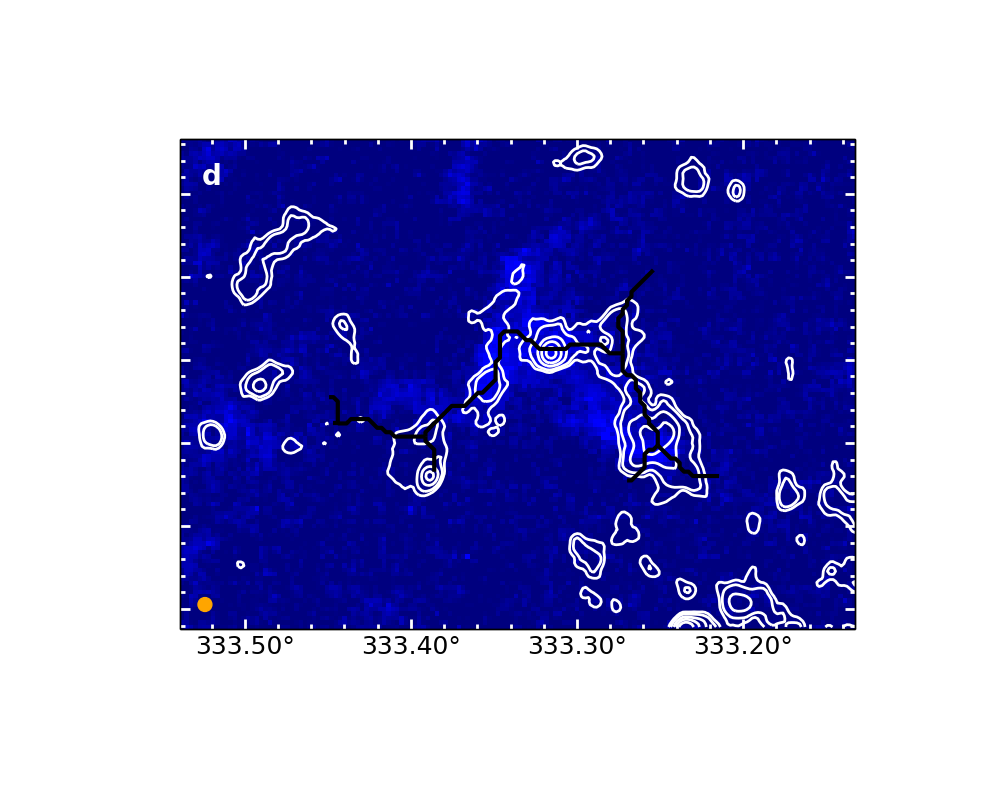}
\end{minipage}
\caption{Integrated \ccol emission of the four velocity components of the filament candidate G333.297+00.073. The intensity was integrated over the velocity ranges $\rm -55.0$ to $\rm -40.5~km\,s^{-1}$ (top left), $\rm -74.0$ to $\rm -66.0~km\,s^{-1}$ (top right), $\rm -96.0$ to $\rm -86.0~km\,s^{-1}$ (bottom left) and $\rm -82.5$ to $\rm -76~km\,s^{-1}$ (bottom right). The beamsize is indicated by the circle in the lower left and the contours show the ATLASGAL [0.1, 0.2, 0.5, 1.0, 2.0 ] Jy/beam levels. The letters in the upper right refer to the marked intervals in Fig. \ref{spectrum}.}
\label{integrated}
\end{figure*}

In the next step we use the same method as for the average spectrum on every spectrum/pixel along the skeleton within the identified velocity ranges. In this way we identify which part of the skeleton is detected in the different velocity components. Velocity components in which less than ten positions of the skeleton are detected in \cco are discarded, as these barely deviate from the noise, and we ensure a minimal elongated shape for all correlated structures (aspect ratio of 3 assuming a width of one beam). Additionally, we are able to detect multiple velocity components within the previously identified velocity range towards individual pixels. In the case where several velocity components are found, we only keep the calculated moments for the one with highest intensity. This is done for the \cco and \coo data. The separation of these subcomponents is limited by the smoothed velocity resolution. Also, we calculate the zeroth, first, and second `order' moments of skeleton pixels in each detected velocity component. With the first order measurements we potentially trace velocity gradients along the spine, but this is beyond the scope of this paper. However, the second order measurements, hence the velocity dispersion, of one pixel includes only the velocity gradient within one beam, which can be considered to be small. Therefore, the average over the skeleton pixels is a better estimate of the velocity dispersion of a filament than the value derived from the average spectrum and used in the further analysis. To check the results of these calculations we plot the derived moments on top of the position-velocity diagram of the filament candidate skeleton (Fig. \ref{pv-plot}). Additionally, we integrate over the velocity ranges of each velocity component; see Fig. \ref{integrated}. 

\subsection{Gas-dust correlation}
\label{correlation}

\begin{figure}[tbh]
\centering
\includegraphics[width=0.45\textwidth, clip=true, trim= 0.5cm 1cm 0cm 0cm]{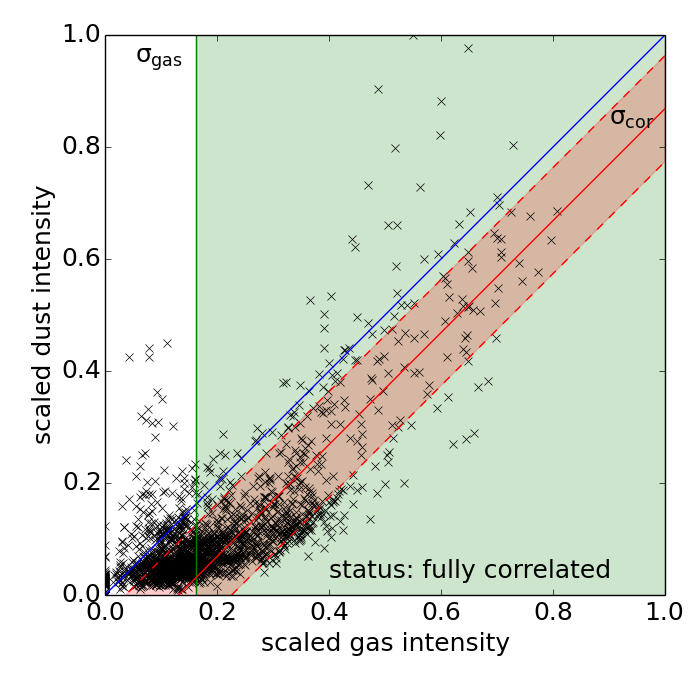}
\caption{Gas-dust correlation plot of the brightest velocity component (``a'' in the spectrum shown in Fig. \ref{spectrum}) of the filament candidate G333.297+00.073. The blue line gives the one-to-one correlation. The green area indicates values above the $\sigma_\text{gas}$ limit ($p_\text{gas}=0.51$). The red line shows the fitting result, and the area within the dashed red lines marks the $\pm \sigma_\text{cor}$ surrounding ($p_\text{cor}=0.80$). $p_\text{cor, gas}=0.70$ is estimated from the overlap of these areas.} 
\label{gas-dust-correlation}
\end{figure}

We further estimate how much each velocity component contributes to the overall dust emission from a given filament candidate. We smooth and re-grid the ATLASGAL maps, which were used for the candidate identification, to the resolution ($30.1\arcsec$) and pixel-grid ($9.5\arcsec$) of the SEDIGISM cubes. 
For comparison with the \cco integrated emission we restrict the maps (ATLASGAL and integrated \cco intensities per velocity component) to an area within a dilation box around the skeleton with a width of 3 beams (9 pixels), which covers the emission seen in ATLASGAL. We scale the ATLASGAL intensities with the minimum and maximum value to an interval of [$0$--$1$], using 
\begin{equation}
I^\text{dust}_s=\frac{I^\text{dust}-\min(I^\text{dust})}{\max(I^\text{dust})-\min(I^\text{dust})} $   ,$
\end{equation}
where $I^\text{dust}_s$ are the scaled ATLASGAL intensities, and $I^\text{dust}$ the original ones, and $\min()$ and $\max()$ describe the minimal and maximal value of the pixel intensities within the dilation box. As the ATLASGAL emission traces only the small scale high density gas, the minimun value is typically around $\rm 0~Jy/beam$. For the \cco data we define the intensity integrated over the velocity range of component $i$ as $I^\text{gas}(v_i)=\int_{v_i} T_\text{mb}(v)\, dv$, where $T_\text{mb}(v)$ is the main beam temperature, and $v_i$ is the velocity interval of one component as defined before. Also, for the correlation we do not apply any threshold. We use for the scaling the maximum and minimum value within the dilation box of the sum over the integrated intensity maps of all velocity components,
\begin{equation}
I^\text{gas}_s(v_i)=\frac{I^\text{gas}(v_i)-\min(\sum_i I^\text{gas}(v_i))}{\max(\sum_i I^\text{gas}(v_i))-\min(\sum_i I^\text{gas}(v_i))} $   ,$
\end{equation}
where $I^\text{gas}_s(v_i)$ are the scaled \cco intensities of the velocity component $i$, and $I^\text{gas}(v_i)$ are the original ones. Assuming a constant gas-to-dust ratio, the ATLASGAL maps should correlate with the molecular line emission integrated over all velocity components. We perform pixel-to-pixel correlation between the scaled ATLASGAL maps and the scaled integrated \cco maps of one velocity component using the same dilation box mask. Therefore, in cases of multiple components per candidate we will not find a one-to-one correlation, but we identify which velocity component shows the filamentary behaviour seen in dust emission. Additionally, noise in the observations and effects like CO depletion will affect the correlation plot. See Fig. \ref{gas-dust-correlation} and appendix \ref{cor_examples} for examples.

We analyse these correlation diagrams as follows. We calculate the standard deviation of the total (scaled) integrated intensities, $\sigma_\text{gas}=\sigma(\sum_i I^\text{gas}_s(v_i))$, presenting an upper limit of the noise in the gas emission, and the standard deviation of the correlation, $\sigma_\text{cor}=\sigma(I^\text{dust}_s-\sum_i I^\text{gas}_s(v_i))$. To estimate the intensity contribution of one velocity component, we perform a linear fit with a slope of $1$ on the data points with $I^\text{gas}_s(v_i) \geq \sigma_\text{gas}$. We then derive the percentage of data points, which are above the $\sigma_\text{gas}$ noise , $p_\text{gas}$, which are within $\pm \sigma_\text{cor}$ from the linear fit (red in Fig. \ref{gas-dust-correlation}), $p_\text{cor}$, and which meet both conditions, $p_\text{cor, gas}$. We then use these three parameters to characterize the different velocity components with the limiting values shown in Table \ref{parameter limits}. 

\begin{table}[tbh]
\caption{Limiting characterization parameters $p_\text{gas}$, $p_\text{cor}$, and $p_\text{cor, gas}$ as described in Section \ref{correlation}. These parameters describe the population of three different areas in the correlation plots, see Fig. \ref{gas-dust-correlation} and appendix \ref{cor_examples}.}
\label{parameter limits}
\centering
\begin{tabular}{cccc}
\hline\hline
Status & $p_\text{gas}$ & $p_\text{cor}$ & $p_\text{cor, gas}$ \\
\hline 
fully correlated & $\geq 0.4$ &  & $\geq 0.4$ \\ 
partially correlated & $\geq 0.05$ &  & $\geq 0.4$ \\ 
diffuse correlated & $< 0.05$ & $\geq 0.4$ &  \\ 
uncorrelated (1) & $\geq 0.05$ &  & $< 0.4$\\ 
uncorrelated (2) & $< 0.05$ & $< 0.4$ &  \\ 
\hline
\end{tabular}

\end{table}

The limiting values for the characterization were obtained from the manual analysis of a test sample of filament candidates. This introduces a slight bias to the general analysis. The uncertainty of the characterization is difficult to determine, as the filament definition given by \cite{Li2016b} can not be applied in a systematic way. Therefore, to get an objective, reproducible result, we decided to use this quantitative characterization over a visual approach as used by \cite{Li2016b}. A later visual inspection and different characterization will be possible, as all velocity components of all filament candidates, which are not uncorrelated, are handled in the same way in the subsequent analysis. Therefore, the characterization bias alters only the statistics of the characterization itself. However, the reliability of a filamentary shape decreases from fully correlated over partially correlated to diffuse structures, where partially correlated structures might not be continuous in position-position space, and diffuse structures might not show clearly enhanced emission from the surrounding. For simplicity we will still refer to all correlated structures as filaments. In total, we identified 422 filaments within the 283 filament candidates. More statistics of the characterization will be presented in Section \ref{finalcat-sec} and \ref{multi-sec}.

\subsection{Thermal and non-thermal motions}
\label{motions}

The unresolved kinematic motions in a molecular cloud can be estimated by the observed linewidth. The total velocity dispersion can be separated into a thermal and a non-thermal component. The thermal motions depend on the observed molecule and gas temperature. For the molecular gas temperature we assume a typical value of $T=\rm 15~K$ \citep{Pillai2006a,Wang2012,Wang2014}, which is also in agreement with measured temperatures of ATLASGAL clumps \citep{Urquhart2018}. The non-thermal motions describe statistical motions of the gas, which are independent of the kinetic temperature of the the gas. The non-thermal velocity dispersion, $\sigma_\text{nt}$, is given by
\begin{equation}
\sigma_\text{nt}^2 = \sigma_\text{x}^2-\frac{k_B T}{m_\text{p} m_\text{x}}  $ ,$
\end{equation}
where $\sigma_\text{x}$ is the measured second moment of \cco or $\rm C^{18}O$, $k_B$ the Boltzmann constant, $m_\text{p}$ is the proton mass,  
and $m_\text{x}$ is the molecular weight of the observed gas, here $m_\text{\cco}=29$ and $m_\text{\coo}=30$.
To derive the non-thermal velocity dispersion of the filament, $\sigma_\text{nt}$, we average the measurements of each pixel along the skeleton, where we choose \coo if it is detected and \cco otherwise. Therefore, this value is independent of velocity gradients along the skeleton, neglecting gradient effects within the beam, and it is less likely to be influenced by optical depth effects. 

The thermal motion of the interstellar medium is given by the sound speed,
\begin{equation}
c_s = \frac{k_B T}{m_\text{p} \mu}   $   ,$
\label{sound_speed_equ}
\end{equation}
where $\mu=2.8$ is the mean molecular weight of the mean free particle \citep{Kauffmann2008}, and other parameters as previously defined. The total velocity dispersion of a filament is given by $\sigma_\text{v} = \sqrt{c_s^2 + \sigma_\text{nt}^2}$.

\subsection{Mass and length of filaments}
\label{mass-sec}

To calculate physical parameters of the 422 filaments it is crucial to estimate the distance towards them. However, estimating distances towards extended and diffuse structures, like these filaments, is difficult. Especially, solving the ambiguity of kinematic distances. Therefore, we use a method similar to that discussed by \cite{Li2016b}, but including additional measurements.

As a first step we identify all ATLASGAL clumps \citep{Contreras2013,Urquhart2014} associated with the filaments. The distances towards these clumps have been estimated in \cite{Urquhart2018}. As these estimates are based on kinematic distances, we must exclude the Galactic centre region ($|\ell| < 5\degree$), because of the large uncertainties. For filaments with an associated clump within the defined limits of the filament in position-position-velocity space) we simply assume the same distance. This provides distances for 222 filaments. In a second step we use friends-to-friends analysis to find adjacent clumps and adopt their distances for the filaments. This adds distances for another 114 filaments, but note the larger uncertainty for the distance estimate. For the friends-of-friends analysis we allow sources with a spatial offset of at most $10\arcmin$ (with $90\%$ closer than $5\arcmin$) and a kinematic offset of at most $\rm 10~km\, s^{-1}$ (with $90\%$ closer than $\rm 4~km\, s^{-1}$). In total, we are able to assign a distance to 336 out of 422 filaments, including diffuse, partially correlated, and fully correlated ones. Additionally, we tested these estimates to be in agreement with (one of) the kinematic distances.

After we obtained the distances towards the filaments we can calculate their physical length. Here we take all pixels along the skeleton into account, towards which \cco was detected for the single filament. This allows us to get accurate measurements of the angular length for complex structures or partially correlated structures, including the detected branches, by adding up the length over each relevant pixel. However, because of a possible inclination of the structures, the derived physical measurements represent lower limits to their true length. 

For calculating the area and the mass of the filaments we again use a dilation box. With the distances in hand we are able to use a physical box-diameter. Here we take the box-diameter as a free parameter and in Section \ref{Filament-profile} we discuss the dependency between the filament mass and box-diameter. For calculating the filament mass we assume its diameter to be of the order of the star-forming size scale. For a first order approximation of the star-forming size scale we use the previously measured velocity dispersion within a filament, $\rm \sigma_v$, and assume a star formation time of $T_\text{sf}=\rm 2~Myr$  \citep{Evans2009}. The size scale is then given by $s_\text{sf}= \sigma_\text{v} \cdot T_\text{sf}$. As box-diameters are limited to discrete multiples of pixels, we interpolate linearly the values from measurements of the next bigger and smaller box size.

We estimate the area covered by the filament, by summing over all pixels within the dilation box of the integrated intensity maps, within which we detect \cco emission (SNR $> 5$). The same positions are used to estimate the filament mass. Here we follow two different approaches: First we use the integrated \cco emission, $W({\rm ^{13}CO})$, in combination with the \cco $X$-factor, $X_{\rm ^{13}CO(2-1)}=1^{+1}_{-0.5} \times 10^{21} \rm cm^{-2} (K\,km\,s^{-1})^{-1}$ derived by \cite{Schuller2017}. This has the advantage of tracing only the emission within the specific velocity component. The molecular hydrogen column density, $N_i(\rm H_2)$, in pixel $i$ was then calculated by $N_i(\text{H}_2)=W_i({\rm ^{13}CO})  X_{\rm ^{13}CO(2-1)}$. We then computed the mass using the equation
\begin{equation}
M(\text{H}_2) = \sum_i N_i(\text{H}_2) A_i \mu m_\text{p}   $   ,$
\label{gas-mass-equ}
\end{equation}
where $N_i(\rm H_2)$ is the $\rm H_2$ column density computed for pixel $i$, $A_i$ its area, $\mu =2.8$ the mean molecular weight per $\rm H_2$ molecule, and $m_p$ the proton mass.

Second, we estimate the mass from dust emission maps of different surveys (ATLASGAL, ATLASGAL+PLANCK) using basic assumptions like a gas-dust ratio of $R=100$, and a dust temperature of $T_D= 15\, \rm K$ \citep{Urquhart2018}. The mass of the filament candidate is then computed through the equation
\begin{equation}
M_\nu (\text{H}_2)=\frac{S_\nu \, d^2 R}{B_\nu(T_D) \kappa_\nu}   $   ,$
\label{dust-mass-equ}
\end{equation}
where $S_\nu$ is the integrated flux density at the frequency $\nu$ of the used survey, $d$ is the distance towards the structure, $B_\nu(T_D)$ is the Planck function at the given dust temperature, and $\kappa_\nu$ is the dust absorption coefficient, which is $\kappa_{\rm 870\mu m} = 1.85 \rm \, cm^2\,g^{-1}$ for the ATLASGAL emission \citep{Schuller2009,Ossenkopf1994}. 

However, because of the contribution of the PLANCK data, the ATLASGAL+PLANCK survey traces not only the filament and the low column density gas around the filament, but also the diffuse Galactic dust emission, which ideally should be removed. To do so, we exclude the filament area of the maps using the inverse of the filament dilation box used previously. The remaining pixels should be dominated by non-filament emission. However, as the box has a width of three beams, we find a few cases in the most nearby ($\rm < 2 kpc$) filaments where the emission extends clearly beyond the mask. Therefore, we use the 20th percentile value of the non filament pixels as estimate of the diffuse Galactic dust emission. We then correct the ATLASGAL+PLANCK maps for the diffuse emission and estimate the masses as shown previously. In Section \ref{gas-dust} we discuss the differences of the these three mass estimates based on dust continuum emission and compare them to the \cco emission estimate.

\section{Results}
\label{results}
\subsection{Final Catalogue}
\label{finalcat-sec}
Using the methods described in the previous section we derived a large set of filament parameters. With these parameters we created a catalogue of velocity coherent structures. The derived parameters of the catalogue are shown in Table \ref{cat-example}, and the complete catalogue is shown in Tables \ref{final-cat1} and \ref{final-cat2}. However, as several parameters are distance dependent, they cannot be calculated for the whole filament catalogue. The following description of the derived parameters includes only the structures with a distance estimate. The catalogue contains all 422 filaments of the 283 observed filament candidates, which show some correlation with the dust emission. In Table \ref{fil_statistics_tab} we show the statistics of the characterization and the subsample for which we have distance estimates. The names of the structures are based on the initial filament candidate name from \cite{Li2016b} and are extended by an integer starting from $0$, indicating the velocity component. 

\begin{table*}[tbh]
\caption{Descriptions of the derived parameters. The complete catalogue is shown in Tables \ref{final-cat1} (top) and \ref{final-cat2} (bottom).}
\label{cat-example}
\centering
\begin{tabular}{ccl}
\hline\hline
Parameter & Unit & Description \\
\hline
\multicolumn{3}{c}{\textit{Table of measured parameters}}\\
Filament ID &  &  \\
$\ell$ & $\degree$ & Galactic longitude of the centre of the filament\\
$b$ & $\degree$ & Galactic latitude of the centre of the filament\\
Status &  & correlation with the ATLASGAL emission \\
$N_\text{c}$ &  & number of detected velocity components in the original filament candidate \\
$v_\text{lsr}(\text{\cco})$ & $\rm km\,s^{-1}$ & peak velocity derived from the \cco average spectrum \\
$v_\text{lsr}(\text{\coo})$ & $\rm km\,s^{-1}$ & peak velocity from the \coo average spectrum \\
$\sigma(v_\text{\cco})$ & $\rm km\,s^{-1}$ & dispersion of the \cco peak velocities along the skeleton \\
$\sigma(v_\text{\coo})$ & $\rm km\,s^{-1}$ & dispersion of the \coo peak velocities along the skeleton \\
$\sigma_\text{v}$ & $\rm km\,s^{-1}$ & average total velocity dispersion along the skeleton (derived from \cco and \coo) \\
$\sigma_\text{v}(\text{\cco})$ & $\rm km\,s^{-1}$ & average \cco velocity dispersion along the skeleton \\
$\sigma_\text{v}(\text{\coo})$ & $\rm km\,s^{-1}$ & average \coo velocity dispersion along the skeleton \\
$\sigma_\text{v,t}(\text{\cco})$ & $\rm km\,s^{-1}$ & \cco velocity dispersion derived from the average spectrum \\
$\sigma_\text{v,t}(\text{\coo})$ & $\rm km\,s^{-1}$ & \coo velocity dispersion derived from the average spectrum \\
\hline
\multicolumn{3}{c}{\textit{Table of derived parameters}}\\
Filament ID &  &  \\
$d$ & kpc & distance from the Sun \\
$l$ & $\degree$ & angular length of the detected skeleton \\
$l(d)$ & pc & physical length of the detected skeleton \\
$M(\text{ATG})$ & $\rm M_\odot$ & filament mass derived from ATLASGAL emission \\
$M(\text{ATG+P})$ & $\rm M_\odot$ & filament mass derived from ATLASGAL+PLANCK emission \\
$M(\text{dust})$ & $\rm M_\odot$ & filament mass derived from corrected ATLASGAL+PLANCK emission \\
$M(^{13}\text{CO})$ & $\rm M_\odot$ & filament mass derived from integrated \cco emission \\
$m_\text{crit,nt}$ & $\rm M_\odot \, pc^{-1}$ & critical, non-thermal line-mass \\
$m_\text{obs}$ & $\rm M_\odot \, pc^{-1}$ & observed line-mass \\
det. \cco &  & fraction of the skeleton detected in \cco \\
det. \coo &  & fraction of the skeleton detected in \coo \\
edge flag &  & skeleton truncated because of the edge of SEDIGISM \\
$d$ flag &  & indicating the method for the distance estimate: 0 no distance; 1 inside ATLASGAL source;\\
 & & 2 nearby ATLASGAL source \\
\hline
\end{tabular}
\end{table*}

\begin{table}[tbh]
\caption{Number of sources (with distance estimate) separated in different groups.}
\label{fil_statistics_tab}
\centering
\begin{tabular}{lcc}
\hline\hline
Groups & total & with dist. \\
\hline
filament candidates \citep{Li2016b} & $517$ & \\
in SEDIGISM area & $283$ & \\
$\geq 1$ correlated velocity component & $260$ & \\
\hline
total velocity components & $812$ & $336$ \\
uncorrelated components & $390$ &  \\
filaments & $422$ & $336$ \\
\hline
fully correlated filaments & $180$ & $151$ \\
partially correlated filaments & $191$ & $148$ \\
diffuse component filaments & $51$ & $37$ \\
\hline
\end{tabular}
\end{table}

\subsection{Detection of filaments in $\rm ^{13}CO$ and $\rm  C^{18}O$}
\begin{figure}[tbh]
\centering
\includegraphics[width=0.5\textwidth]{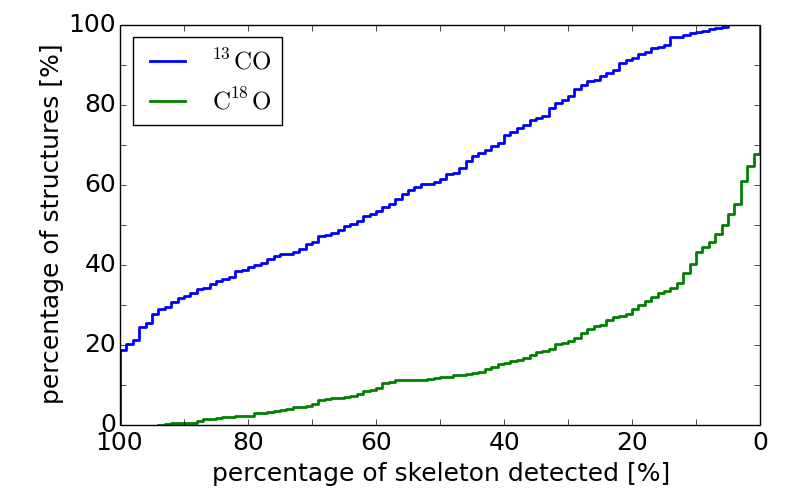}
\caption{Cumulative histogram of the percentage of filament candidate skeletons detected in \cco (blue) and \coo (green).}
\label{detection_cum}
\end{figure}

Out of the $283$ ATLASGAL filament candidates within the SEDIGISM survey we detect correlated \cco emission for $260$ filament candidates, which then show $422$ velocity coherent (continuous kinematic structure, which cannot be resolved into separate components) filaments. We do not find a correlated \cco velocity component for $23$ filament candidates, which is partially because of the sensitivity of the SEDIGISM survey, and partially because the candidates result from line-of-sight alignments of diffuse gas clouds. About $20 \%$ of the detected filaments show \cco emission at every position of the skeleton and for about $60 \%$ we detected \cco emission over at least half of the length of the skeleton. About $32 \%$ of the \cco detected filaments show no detection of \coo on the skeleton, about $13 \%$ have \coo detected over at least half of the skeleton, and for no filament C$^{18}$O is detected over its entire length; see Fig. \ref{detection_cum}.

This difference in the detection rate is very likely due to the different abundances of the molecules \citep[$\rm ^{13}CO / C^{18}O = 8.3$; ][]{Miettinen2012}. The \coo line is expected to be weaker, resulting in a lower signal-to-noise ratio and the observed lower detection rate.

\subsection{Galactic distribution}

\begin{figure}[h]
\centering
\includegraphics[width=0.5\textwidth,trim={1cm 0 1.5cm 0cm},clip]{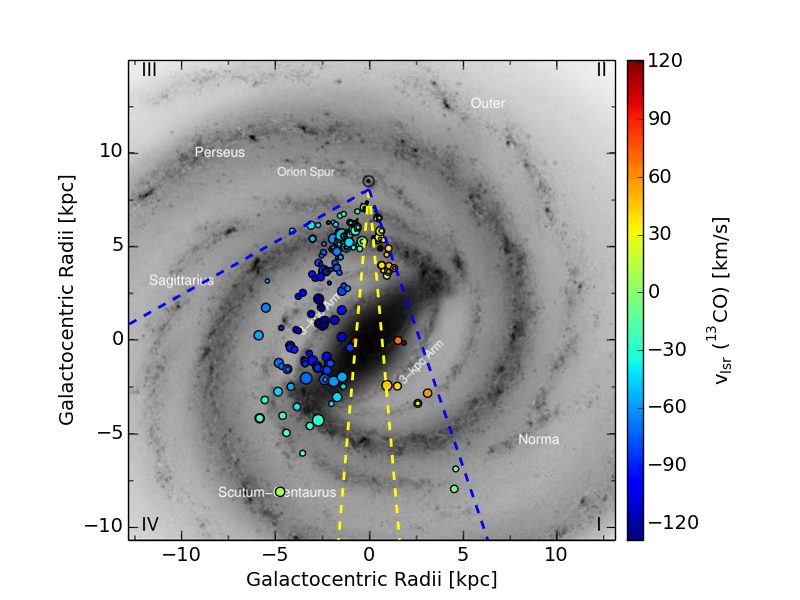}
\caption{The filaments with distance estimated are plotted onto an artist's impression of the Milky Way (Robert Hurt), the size is indicating the length of the filament, and the colour indicates the \cco velocity. The yellow lines mark the range $|\ell| \leq 5\degree$, where distances are uncertain and the blue lines mark the SEDIGISM survey limits.}
\label{galaxy}
\end{figure}

Using the distance estimates we can derive the Galactocentric coordinates and plot the positions onto a face-on artist's impression of the Milky Way (Fig. \ref{galaxy}). We find that a large fraction of the filaments are likely to be associated with the near Scutum-Centaurus arm. We also find some filaments located in the near Sagittarius arm, the near and far 3-kpc arm, and the near Norma arm, but also in some inter-arm regions. Note, we do not have distance estimates for the Galactic Centre region ($|\ell| \leq 5\degree$).

\begin{figure}[h]
\centering
\includegraphics[width=0.5\textwidth]{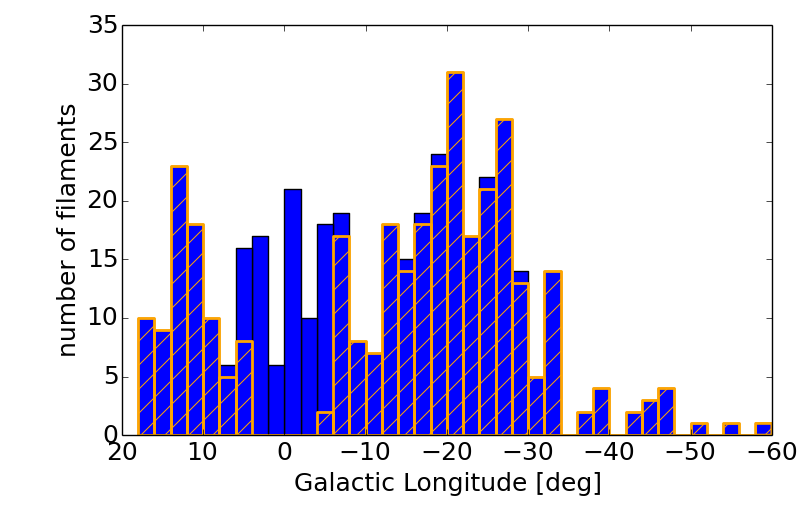}
\caption{Distribution of the filament positions in Galactic Longitude. The orange hatch marks the filaments with distance estimate.}
\label{gl}
\end{figure}

\begin{figure}[h]
\centering
\includegraphics[width=0.5\textwidth]{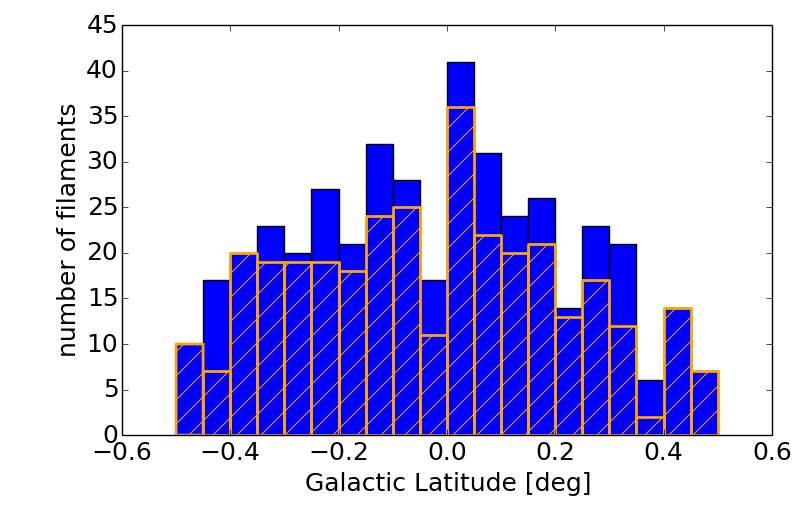}
\caption{Distribution of the filament positions in Galactic Latitude. The orange hatch marks the filaments with distance estimate.}
\label{gb}
\end{figure}

Histograms representing positions of the detected filaments with Galactic longitude and latitude are shown in Figs: \ref{gl} and \ref{gb}, respectively. The distribution with Galactic longitude shows a peak around $\rm \textit{l} = -21 \degree$ with a strong decrease towards the outer Galaxy (only $14$ structures for $\rm \textit{l} < -45 \degree$), and a decrease towards the Galactic centre. As the filament candidates were identified in the ATLASGAL maps that trace only high column density dust emission, it is more unlikely to find filaments towards the outer Galaxy, which contains fewer dense molecular cloud regions. However, the number of filaments is also suppressed in the direction of the Galactic centre, where identification is difficult because many structures along the line-of-sight are confused, and were categorized as networks, complexes or unclassified, such as the Galactic centre region \citep{Li2016b}. Nevertheless, comparing the distribution of filaments to the distribution of ATLASGAL clumps presented by \cite[][Fig. 3]{Beuther2012} and \cite[][Fig. 16]{Csengeri2014} shows similarities for the location of peaks, indicating a possible correlation with active star-forming regions.

The distribution of the detected filaments with Galactic latitude (Fig. \ref{gb}) shows a broad, almost flat behaviour similar to the findings of \cite{Li2016b}. However, we find the peak and mean ($\rm \langle \textit{b}\rangle = +0.02 \degree$) of the distribution aligned with the Galactic mid-plane, which is in contrast to the general finding of more sources for $\rm \textit{b} < 0.0 \degree$ than for $\rm \textit{b} > 0.0 \degree$ \citep{Beuther2012, Li2016b}. Note, our sample is not identical with that of \cite{Li2016b}, as we use only a sub-sample of their filament candidates and split some of these candidates in different velocity components, hence filaments.

\subsection{Distributions of velocity dispersion, mass, length and distance}
\label{properties}
In the following, we give a short overview on the most interesting measured properties of the filaments, which are the non-thermal velocity dispersion along the skeleton, the mass derived from the \cco emission, and the length of the filament, which we define as the sum over the detected parts of the skeleton.

\begin{figure}[tbh]
\centering
\includegraphics[width=0.5\textwidth]{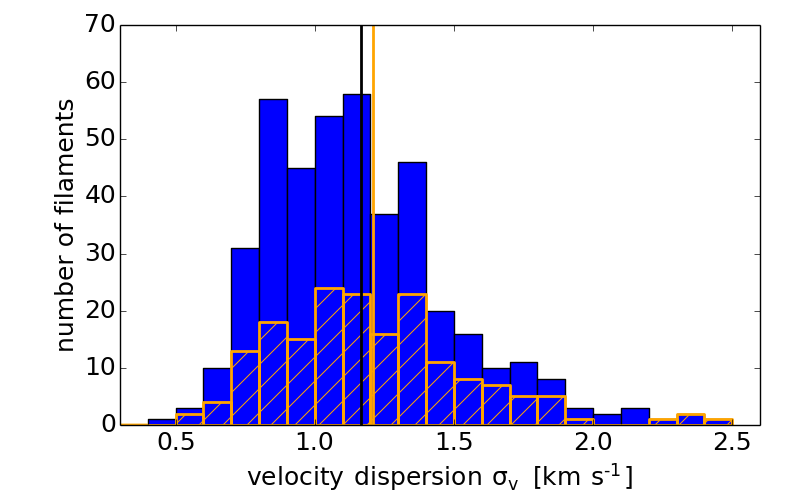}
\caption{Distribution of the measured total velocity dispersion of all filaments (blue). The fully correlated filaments are marked by the orange-coloured hatched area. The vertical lines indicate the mean value for the complete (black) and the sub-sample (orange).}
\label{sigma_hist}
\end{figure}

For the calculation of the total velocity dispersion we assumed an isothermal medium of $\rm 15~K$, see Section \ref{motions}. The distribution of the resulting values is shown in Fig. \ref{sigma_hist}. We find values reaching from about $\rm 0.5~km\, s^{-1}$ to $\rm 2.5~km\, s^{-1}$ with a relatively flat center between $\rm 0.8~km\, s^{-1}$ and $\rm 1.4~km\, s^{-1}$, and a mean of about $\rm 1.17~km\, s^{-1}$. Concentrating on the 180 fully correlated, hence, the most reliable filaments, we find a similar distribution with the mean at about $\rm 1.20~km\, s^{-1}$. In general, these values are higher than what \cite{Arzoumanian2013} find in nearby filaments ($\sigma_\text{v} \approx \rm 0.3~km\, s^{-1}$), but in agreement with studies of similar (more distant and more massive) objects like the DR21 filament \citep{Schneider2010}.

\begin{figure}[tbh]
\centering
\includegraphics[width=0.5\textwidth]{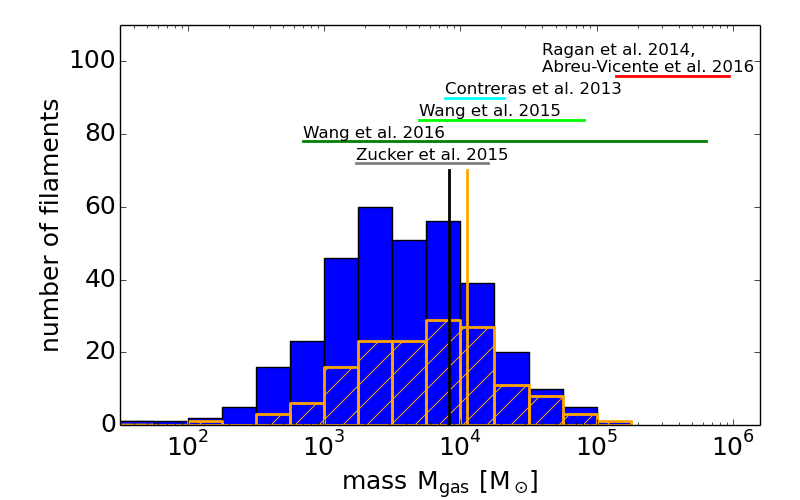}
\caption{Distribution of the measured mass based on \cco of all filaments (blue). The fully correlated filaments are marked by the orange hatch. The vertical lines indicate the mean value for the complete (black) and the sub-sample (orange). The horizontal lines mark the mass ranges measured by the studies mentioned above the lines.}
\label{mass_hist}
\end{figure}

For the logarithmic distribution of the calculated masses, Fig. \ref{mass_hist}, we find a flat part between $\rm 1800~M_\odot$ and $\rm 18000~M_\odot$ with a mean mass of $\rm 8600~M_\odot$. Again the distribution of the fully correlated filaments (151 with distance estimate) is similar to the complete distribution with a mean of $\rm 11000~M_\odot$. For comparison we also show the mass ranges covered by other filament catalogues \citep{Ragan2014,Wang2015,Zucker2015,Abreu-Vicente2016,Wang2016} and the study of \cite{Contreras2013a}. These studies report filaments with similar or higher masses. However, several of these studies have tried to identify the largest structures in the Galaxy and therefore, are biased to larger structures. As a result, some filaments mentioned in this study are only parts of structures in the other catalogues. Also, we find filaments that are almost identical in several catalogues, like G11.046-00.069\_2 (Snake) or G332.370-00.080\_1.

\begin{figure}[tbh]
\centering
\includegraphics[width=0.5\textwidth]{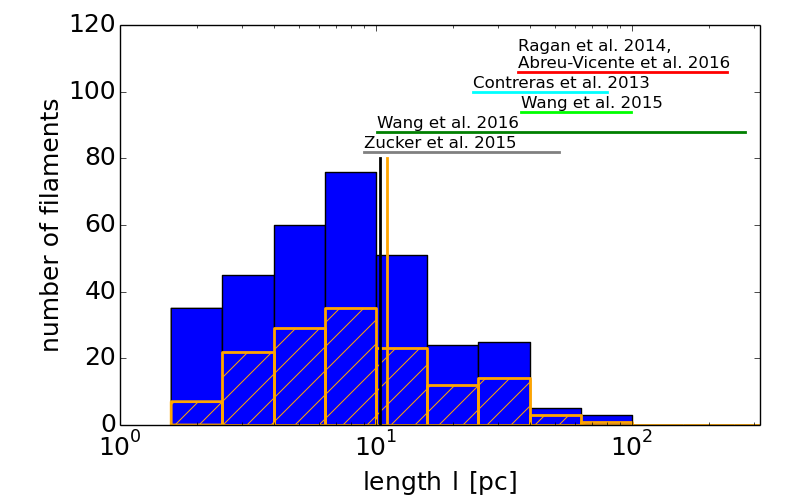}
\caption{Distribution of the measured length over the detected skeleton of all filaments (blue). The fully correlated filaments are marked by the orange hatch. The vertical lines indicate the mean value for the complete (black) and the sub-sample (orange). The horizontal lines mark the length ranges measured by the studies mentioned above the lines.}
\label{length_hist}
\end{figure}

We also find overlap between the catalogues for the lengths of the filaments (Fig. \ref{length_hist}), where the shortest filaments of the other studies are as long as the mean of our sample ($\rm 10.3~pc$ all, $\rm 11.1~pc$ fully correlated). In general, we cover the range from $\rm 2~pc$ to $\rm 100~pc$ with a peak around $\rm 8~pc$.

\begin{figure}[tbh]
\centering
\includegraphics[width=0.5\textwidth]{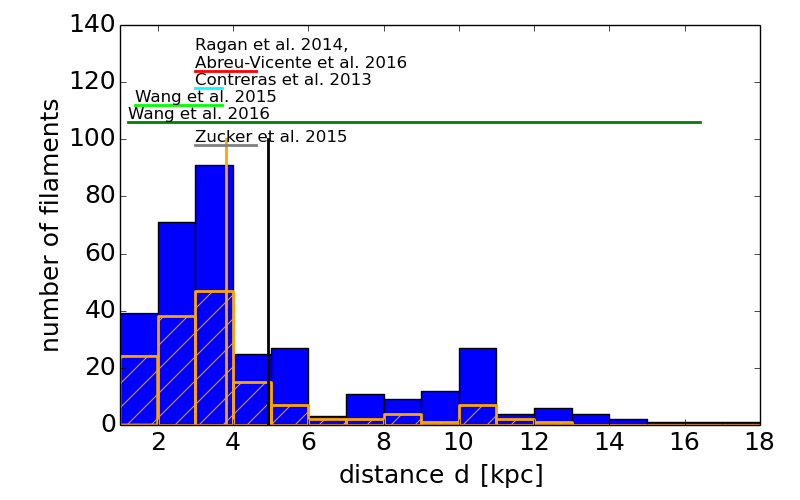}
\caption{Distribution of the estimated distances of all filaments (blue). The fully correlated filaments are marked by the orange hatch. The vertical lines indicate the mean value for the complete (black) and the sub-sample (orange).}
\label{dist_hist}
\end{figure}

Most filaments are found within $\rm 5~kpc$ from the Sun (Fig. \ref{dist_hist}), which is also the area where the other surveys found the long filaments. This area includes parts of the nearby Sagitarius and Scutum-Centaurus spiral arms. Another peak in the distance distribution is found at around $\rm 10~kpc$, which is about the distance of the connection point of the Galactic bar with the Perseus spiral arm (see also Fig. \ref{galaxy}).

\begin{figure}[tbh]
\centering
\includegraphics[width=0.5\textwidth]{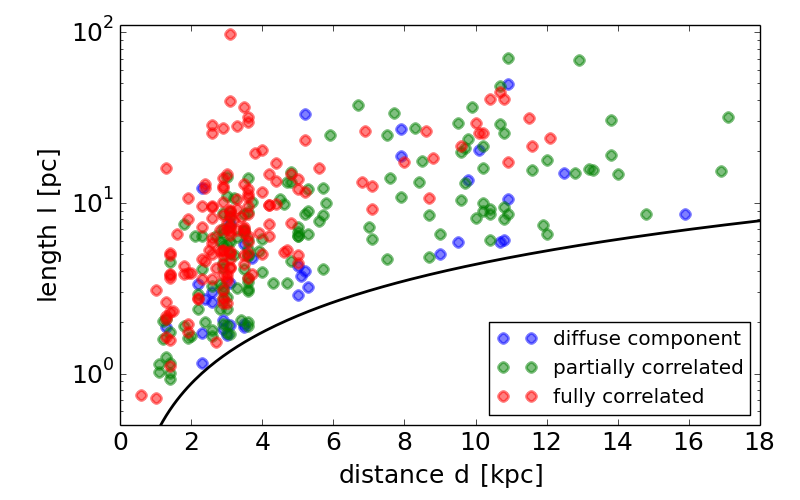}
\caption{Filament length plotted against the estimated distance. The three filament categories are indicated by blue, green, and red for diffuse component, partially correlated, and fully correlated, respectively. The black line shows the criteria of a minimum length of 10 pixels.}
\label{dist-length}
\end{figure}

Plotting the filament lengths against the estimated distances separated by the categories (Fig. \ref{dist-length}), we find that especially the fully correlated filaments follow the distance distribution, while the others are more equally distributed. Also, we find no correlation between the longest filaments and the distance. This results in a larger scatter of lengths for a given distance. The shortest filaments reproduce our minimal length criteria of at least 10 pixels.

\subsection{$\rm ^{13}CO$ -- $\rm  C^{18}O$ velocity comparison}
\begin{figure}[tbh]
\centering
\includegraphics[width=0.5\textwidth]{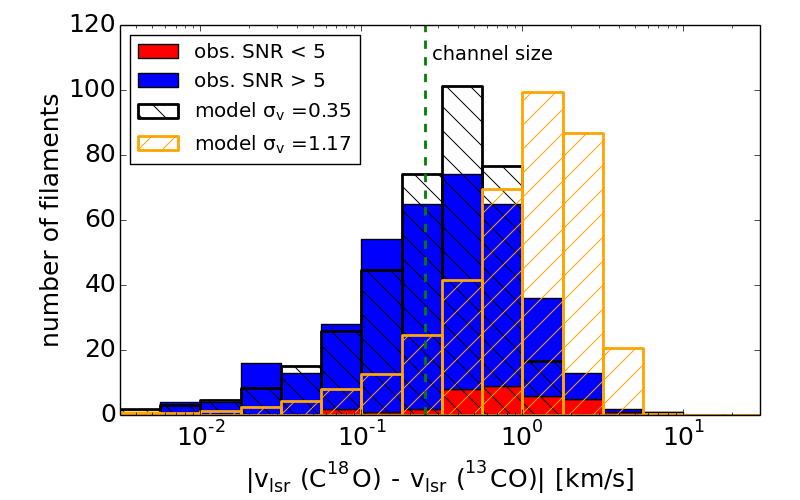}
\caption{Histogram of the absolute velocity difference. Sources with $\rm SNR < 5$ are shown in red. On top the model difference distributions given by a underlying Gaussian velocity distribution with a dispersion of $\rm 1.17~km\,s^{-1}$ (orange) and $\rm 0.35~km\,s^{-1}$ (black) are shown. The green dashed line indicates the velocity channel size of the data.}
\label{velocity-dif}
\end{figure}

As mentioned before, \coo is less abundant than \cco and traces mainly the bright, dense parts of the filaments, where \cco is likely to be optically thick. However, to combine the kinematics of the two lines we need to be sure that both trace the same gas. Therefore, Fig. \ref{velocity-dif} shows the distribution of the absolute difference between the \cco and \coo peak velocities of each filament derived from the average spectrum along the full skeleton, which is supposed to be zero if both istotopologues trace the same gas. The logarithmic distribution shows a plateau between $\rm 0.17~km\, s^{-1}$ and $\rm 1.0~km\, s^{-1}$ and decreases steeply to both sides. Additionally, we find the largest difference in filaments with a signal-to-noise of the \coo average spectra of $\rm SNR < 5$. In general, low SNR filaments peak at higher velocity differences (red).

We compare the observed distribution to a model distribution given by the mean velocity dispersion along the filament $\rm \overline{\sigma}_v = 1.17~km\,s^{-1}$ (see Section \ref{properties} and Fig. \ref{sigma_hist}). Given the wide distribution of velocity dispersions this model gives only a first order impression of the expected distribution. We model the absolute difference between two velocities drawn from two Gaussian distributions. The dispersion of the differences is then given by $\sigma_{\delta v}\rm =\sqrt{\overline{\sigma}_v^2 + \overline{\sigma}_v^2}=\sqrt{2}\,\overline{\sigma}_v$. We generate an artificial difference distributions, using $10,000$ draws to avoid statistical noise, bin the absolute values of the sample like the observed differences, and scale the height by $0.0373$ to get a comparable total number of filaments as our sample. The resulting distribution (orange hatched in Fig. \ref{velocity-dif}) does not agree with the observed one, as it is shifted to larger differences. 

To further investigate this trend, we reduced the dispersion of the underlying velocity distribution until we found a distribution that matched the differences (black hatched) area. Its velocity dispersion is $\rm \sigma_v(model) = 0.35~km\,s^{-1}$, which is about $\sqrt{2}$ times the channel width ($\rm 0.25~km\,s^{-1}$). We speculate therefore, that this distribution is likely to arise from the sampling of the spectra. 

\begin{figure}[tbh]
\centering
\includegraphics[width=0.5\textwidth]{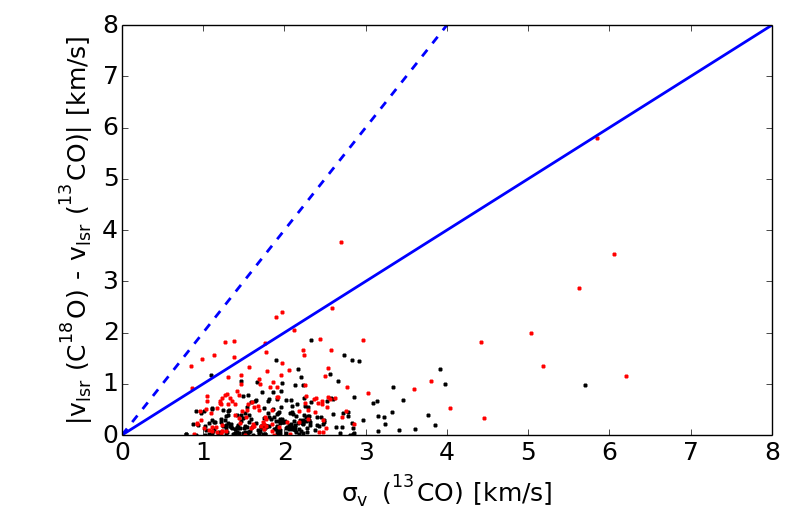}
\caption{Plot of the absolute velocity difference against the velocity dispersion of the filament, where filaments with a \coo detection rate below $10\%$ are indicated in red. The blue lines show the one-to-one ($1\sigma$, solid) and two-to-one ($2\sigma$, dashed) relations.}
\label{velocity-dis}
\end{figure}

However, we also see some filaments that show a larger difference between the \cco and \coo peak velocities than can be expected by the channel-width introduced sampling issues. For these filaments we speculate that they show a gradient along the skeleton and are only partially detected in C$^{18}$O. To investigate this, we plot the velocity difference against the \cco velocity dispersion of the average spectrum (Fig. \ref{velocity-dis}), as gradients along the skeleton result in a higher velocity dispersion. Additionally, filaments for which less than $10\,\%$ of the skeletons are detected in \coo are marked in red. We find that almost all filaments fall below the one-to-one line and that all velocity differences are smaller than $\rm 2 \sigma_v (^{13}CO)$. We also see that 31 out of 43 filaments with a velocity difference larger than $\rm 1~km\,s^{-1}$ show low \coo detection rates.

In summary we rule out systematic differences between the kinematic of the isotopologues. The observed differences are based on observational limitations, like the velocity resolution and sensitivity.

\subsection{Multiplicity in velocity space}
\label{multi-sec}
Filamentary structures are often identified in continuum emission maps. But it is unknown whether these structures are actual continuous filaments or only an effect of line-of-sight projection of multiple velocity components. We address this question with our data.

\begin{figure}[htb]
\centering
\includegraphics[width=0.5\textwidth]{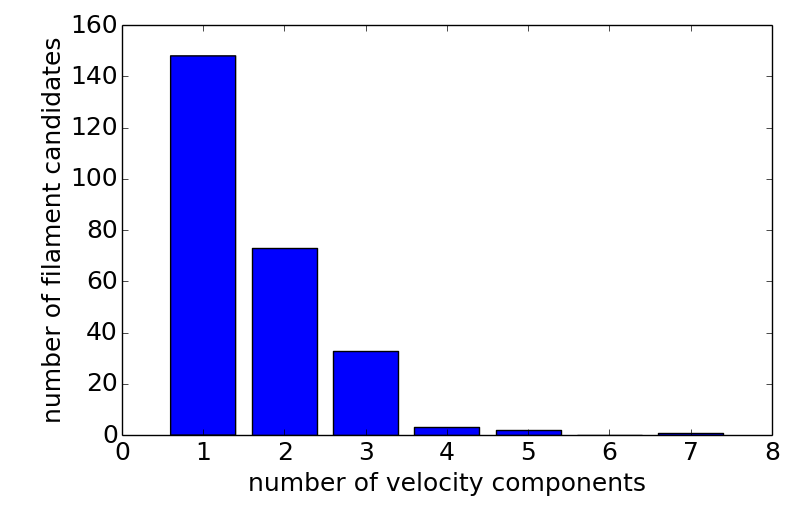}
\caption{Histogram of the number of velocity components per filament.}
\label{multiplicity}
\end{figure}

The $260$ detected filament candidates split up in $422$ velocity coherent filaments in total. Kinematic subcomponents are identified in single spectra for $14$ of the filaments, but will not be discussed any further as more detailed studies will be needed. Analysis of the velocity components shows that about $58\%$ of the filament candidates exhibit one velocity component. Another significant fraction of the filaments, $27\%$ and $12\%$, have 2 or 3 components, respectively. 6 filaments have 4 or more velocity components with a maximum of 7 components, seen in only one filament (Fig. \ref{multiplicity}). 

\begin{figure}[htb]
\centering
\includegraphics[width=0.5\textwidth]{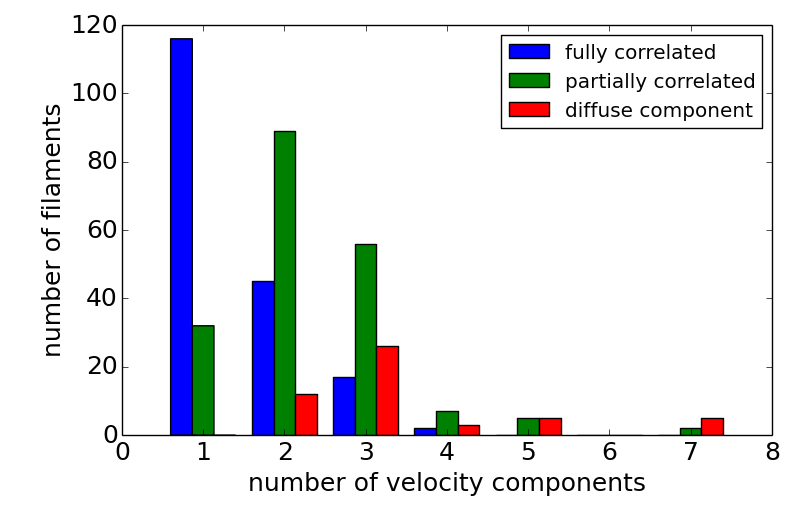}
\caption{Histogram of the number of velocity components for fully correlated (blue), partially correlated (green), and diffuse (red) structures.}
\label{status-multi}
\end{figure}

The categorization of the velocity components shows that a filament candidate can have several velocity components even in the case of one component being fully correlated. This is shown in Fig. \ref{status-multi}. However, a filament candidate with a single component does not necessarily have a fully correlated structure. In general, we find that filament candidates with fewer velocity components are more likely to have a fully correlated component, and candidates with an increasing number of velocity components are more likely to have partially correlated and diffuse components.

\begin{figure}[tbh]
\centering
\includegraphics[width=0.5\textwidth]{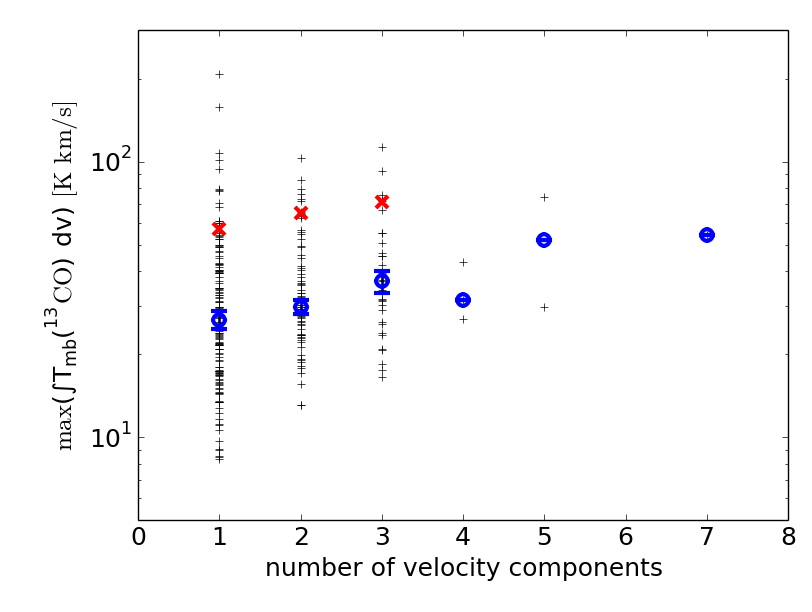}
\includegraphics[width=0.5\textwidth]{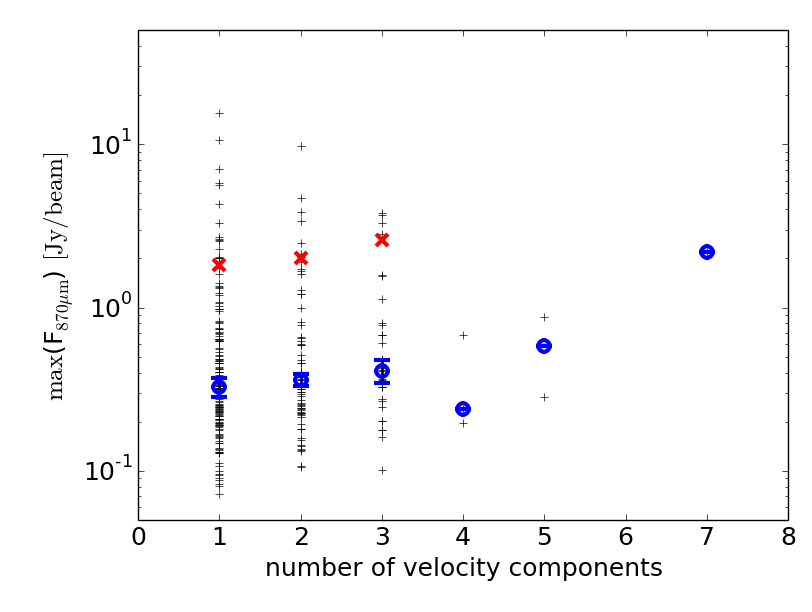}
\caption{The peak integrated \cco intensity (\emph{top}) and the  peak ATLASGAL dust intensity (\emph{bottom}) of a filament are plotted against the number of velocity components. The blue circles mark the median, and the red crosses mark the 90th percentile for each number of components. The error-bars show the uncertainty of the median derived by bootstrapping.}
\label{int-comp-gas-peak}
\end{figure}

Many structures are identified on continuum data that does not provide information about the velocity coherence. Therefore, we test whether there a correlation exists between the intensities of 2D-data and the number of velocity components. With the known multiplicity of the filament candidates, we are able to show that filament candidates with several velocity components tend to be brighter (see Fig. \ref{int-comp-gas-peak}). We do so as follows: we derive the mean and maximum intensity of the ATLASGAL dust emission and the \cco emission integrated over all velocity components. Because the statistical scatter of the intensity values shows a non-Gaussian distribution, we take the median and the 90th percentile value of the intensity distributions of each filament category (i.e. separated according to the number of components) as a qualitative measure. We also estimate the uncertainty of the median using a bootstrapping method. However, only the sample sizes of the filament categories with $1$, $2$, and $3$ components are sufficient to use the bootstrapping method. In this method we draw new, random samples of intensities from among the observed values. We then calculate the median of these new, simulated samples. The resulting distribution of the median values then estimates the sampling function of the observed median and is used to estimate the uncertainty using the standard deviation. We find, that the medians of the \cco peak intensities increase outside their uncertainties as the number of velocity components increases. The same increase is also seen for the 90th percentile values (Fig. \ref{int-comp-gas-peak}). Our data suggests a similar increase for the ATLASGAL peak intensities, but a flat behaviour is also consistent with the data. We could not find such a behaviour for the mean intensities of the filaments (not shown in figures).

\begin{figure}[htb]
\centering
\includegraphics[width=0.5\textwidth]{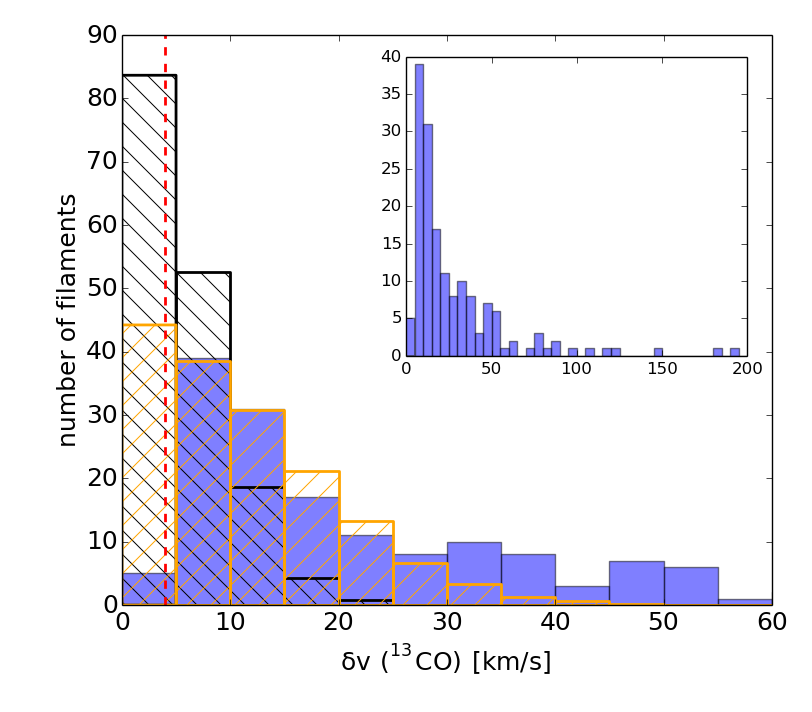}
\caption{Histogram of the absolute difference in velocity between the neighbouring velocity components of a filament. The upper right panel shows the complete distribution, while the main panel shows only the lower velocity separations. The black and orange hatched distributions indicate the models for spiral arm velocity dispersions of $\rm 5~km\,s^{-1}$ and $\rm 10~km\,s^{-1}$. The dashed red line indicates the average velocity separation limit of $\rm 3.5~km\,s^{-1}$ (see Eq. \ref{eq-separation} and mean velocity dispersion of $\rm 1.17~km\,s^{-1}$)}
\label{dv-CO-near}
\end{figure}

For filament candidates with multiple components we investigate whether two physically separate filaments can be located within the same spiral arm. Therefore, we show in Fig. \ref{dv-CO-near} a histogram of the absolute peak-to-peak velocity difference (blue). The bins up to $\delta v = \rm 10~km\,s^{-1}$ are likely to be incomplete because of the component separation limit, as shown before (Fig. \ref{vel_separation}, Eq. \ref{eq-separation}). We compare the distribution of the observed velocity differences with model distributions (hatched) of expected velocity difference from a spiral arm. We assume velocity dispersions of $\rm \sigma_{v_1}(arm) = 5~km\,s^{-1}$ and $\rm \sigma_{v_2}(arm) = 10~km\,s^{-1}$ following \cite{Reid2016} and \cite{CalduPrimo2013}. As we measure the absolute difference between two velocities drawn from a Gaussian distribution, the dispersion of the differences is given by $\sigma_{\delta v}\rm =\sqrt{\sigma_v(arm)^2 + \sigma_v(arm)^2}=\sqrt{2}\sigma_v(arm)$. We sample the difference distributions with $10,000$ draws to avoid statistical noise, bin the absolute values of the sample like the observed differences, and scale by $0.016$ to get a comparable total number of filaments. We find, that the observed and the model distribution of $\rm \sigma_{v_2}(arm)$ are similar for $\delta v \rm \leq 30~km\,s^{-1}$, but we see more observed filaments for larger velocity separations. The model distribution for $\rm \sigma_{v_1}(arm)$ does not describe the observed one. Therefore, we can conclude that a large fraction of separated filaments might be located in the same spiral arm with a velocity dispersion of $\rm \sigma_{v_2}(arm)$, but we also see filaments from different Galactic structures along the line-of-sight. However, because of the kinematic distance ambiguity filaments located in different spiral arms can have similar line-of-sight velocities at specific Galactic Longitudes.

\subsection{Comparison of masses derived from gas and dust }
\label{gas-dust}

Calculating the masses of the filaments is an important part of the analysis. However, it comes with some difficulties. Commonly, dust emission or dust extinction is used to calculate the mass of objects. However, because of the line-of-sight projection several filaments that may appear at the same position, and cannot be separated in the continuum data, this is not applicable here. Therefore, we need to use the CO emission to disentangle the projected emission from different structures in velocity space. For the mass estimate we then use the emission integrated over the velocity range of the filament. Specifically, we use the \cco emission as it has a higher signal-to-noise and traces the lower column density gas around the skeleton, and calculate the mass like described in Section \ref{mass-sec}.

\begin{figure}[htb]
\centering
\includegraphics[width=0.5\textwidth]{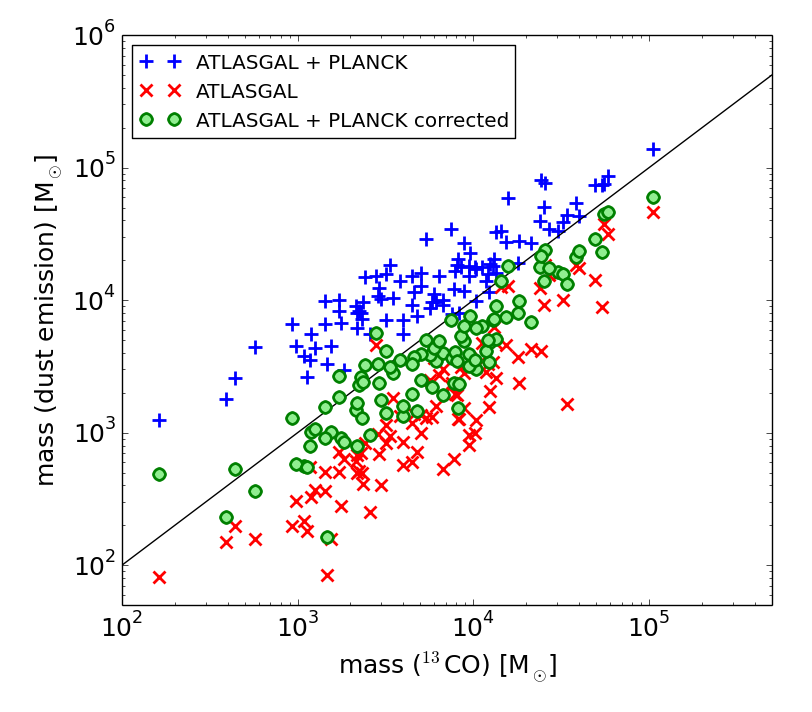}
\caption{Mass per fully correlated filament derived from dust versus the mass derived from integrated $^{13}$CO using an $X_{\rm ^{13}CO(2-1)}$ factor from \cite{Schuller2017}. The black line indicates the one-to-one ratio.}
\label{dust-gas-mass}
\end{figure}

We first have to test whether this $X$-factor is a good approximation for the whole survey area. To do so we take a sample of filaments, which show only one velocity component and are fully correlated with the ATLASGAL dust emission. We calculate the masses for this sample using the integrated \cco emission with Equation \ref{gas-mass-equ} and using the ATLASGAL and ATLASGAL+PLANCK dust emission with Equation \ref{dust-mass-equ}. For all three data-sets we use the same mask around the skeleton. The comparison of the resulting masses is shown in Fig. \ref{dust-gas-mass}.

We find that masses derived from \cco are systematically larger than the masses derived from ATLASGAL, but systematically smaller than the masses derived from ATLASGAL+PLANCK. This behaviour is expected as the ATLASGAL maps are sensitive to the small scale ($2.5 \arcmin$), high column density dust structures and extended emission from the diffuse surrounding gas is filtered out because of the observing technique (see \cite{Schuller2009}). Therefore, \cite{Csengeri2016} combined the ATLASGAL data with the PLANCK data, which traces the diffuse, large scale structures, but does not resolve the small scales because of the low resolution ($4.8 \arcmin$). However, the combined data also traces the dust emission along the line-of-sight, i.e foreground and background. Thus, masses derived from ATLASGAL are likely to be underestimated while masses derived from ATLASGAL+PLANCK are likely to be overestimated. 

As shown in Section \ref{mass-sec}, we corrected the ATLASGAL+PLANCK data for the line-of-sight emission towards every filament and used this data to derive another mass estimate. On average we find a mean Galactic emission of $\rm 0.52~Jy/beam$ (beam size of $21\arcsec$). These corrected dust masses are in agreement with the \cco derived masses within a factor of 2. Therefore we conclude that the \cco $X$-factor derived from the SEDIGISM science demonstration field \citep{Schuller2017} is a good approximation for the whole survey area.

\section{Discussion}
\label{discussion}
\subsection{Radial filament profiles}
\label{Filament-profile}

Nearby ($\rm < 500~pc$), low line-mass ($\rm < 100~M_\odot \, pc^{-1}$) filaments have been found to have an FWHM size of the order of $\rm 0.1~pc$ \citep{Arzoumanian2011}. The corrected ATLASGAL+PLANCK and \cco data trace the wide range of column densities that is needed to study the filament profile. To ensure that we are looking only at true filaments we restrict our sample to the 151 fully correlated filaments with a distance estimate, but for completeness we show the results of the other filaments in Appendix \ref{other_profiles}. However, measuring the filament profile is challenging as most filaments are not homogeneous, linear structures, but show branches and varying central densities.

Therefore, we do not extract the radial column density distribution directly from the data, but estimate the mass of the filaments within filament masks with increasing diameter, $s_\text{box}$, using the same equations and assumptions as in Section \ref{mass-sec}. The mass, $M(R)$, is then given by
\begin{equation}
M(R)=2 l \int_0^R \Sigma(r) \,dr   $   ,$
\label{mass-equ}
\end{equation}
where $l$ is the length of the filament, $\Sigma(r)$ is the column density of the gas at distance $r$ from the skeleton (to both sides, with the skeleton at $r=\rm 0\,pc$), and $R = 0.5 \, s_\text{box}$ the maximum radius. We normalize the values with the mass from a box-diameter of $s_\text{max} = \rm 4~pc$, where the typically found radial profile is almost flat \citep{Arzoumanian2011}. The smallest box-diameter is given by the pixel size. The resulting mass curves of the \cco emission are shown in Fig. \ref{mass-width}, and of the continuum emission and less correlated filaments in Appendix \ref{other_profiles}. As the physical resolution is changing with the distance towards the source, we group the filaments in four distance intervals $d_i$ ($d_1< 2~\text{kpc}<d_2<4~\text{kpc}<d_3<8~\text{kpc}<d_4$) and average the mass curves within these intervals (see Fig. \ref{mass-width-mean}).

\begin{figure*}
\centering
\begin{minipage}{0.505\textwidth}
\includegraphics[width=\textwidth, clip=true, trim= 1.cm 1.5cm 0.5cm 0.3cm]{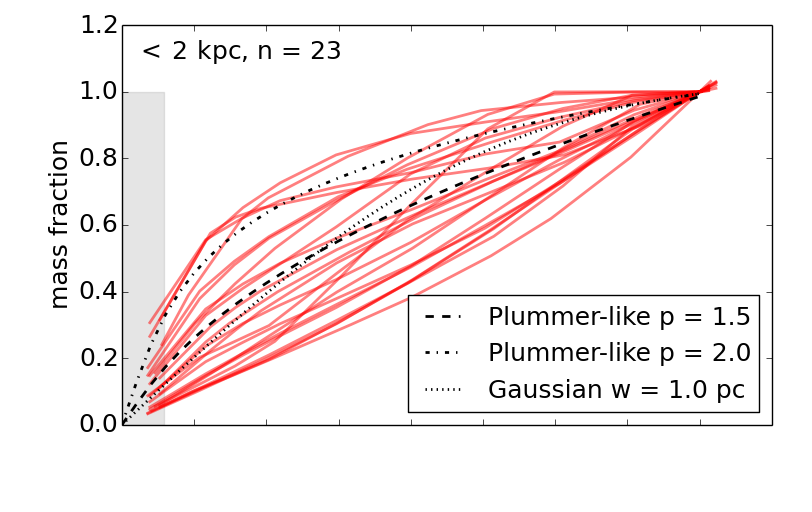}
\end{minipage}
\begin{minipage}{0.48\textwidth}
\includegraphics[width=\textwidth, clip=true, trim= 2.cm 1.5cm 0.5cm 0.3cm]{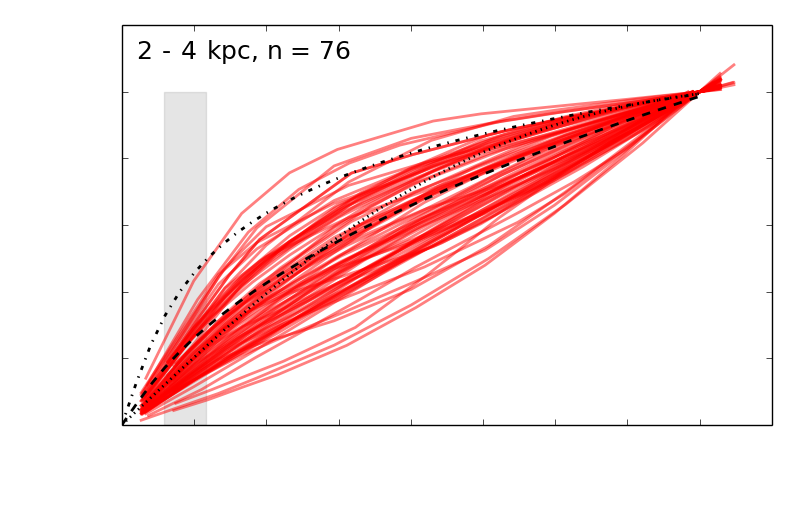}
\end{minipage}

\begin{minipage}{0.505\textwidth}
\includegraphics[width=\textwidth, clip=true, trim= 1.cm 0.5cm 0.5cm 0.3cm]{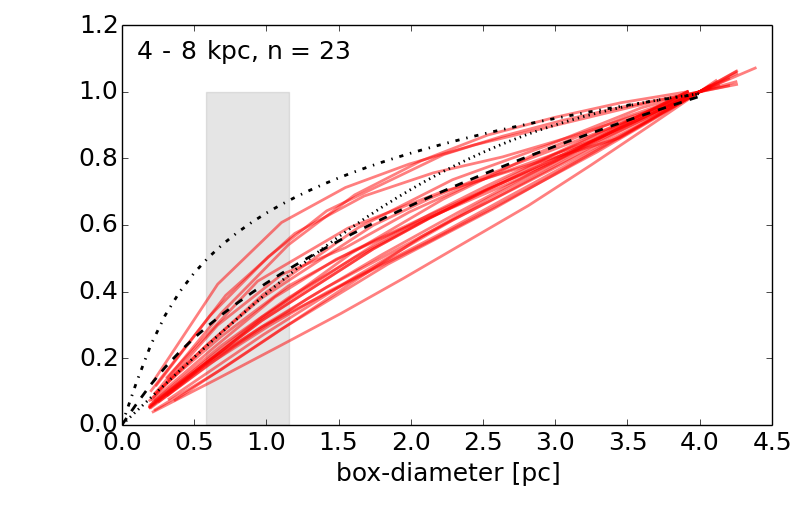}
\end{minipage}
\begin{minipage}{0.48\textwidth}
\includegraphics[width=\textwidth, clip=true, trim= 2.cm 0.5cm 0.5cm 0.3cm]{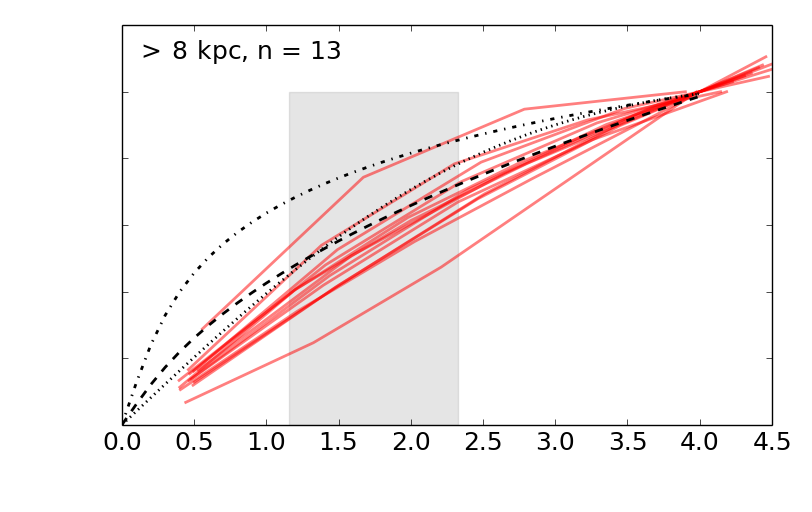}
\end{minipage}

\caption{Fraction of the filament mass derived from \cco emission dependent on the box-diameter of the mask separated by distances. \emph{Top left:} $d < 2~\text{kpc}$, \emph{Top right:} $2~\text{kpc} < d < 4~\text{kpc}$, \emph{Bottom left:} $4~\text{kpc} < d < 8~\text{kpc}$, \emph{Bottom right:} $d > 8~\text{kpc}$. One curve describes one fully correlated filament at its distance estimate. The grey lines indicate the physical beamsize at distances of 2 kpc, 3 kpc, 6 kpc, and 8 kpc. The black lines show the integrated theoretical radial profiles, which describe a Plummer-like distribution $p=1.5$ (dashed) or $p=2.0$ (dash-dotted), and a Gaussian distribution with a dispersion of $w = 1.0$ (dotted).}
\label{mass-width}
\end{figure*}

\begin{figure}[htb]
\centering
\includegraphics[width=0.5\textwidth]{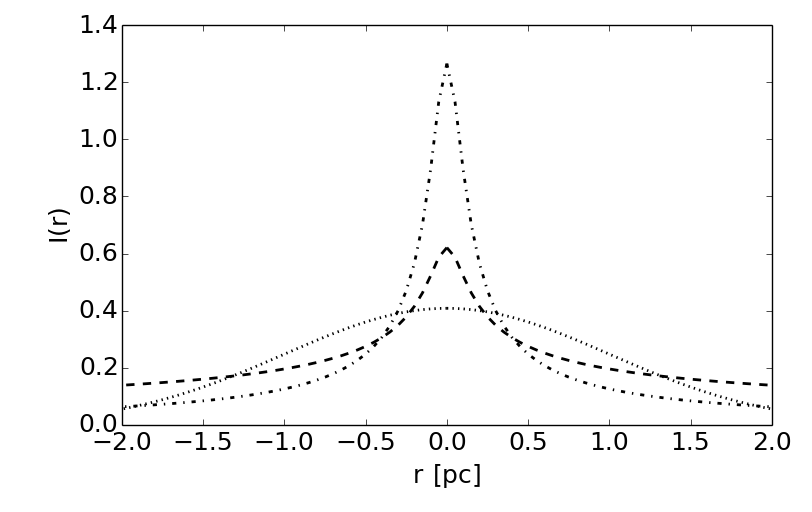}
\caption{Theoretical filament profiles normalized to an integrated intensity of 1, which describe a Plummer-like function with $R_\text{flat} = \rm 0.1~pc$ and $p=1.5$ (dashed) or $p=2.0$ (dash-dotted), and a Gaussian with a dispersion of $w =\rm 1.0~pc$ (dotted). The radial integration of these profiles is shown in Figs. \ref{mass-width} and \ref{mass-width-mean}.}
\label{profiles}
\end{figure}

\begin{figure}[htb]
\centering
\includegraphics[width=0.5\textwidth]{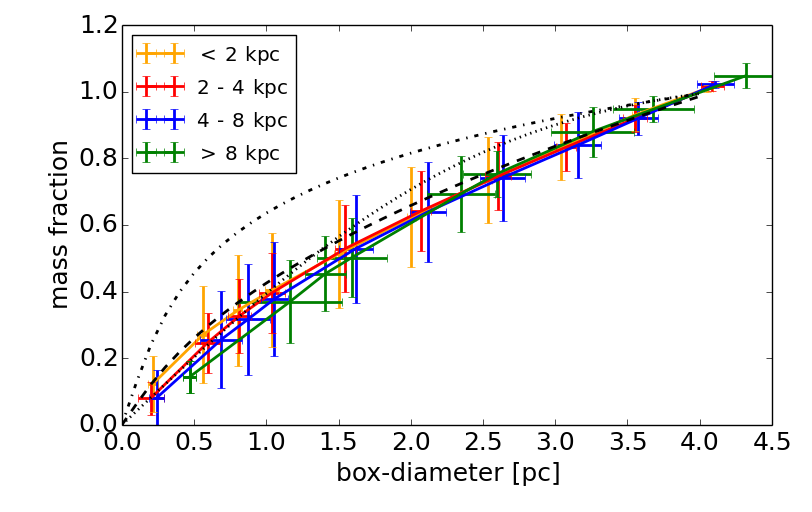}
\caption{Average fraction of the filament mass derived from \cco emission dependent on the box-diameter of the mask. The color indicates the distance of the filament with $d_1$ orange, $d_2$ red, $d_3$ blue, and $d_4$ green. The errorbars indicate the dispersion of the measured mass fraction and box-diameter. The black lines are same as Fig. \ref{mass-width}.}
\label{mass-width-mean}
\end{figure}

The profiles of filamentary structures are found to be well described by a Plummer-like density distribution \citep{Nutter2008,Arzoumanian2011,Contreras2013a}, which is given by 
\begin{equation}
\rho(r)=\rho_c \left(1+\left(\frac{r}{R_\text{flat}}\right)^2 \right)^{-p/2}   $   ,$
\label{density-equ}
\end{equation}
where $\rho_c$ is the central density of the filament, and $R_\text{flat}$ the characteristic radius of the flat inner part. The column density profile of the filament \citep{Arzoumanian2011, Panopoulou2014} then is described by
\begin{equation}
\Sigma_p(r)=A_p \frac{\rho_c R_\text{flat}}{\left[1+(r/R_\text{flat})^2\right]^{\frac{p-1}{2}}}   $   ,$
\end{equation}
where $\Sigma=N(\text{H}_2) \mu m_p$, $\mu$ and $m_p$ as previously defined, and $A_p=(\cos i)^{-1}\int_{-\infty}^{\infty} (1+u^2)^{-p/2} du$, a finite constant for $p>1$. Other studies \citep[e.g. ][]{Arzoumanian2011} have shown that the inner part of the radial profile of a filament can also be well described by a Gaussian column density distribution. These two models are shown in Fig. \ref{profiles}, where they are normalized to an integrated intensity of 1.

The mass within a box around the theoretical filament is then given by Eq. \ref{mass-equ}. We plot the measured mass curves as well as the theoretical ones as $M(R)/M(2~\text{pc})$ and test different values for the exponent, $p$, and the inner radius, $R_\text{flat}$, of the Plummer-like distribution and for the dispersion, $w$, of the Gaussian (see Fig. \ref{mass-width-mean}). 

A detailed analysis of the density structure of filaments is beyond the scope of this paper. Still, we perform a rough visual comparison between the observed radial column density profiles and modeled ones. We find that the Plummer-like distribution is in agreement with the average profile of the \cco and dust observations for $p \approx 1.5 \pm 0.5$ and $R_\text{flat} \approx \rm 0.1~pc$. Also the Gaussian column density distribution with a dispersion of $w=\rm 1~pc$ describes the observation within the uncertainties. However, the two fitting models lead to different $FWHMs$ for the filaments. While for a Plummer-like function the $FWHM_P=(2^{2/(p-1)}-1)^{1/2} R_\text{flat}$ ($FWHM_{2.0}\approx \rm 0.17~pc$, $FWHM_{1.5}\approx \rm 0.39~pc$ \citep{Heitsch2013}, for the Gaussian $FWHM_G = \sqrt{8 \ln{2}}\, w \approx \rm 2.36~pc$.

One possible interpretation is that the Gaussian traces only the low column density surrounding of the filament, but not the dense inner part, hence the actual filament. From previous studies \citep[see][]{Arzoumanian2011, Panopoulou2014} and the Plummer-like function we see measurements of the FWHM between $\rm 0.1~pc$ and $\rm 0.6~pc$. The physical beam size of the SEDIGISM data at a distance of $\rm 2~kpc$ is about $\rm 0.3~pc$ and therefore, we are at the resolution limit for the dense filament spine, but note that the mass curves (integral over the radial profile) have a dependency on $R_\text{flat}$ in the Plummer-like case. However, we find that small changes of $R_\text{flat}$ do not significantly change the agreement with the observation. Therefore, we conclude that the mass of the filament is dominated by the low column density gas surrounding the spine, and that the resolution of the SEDIGISM data is not sufficient to meaningfully fit the inner spine with a Gaussian radial profile.

The exponent of the average density profile is $p \approx 1.5$, which is in agreement with the lower limit found by \cite{Arzoumanian2011}. The single profiles scatter between $p \approx 1.0$ and $p \approx 2.0$, with the scatter decreasing with more distant filaments most likely because of the smaller sample. Also, we tested the effect of the beam size to the theoretical radial profiles by convolving the profile with a Gaussian beam. The resulting theoretical mass curves are shallower with increasing distance, but not significantly, given the scatter of the single observed mass curves. The study of \cite{Arzoumanian2011} analyses prominent filaments in nearby molecular clouds using dust continuum emission. This selection of prominent filaments might give a bias towards higher exponents. Theoretically, an isolated, isothermal, cylindrical filament in hydrostatic equilibrium would be expected to have an exponent of $p=4$ \citep{Ostriker1964}. However, this exponent is typically not found in observations and models \citep{Juvela2012,Kainulainen2015}. Low resolution and signal-to-noise data explain only partially the observed exponents. Therefore, observations suggest that filaments are embedded in a surrounding molecular cloud \citep{Fischera2012a}, not isothermal \citep{Recchi2013} and/or not in hydrostatic equilibrium \citep{Heitsch2013a,Heitsch2013}.

It is important to mention that for an exponent of $p<2$ mathematically the mass diverges with an increasing radius. This can be seen in Fig. \ref{profiles}. As a result the mass, $M$, and therefore, also the line-mass (mass per unit length), $m=M/l$, are not well-defined. However, filaments are not isolated structures, but surrounded by low density gas, which sets boundary conditions that are not considered in this model. As mentioned before, we decided to use a radius dependent on the velocity dispersion of the filament to estimate the mass of the filaments in this study.

\subsection{Stability against Collapse}

Thermally supercritical filaments are commonly seen as star formation sites. Therefore, they need to build the connection between the diffuse gas of the molecular cloud and the dense gas in the star-forming cores. \cite{Inutsuka1992} showed that isothermal, infinitely long, self-gravitating cylinders will collapse radially if their line-mass (mass per unit length) exceed a critical value, and fragment along the axis in the sub-critical and equilibrium case. The critical line-mass is given by 
\begin{equation}
m_\text{crit,th} = \frac{2 c_s^2}{G}   
\label{mcrit_equ}
\end{equation}
\citep{Ostriker1964}, where $G$ is the gravitational constant and $c_s$ is the sound speed of the medium, which is dependent on the gas temperature $T$ (Eq. \ref{sound_speed_equ}). Assuming a typical gas temperature of $T = \rm 15~K$ the critical line-mass is $m_\text{crit} =\rm 20~M_\odot\, pc^{-1}$.

Based on our observations and analysis we can estimate the line-mass for all the filaments with a distance estimate by $m_\text{obs} = M/l$, where $M$ is the mass estimated from the \cco emission, and $l$ is the length along the velocity coherent skeleton. Because of the separation of the velocity components this length does no longer securely describe the linear shape (with small branches) of the original filament candidate sample, especially for the not fully correlated filaments. Therefore, we concentrate this discussion on the fully correlated filaments, but also perform the calculations for the others. 

\begin{figure}[htb]
\centering
\includegraphics[width=0.5\textwidth]{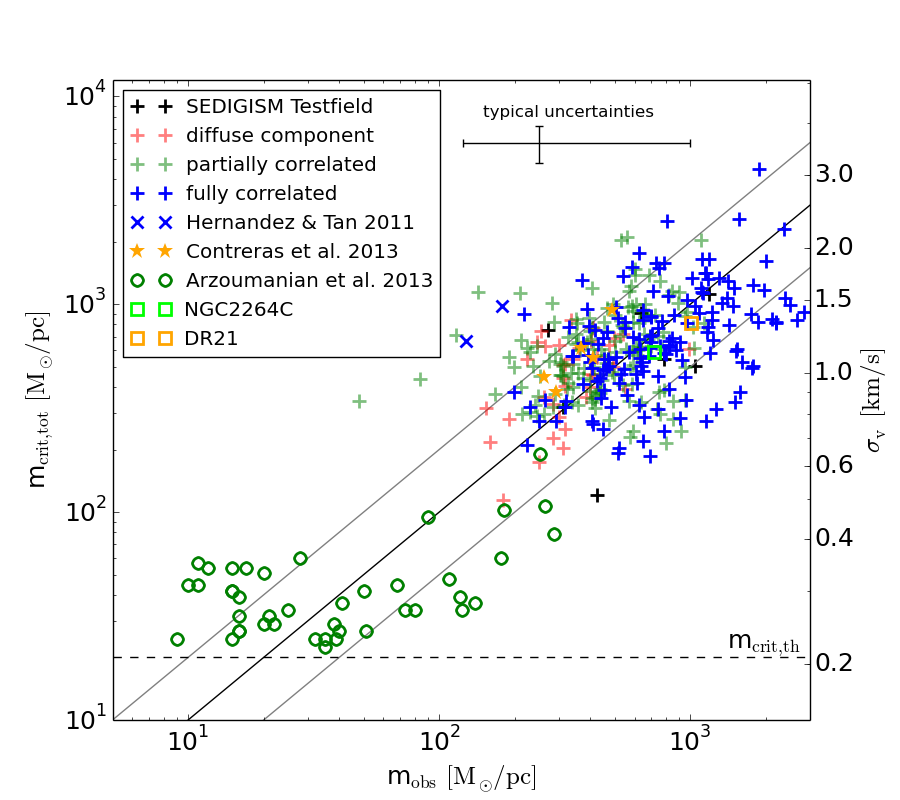}
\caption{Critical, non-thermal line-mass derived from the velocity dispersion versus observed line-mass using integrated $^{13}$CO. The fully correlated filaments are indicated in blue, and the other filaments of this study in a shaded green and red. The black solid line shows the one-to-one correlation, the grey lines indicate a factor 2 uncertainty, and the dashed line shows the critical thermal line-mass.}
\label{line-mass}
\end{figure}

The line-masses we observe with our resolution are significantly above the critical thermal value (see Fig. \ref{line-mass}). This leaves us with two possible conclusions: either the filaments are collapsing radially or they have a supporting mechanism additional to the thermal pressure. Moreover, we find that the linewidth of the molecular gas is significantly larger than the sound speed, $c_s =\rm 0.21~km\,s^{-1}$. This increased linewidth can support both theories, as it can be interpreted as either structured motions, like collapse, or turbulent motions within the gas. 

Assuming that non-thermal motions contribute to the supporting mechanism, equation \ref{mcrit_equ} can be modified to
\begin{equation}
m_\text{crit,tot} = \frac{2 (c_s^2 + \sigma_\text{nt}^2)}{G}
\label{mcrit_non_equ}
\end{equation}
\citep{Fiege2000a}, where $\rm \sigma_\text{nt}$ is the non-thermal velocity dispersion of the filament, and $m_\text{crit,tot}$ is the critical, total (thermal and non-thermal) line-mass. After determining the velocity dispersion for all filaments, we can calculate the critical non-thermal line-mass and compare it with the observed one (Fig. \ref{line-mass}).

The uncertainty of the critical line-mass is given by the observed velocity dispersion and therefore depends on the velocity resolution and the quality of the signal. The main contributions for the uncertainties of the mass estimates are the $X$-factor (factor of $0.5$--$2$), optically thick \cco emission (factor of $1$--$2$), and the distance. As the length is also dependent on the distance estimate, the line-mass is only linearly dependent on the distance, which adds another factor of $0.8$--$1.2$ to the uncertainty. Additionally, the length is measured as projection on the sky and therefore, the observed line-mass is an upper limit considering possible inclinations. The typical uncertainty is given by the black cross in Fig. \ref{line-mass}. Additionally, it needs to be mentioned that based on the resolution of our data we are only able to derive global parameters. Higher resolution data (spatial and kinematic) could reveal substructures, which might lead to different results.

We find that the critical, non-thermal line-mass is, within the uncertainties, in agreement with the observed line-mass. Therefore, Eq. \ref{mcrit_non_equ} seems to describe a common relation between the observed linewidth and line-mass in the form $m \propto (c_s^2 + \sigma_\text{nt}^2)$. The sound speed, $c_s$, depends only on the temperature of the ISM, which can be assumed to be about constant. Hence, the line-mass is proportional to the non-thermal motion. We also find that partially correlated filaments and diffuse components follow the same relation as the fully correlated filaments, but with a slightly wider spread.

We now want to investigate where this relation comes from. As we discussed before, one explanation might be infall motion. \cite{Inutsuka1992} showed that infinitely long, isothermal filaments with a line-mass above the critical value collapse radially. \cite{Heitsch2009} and \cite{Heitsch2013} determined the accretion velocity profile, $v(R)$, for gas in a steady-state free-fall onto the filament axis as
\begin{equation}
v(R) = 2 \left(G m \ln{\frac{R_\text{ref}}{R}}\right)^{1/2}   $   ,$
\label{velocity_equ}
\end{equation}
where $G$ is the gravitational constant, $m$ is the line-mass of the filament, $R_\text{ref}$ is the limiting, outer radius of the filament, and $R$ is the radial position of the gas. 

\begin{figure}[htb]
\centering
\includegraphics[width=0.5\textwidth]{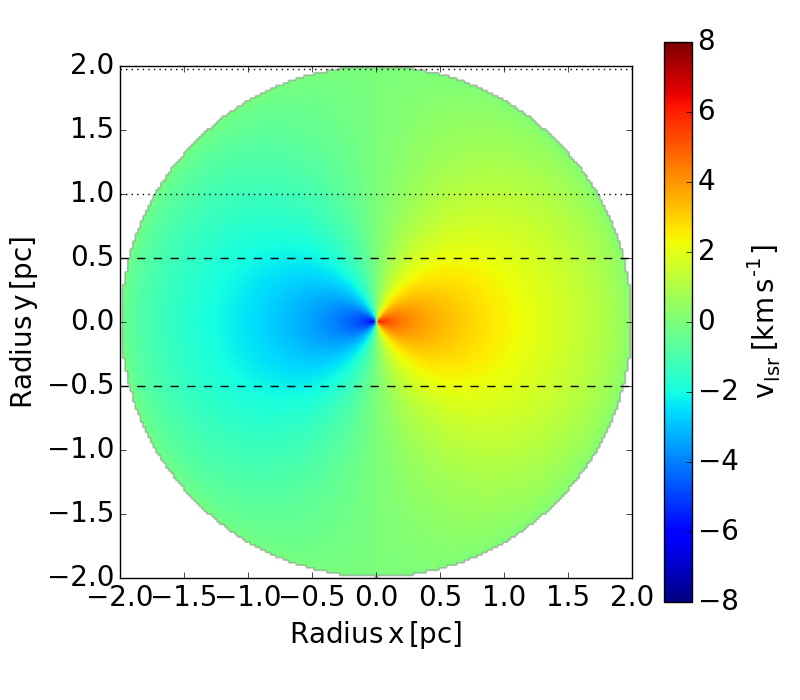}
\caption{Line-of-sight velocity distribution across a slice of a filament base on Eq. \ref{velocity_equ} using a line-mass of $m =\rm 500~M_\odot\,pc^{-1}$ and $R_\text{ref}=\rm 2.0~pc$. The observer is located on the right side, observing the whole slice, the inner part (dashed lines), and outer part (dotted lines), which where analysed separately.}
\label{rad_vel}
\end{figure}

We use this radial velocity distribution, $v(R)$, to estimate the signal which would be observed from a collapsing filament, similar to \cite{Heitsch2013a}. First, we derive the line-of-sight velocities, $v_{lsr}$, across the filament for an observer looking edge on (see Fig. \ref{rad_vel}),
\begin{equation}
v_{lsr} = v(R) \cdot \frac{x}{R}   $   ,$
\end{equation}
where $R$ is the radial distance to the center, and $x$ is the position  in the x-axis direction of the Cartesian coordinate system. Second, we draw for each position in the filament 50 values from a Gaussian distribution centered on the derived velocities with a thermal velocity dispersion of $c_s =\rm 0.21~km\,s^{-1}$. Third, we plot a weighted histogram of the velocities with bins identical to the SEDIGISM channel width of $\rm 0.25~km\,s^{-1}$, where we use the density at the position of the filament as weight. The density is given by a Plummer-like distribution (see Section \ref{Filament-profile}). From the weighted histogram we calculate the standard deviation, hence the theoretically observed velocity dispersion.

\begin{figure*}[t]
\includegraphics[width=\textwidth, clip=true, trim= 0cm 2.cm 0.0cm 1.5cm]{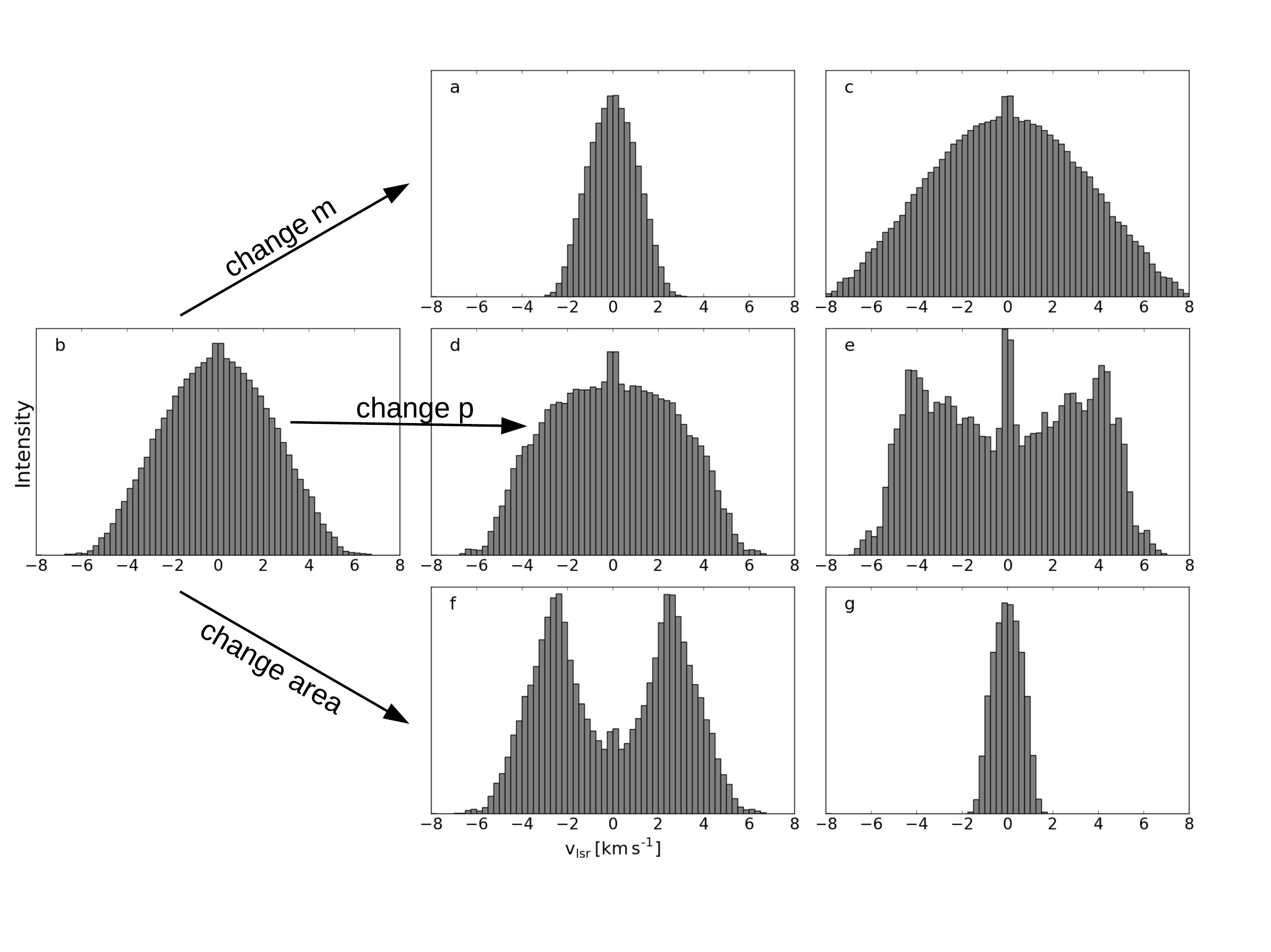}
\caption{Theoretical molecular line profiles with varying parameters, see Table \ref{theo-line-tab}.}
\label{line-profiles}

\end{figure*}

Within this template we vary the line-mass, $m$, the exponent of the density profile, $p$, and the area of the filament we observe (complete, middle, edge) to study their effect on theoretical signal. We choose the outer radius $R_\text{ref}=\rm 2.0~pc$ and the characteristic radius of the density distribution $R_0=\rm 0.1~pc$, see Section \ref{Filament-profile}. The results are shown in Table \ref{theo-line-tab} and Fig. \ref{line-profiles}.

\begin{table}[tbh]
\caption{The three input parameters, the resulting velocity dispersion $\sigma_\text{v}$, derived from the modelled collapse spectra, and the expected total velocity dispersion, following the found relation with $m$.}
\label{theo-line-tab}
\centering
\begin{tabular}{cccccc}
\hline\hline
Fig. \ref{line-profiles} & $m$ & $p$ & area & $\sigma_\text{v}$ & $\sigma_\text{crit,nt}$\\
 & $\rm M_\odot\,pc^{-1}$ &  &  & $\rm km\,s^{-1}$ & $\rm km\,s^{-1}$\\
\hline 
a & $100$  & $1.5$ & all    & $1.03$ & $0.46$ \\ 
b & $500$  & $1.5$ & all    & $2.29$ & $1.04$ \\ 
c & $1000$ & $1.5$ & all    & $3.23$ & $1.47$ \\ 
d & $500$  & $2.0$ & all    & $2.65$ & $1.04$ \\ 
e & $500$  & $3.0$ & all    & $3.24$ & $1.04$ \\ 
f & $500$  & $1.5$ & middle & $2.86$ & - \\
g & $500$  & $1.5$ & edge   & $0.62$ & - \\
\hline
\end{tabular}
\end{table}

We find that the infall motions indeed show the observed relation between line-mass and velocity dispersion, $m \propto \sigma_\text{v}^2$ (Table \ref{theo-line-tab}, models a, b, c). However, the theoretical velocity dispersion is higher by a factor of 2 than the observed values. Additionally, we did not take turbulent motions into account, which would be caused by the infall \citep{Heitsch2013}, and give a wider theoretical signal. To fit the observations, the collapse needs to be slowed down, which can be caused by the turbulence created by the collapse. Finding a combination of collapse and turbulent motions that would reproduce the observed kinematics is beyond the scope of this study.

Another way to identify ongoing collapse could be an analysis of the shape of the emission lines \citep{Schneider2010}. The theoretical line profiles show double-peaked shapes towards the centre of the filament and for filaments with a steep density profile. However, a comparison of the theoretical line shapes with the observed ones is difficult as several other effects can alter the shape, like spatially unresolved motions of the filament within the beam, or self-absorption, also see \cite{Heitsch2013a} for different inclination angles. In a few filaments we find some evidence for a double-peaked velocity profile (see Section \ref{vel_identification}). But a more detailed analysis of, preferably, geometrically simple filaments with higher spatial and kinematic resolution would be necessary to address the effects of complex kinematics.

Rapid radial collapse would lead to extremely narrow filaments, which rarely have been observed up to now \citep[e.g. ][]{Stutz2016}. However, \cite{Heitsch2013} shows that the fragmentation timescales are shorter than the collapse timescales, which is supported by the fact that fragmentation is seen in almost all filaments \citep{Jackson2010, Kainulainen2013a, Takahashi2013, Wang2014, Beuther2015, Teixeira2016, Kainulainen2017}. Our finding of slowed collapse would increase the difference between the timescales, which is still in agreement with the observations. Also, simulations of filament evolution by \cite{Chira2017} show a start of fragmentation before the filaments reach the critical mass for gravitational collapse.

However, we also cannot rule out the possibility that the observed velocity dispersion is mainly created by turbulent motions. These turbulent motions are discussed to increase the internal pressure and support the filament against gravitational collapse. Therefore, this theory is also in agreement with the observations.

\section{Conclusions}
\label{conclusion}
In this study we studied spectral line emission from 283 filament candidates detected with ATLASGAL continuum dust emission from the catalogue of \cite{Li2016b} in the SEDIGISM \cco and \coo survey. As these candidates can be the result of line-of-sight projection of multiple structures, we tested the candidates for coherence in velocity space and derived the mass, size, and a collection of kinematic properties. To do so we developed an automated analysis tool that finds the different velocity components of a candidate, if existing, separates them and checks for correlation with the original ATLASGAL emission. We found 422 velocity-coherent filaments that correlate completely or partially with the original candidate. For these filaments we find the following:
\begin{itemize}
\item Two-thirds of the filament candidates are single velocity-coherent structures. The other candidates are line-of-sight projections of mainly two and three velocity components, and up to one candidate with seven velocity components. Also, we found a possible indication for a correlation between the maximum intensity within the filament candidate and the number of velocity components for the integrated \cco and ATLASGAL dust emission, but a flat behaviour is within the uncertainties.

\item Comparing the kinematics of the filaments seen in \cco and \coo, we could show that both isotopologues trace the same gas. Differences found in the comparison could be identified as biases arising from low signal-to-noise \coo data.

\item The filament profiles are on average in agreement with a Plummer-like density distribution with an exponent of $p \approx 1.5 \pm 0.5$. The inner radius cannot be constrained exactly because of the limited resolution of the data. This low exponent indicates that filaments are typically located within larger molecular clouds, and therefore, the outer radius of a filament cannot be well defined. For the mass estimates we chose a radius which includes the gas that can take part in star formation within the next $\rm 2~Myr$.

\item The observed line-mass of the filaments is in agreement with the critical non-thermal line-mass and significantly higher than the critical thermal line-mass. However, we do not know the source of the observed velocity dispersion. Comparing the relation we find between velocity dispersion and line-mass with the theoretical infall velocity profile based on \cite{Heitsch2013} generally does not reveal evidence for free-fall collapse. However, radial infall of the gas onto the skeleton can possibly explain the relation.

\end{itemize}

In this study we analysed the kinematics of 283 filament candidates, finding 180 reliable velocity coherent filaments, 151 with distance estimates between $\rm 1~kpc$ and $\rm 13~kpc$, and 242 other velocity coherent filamentary structures in the line-of-sight of the candidates, leading to the largest statistics of filament parameters so far. However, due to the spatial resolution of $30\arcsec$ and velocity resolution of $\rm 0.25~km\,s^{-1}$, the derived parameters generally only describe global behaviour of the filaments. As the evolution and fragmentation of filaments is a hierarchical process it will be necessary to also study the smaller scales. High resolution observations to recover the small scales ($\rm < 0.1~pc$) are essential, and this study can be the starting point for the selection of a representative sample for such higher resolution follow-ups.

\section*{Acknowledgments}
We want to thank the anonymous referee for constructive and very detailed comments.
Also, we thank Guang-Xing Li for supporting this study by offering additional data and information on the filament candidates. Also, we thank Prof. Kostas Tassis, Dr. Dario Colombo, Carsten K\"onig, and the rest of the SEDIGISM consortium for support and productive discussions.

M.M. is supported for this research through a stipend from the International Max Planck Research  School (IMPRS) for Astronomy and Astrophysics at the Universities of Bonn and Cologne.
T.Cs. acknowledges support from the \emph{Deut\-sche For\-schungs\-ge\-mein\-schaft, DFG\/}  via the SPP (priority programme) 1573 'Physics of the ISM'. 
H.B. acknowledges support from the European Research Council under the Horizon 2020 Framework Program via the ERC Consolidator Grant CSF-648505. 
L.B. acknowledges support by CONICYT project PFB-06.
This work of J.K. has received funding from the European Union's Horizon 2020 research and innovation program under grant agreement No 639459 (PROMISE).

\bibliography{./library}{}

\begin{thebibliography}{77}
\expandafter\ifx\csname natexlab\endcsname\relax\def\natexlab#1{#1}\fi

\bibitem[{Abreu-Vicente {et~al.}(2016)Abreu-Vicente, Ragan, Kainulainen,
  Henning, Beuther, \& Johnston}]{Abreu-Vicente2016}
Abreu-Vicente, J., Ragan, S., Kainulainen, J., {et~al.} 2016, \aap, 590, A131

\bibitem[{Andr{\'{e}} {et~al.}(2010)Andr{\'{e}}, Men'shchikov, Bontemps,
  K{\"{o}}nyves, Motte, Schneider, Didelon, Minier, Henning, Royer,
  Mer{\'{i}}n, Vavrek, Attard, Arzoumanian, Wilson, Ade, Aussel, Men, Bontemps,
  K{\"{o}}nyves, Motte, Schneider, Didelon, Minier, Henning, Royer,
  Mer{\'{i}}n, Vavrek, Attard, Arzoumanian, Wilson, Ade, \& Aussel}]{Andre2010}
Andr{\'{e}}, P., Men'shchikov, A., Bontemps, S., {et~al.} 2010, \aap, 518, L102

\bibitem[{Arzoumanian {et~al.}(2011)Arzoumanian, Andr{\'{e}}, Didelon,
  K{\"{o}}nyves, Schneider, Men'shchikov, Sousbie, Zavagno, Bontemps, {Di
  Francesco}, Griffin, Hennemann, Hill, Kirk, Martin, Minier, Molinari, Motte,
  Peretto, Pezzuto, Spinoglio, Ward-Thompson, White, Wilson, Men, Sousbie,
  Zavagno, Motte, Peretto, Pezzuto, Spinoglio, White, \&
  Wilson}]{Arzoumanian2011}
Arzoumanian, D., Andr{\'{e}}, P., Didelon, P., {et~al.} 2011, \aap, 529, L6

\bibitem[{Arzoumanian {et~al.}(2013)Arzoumanian, Andr{\'{e}}, Peretto, \&
  K{\"{o}}nyves}]{Arzoumanian2013}
Arzoumanian, D., Andr{\'{e}}, P., Peretto, N., \& K{\"{o}}nyves, V. 2013, \aap,
  553, A119

\bibitem[{Bally {et~al.}(1987)Bally, Stark, Wilson, Langer, Stark, Wilson, \&
  Langer}]{Bally1987}
Bally, J., Stark, A.~A., Wilson, R.~W., {et~al.} 1987, ApJ, 312, L45

\bibitem[{Beuther {et~al.}(2015)Beuther, Ragan, Johnston, Henning, Hacar, \&
  Kainulainen}]{Beuther2015}
Beuther, H., Ragan, S.~E., Johnston, K., {et~al.} 2015, \aap, 584, A67

\bibitem[{Beuther {et~al.}(2012)Beuther, Tackenberg, Linz, Henning, Schuller,
  Wyrowski, Schilke, Menten, Robitaille, Walmsley, Bronfman, Motte,
  Nguyen-Luong, \& Bontemps}]{Beuther2012}
Beuther, H., Tackenberg, J., Linz, H., {et~al.} 2012, ApJ, 747, 43

\bibitem[{Caldu-Primo {et~al.}(2013)Caldu-Primo, Schruba, Walter, Leroy,
  Sandstrom, de~Blok, Ianjamasimanana, \& Mogotsi}]{CalduPrimo2013}
Caldu-Primo, A., Schruba, A., Walter, F., {et~al.} 2013, \aj, 146, 150

\bibitem[{Chira {et~al.}(2018)Chira, Kainulainen,
  Ib{\`{a}}{\~{n}}ez-Mej{\'{i}}a, Henning, \& {Mac Low}}]{Chira2017}
Chira, R.~A., Kainulainen, J., Ib{\`{a}}{\~{n}}ez-Mej{\'{i}}a, J.~C., Henning,
  T., \& {Mac Low}, M.~M. 2018, \aap, 610, A62

\bibitem[{Contreras {et~al.}(2013{\natexlab{a}})Contreras, Rathborne, \&
  Garay}]{Contreras2013a}
Contreras, Y., Rathborne, J., \& Garay, G. 2013{\natexlab{a}}, \mnras, 433, 251

\bibitem[{Contreras {et~al.}(2013{\natexlab{b}})Contreras, Schuller, Urquhart,
  Csengeri, Wyrowski, Beuther, Bontemps, Bronfman, Henning, Menten, Schilke,
  Walmsley, Wienen, Tackenberg, \& Linz}]{Contreras2013}
Contreras, Y., Schuller, F., Urquhart, J.~S., {et~al.} 2013{\natexlab{b}},
  \aap, 549, A45

\bibitem[{Csengeri {et~al.}(2014)Csengeri, Urquhart, Schuller, Motte, Bontemps,
  Wyrowski, Menten, Bronfman, Beuther, Henning, Testi, Zavagno, \&
  Walmsley}]{Csengeri2014}
Csengeri, T., Urquhart, J.~S., Schuller, F., {et~al.} 2014, \aap, 565, A75

\bibitem[{Csengeri {et~al.}(2016)Csengeri, Weiss, Wyrowski, Menten, Urquhart,
  Leurini, Schuller, Beuther, Bontemps, Bronfman, Henning, \&
  Schneider}]{Csengeri2016}
Csengeri, T., Weiss, A., Wyrowski, F., {et~al.} 2016, \aap, 585, A104

\bibitem[{{Evans II} {et~al.}(2009){Evans II}, Dunham, J{\o}rgensen, Enoch,
  Mer{\'{i}}n, van Dishoeck, Alcal{\'{a}}, Myers, Stapelfeldt, Huard, Allen,
  Harvey, van Kempen, Blake, Koerner, Mundy, Padgett, \& Sargent}]{Evans2009}
{Evans II}, N.~J., Dunham, M.~M., J{\o}rgensen, J.~K., {et~al.} 2009, ApJ, 181,
  321

\bibitem[{Federrath(2016)}]{Federrath2016}
Federrath, C. 2016, \mnras, 457, 375

\bibitem[{Fiege \& Pudritz(2000)}]{Fiege2000a}
Fiege, J.~D. \& Pudritz, R.~E. 2000, \mnras, 311, 85

\bibitem[{Fischera \& Martin(2012)}]{Fischera2012a}
Fischera, J. \& Martin, P.~G. 2012, \aap, 542, A77

\bibitem[{Gonzalez \& Woods(1992)}]{Gonzalez1992}
Gonzalez, R.~C. \& Woods, R.~E. 1992, Reading, MA Addison-Wesley, 1992

\bibitem[{G{\"{u}}sten {et~al.}(2006)G{\"{u}}sten, Nyman, Schilke, Menten,
  Cesarsky, \& Booth}]{Guesten2006}
G{\"{u}}sten, R., Nyman, L.~A., Schilke, P., {et~al.} 2006, \aap, 454, L13

\bibitem[{Hacar {et~al.}(2016)Hacar, Kainulainen, Tafalla, Beuther, \&
  Alves}]{Hacar2015}
Hacar, A., Kainulainen, J., Tafalla, M., Beuther, H., \& Alves, J. 2016, \aap,
  587, A97

\bibitem[{{Hacar} {et~al.}(2018){Hacar}, {Tafalla}, {Forbrich}, {Alves},
  {Meingast}, {Grossschedl}, \& {Teixeira}}]{Hacar2018}
{Hacar}, A., {Tafalla}, M., {Forbrich}, J., {et~al.} 2018, \aap, 610, A77

\bibitem[{Hacar {et~al.}(2013)Hacar, Tafalla, Kauffmann, \&
  Kov{\'{a}}cs}]{Hacar2013}
Hacar, A., Tafalla, M., Kauffmann, J., \& Kov{\'{a}}cs, A. 2013, \aap, 554, A55

\bibitem[{Heitsch(2013{\natexlab{a}})}]{Heitsch2013a}
Heitsch, F. 2013{\natexlab{a}}, ApJ, 769, 115

\bibitem[{Heitsch(2013{\natexlab{b}})}]{Heitsch2013}
Heitsch, F. 2013{\natexlab{b}}, ApJ, 776, 62

\bibitem[{Heitsch {et~al.}(2009)Heitsch, Ballesteros-Paredes, \&
  Hartmann}]{Heitsch2009}
Heitsch, F., Ballesteros-Paredes, J., \& Hartmann, L. 2009, ApJ, 704, 1735

\bibitem[{Henshaw {et~al.}(2014)Henshaw, Caselli, Fontani, Jimenez-Serra, \&
  Tan}]{Henshaw2014}
Henshaw, J.~D., Caselli, P., Fontani, F., Jimenez-Serra, I., \& Tan, J.~C.
  2014, \mnras, 440, 2860

\bibitem[{Hernandez {et~al.}(2012)Hernandez, Tan, Kainulainen, Caselli, Butler,
  Jimenez-Serra, Fontani, Jim{\'{e}}nez-Serra, \& Fontani}]{Hernandez2012}
Hernandez, A.~K., Tan, J.~C., Kainulainen, J., {et~al.} 2012, ApJ, 756, L13

\bibitem[{Hill {et~al.}(2011)Hill, Motte, Didelon, Bontemps, Minier, Hennemann,
  Schneider, Andr{\'{e}}, Men‘shchikov, Anderson, Arzoumanian, Bernard,
  di~Francesco, Elia, Giannini, Griffin, K{\"{o}}nyves, Kirk, Marston, Martin,
  Molinari, {Nguyen Luong}, Peretto, Pezzuto, Roussel, Sauvage, Sousbie, Testi,
  Ward-Thompson, White, Wilson, \& Zavagno}]{Hill2011}
Hill, T., Motte, F., Didelon, P., {et~al.} 2011, \aap, 533, A94

\bibitem[{Inutsuka \& Miyama(1992)}]{Inutsuka1992}
Inutsuka, S. \& Miyama, S.~M. 1992, ApJ, 338, 392

\bibitem[{Jackson {et~al.}(2010)Jackson, Finn, Chambers, Rathborne, \&
  Simon}]{Jackson2010}
Jackson, J., Finn, S., Chambers, E., Rathborne, J., \& Simon, R. 2010, \apjl,
  719, L185

\bibitem[{Johnstone {et~al.}(2003)Johnstone, Jason, Redman, Feldman, Carey,
  Fiege, Redman, Feldman, Carey, Jason, Redman, Feldman, \&
  Carey}]{Johnstone2003}
Johnstone, D., Jason, D.~F., Redman, R.~O., {et~al.} 2003, ApJ, 588, L37

\bibitem[{Juvela {et~al.}(2012)Juvela, Malinen, \& Lunttila}]{Juvela2012}
Juvela, M., Malinen, J., \& Lunttila, T. 2012, \aap, 544, 11 pp

\bibitem[{Kainulainen {et~al.}(2015)Kainulainen, Hacar, Alves, Beuther, Bouy,
  Tafalla, Kainulainen, Hacar, Alves, Beuther, Bouy, \&
  Tafalla}]{Kainulainen2015}
Kainulainen, J., Hacar, A., Alves, J., {et~al.} 2015, \aap, 586, A27

\bibitem[{Kainulainen {et~al.}(2013)Kainulainen, Ragan, Henning, \&
  Stutz}]{Kainulainen2013a}
Kainulainen, J., Ragan, S.~E., Henning, T., \& Stutz, A. 2013, \aap, 557, A120

\bibitem[{{Kainulainen} {et~al.}(2017){Kainulainen}, {Stutz}, {Stanke},
  {Abreu-Vicente}, {Beuther}, {Henning}, {Johnston}, \&
  {Megeath}}]{Kainulainen2016}
{Kainulainen}, J., {Stutz}, A.~M., {Stanke}, T., {et~al.} 2017, \aap, 600, A141

\bibitem[{Kainulainen {et~al.}(2017)Kainulainen, Stutz, Stanke, Abreu-Vicente,
  Beuther, Henning, Johnston, \& Megeath}]{Kainulainen2017}
Kainulainen, J., Stutz, A.~M., Stanke, T., {et~al.} 2017, \aap, 600, A141

\bibitem[{Kauffmann {et~al.}(2008)Kauffmann, Bertoldi, Bourke, Evans, \&
  Lee}]{Kauffmann2008}
Kauffmann, J., Bertoldi, F., Bourke, T.~L., Evans, N.~J., \& Lee, C.~W. 2008,
  \aap, 487, 993

\bibitem[{Kirk {et~al.}(2013)Kirk, Myers, Bourke, Gutermuth, Hedden, \&
  Wilson}]{Kirk2013}
Kirk, H., Myers, P.~C., Bourke, T.~L., {et~al.} 2013, ApJ, 766, 115

\bibitem[{Koch \& Rosolowsky(2015)}]{Koch2015}
Koch, E.~W. \& Rosolowsky, E.~W. 2015, \mnras, 452, 3435

\bibitem[{Lamarre {et~al.}(2010)Lamarre, Puget, Ade, Bouchet, Guyot, Lange,
  Pajot, Arondel, Benabed, Beney, Beno{\^{i}}t, Bernard, Bhatia, Blanc, Bock,
  Br{\'{e}}elle, Bradshaw, Camus, Catalano, Charra, Charra, Church, Couchot,
  Coulais, Crill, Crook, Dassas, de~Bernardis, Delabrouille, de~Marcillac,
  Delouis, D{\'{e}}sert, Dumesnil, Dupac, Efstathiou, Eng, Evesque, Fourmond,
  Ganga, Giard, Gispert, Guglielmi, Haissinski, Henrot-Versill{\'{e}}, Hivon,
  Holmes, Jones, Koch, Lagard{\`{e}}re, Lami, Land{\'{e}}, Leriche, Leroy,
  Longval, Mac{\'{i}}as-P{\'{e}}rez, Maciaszek, Maffei, Mansoux, Marty, Masi,
  Mercier, Miville-Desch{\^{e}}nes, Moneti, Montier, Murphy, Narbonne, Nexon,
  Paine, Pahn, Perdereau, Piacentini, Piat, Plaszczynski, Pointecouteau, Pons,
  Ponthieu, Prunet, Rambaud, Recouvreur, Renault, Ristorcelli, Rosset, Santos,
  Savini, Serra, Stassi, Sudiwala, Sygnet, Tauber, Torre, Tristram, Vibert,
  Woodcraft, Yurchenko, \& Yvon}]{Lamarre2010}
Lamarre, J.-M., Puget, J.-L., Ade, P. A.~R., {et~al.} 2010, \aap, 520, A9

\bibitem[{Li {et~al.}(2016)Li, Urquhart, Leurini, Csengeri, Wyrowski, Menten,
  \& Schuller}]{Li2016b}
Li, G.-X., Urquhart, J.~S., Leurini, S., {et~al.} 2016, \aap, 591, A5

\bibitem[{Mattern {et~al.}(2018)Mattern, Kainulainen, Zhang, \&
  Beuther}]{Mattern2018a}
Mattern, M., Kainulainen, J., Zhang, M., \& Beuther, H. 2018, eprint
  arXiv:1804.02256 [\eprint[arXiv]{1804.02256}]

\bibitem[{Miettinen(2012)}]{Miettinen2012}
Miettinen, O. 2012, \aap, 545, A3

\bibitem[{Molinari {et~al.}(2010)Molinari, Bally, Barlow, Bernard, Martin,
  Moore, Noriega-Crespo, Plume, Testi, Zavagno, Abergel, Ali, Andr{\'{e}},
  Baluteau, Benedettini, Bern{\'{e}}, Billot, Blommaert, Bontemps, Boulanger,
  Brand, Brunt, Burton, Campeggio, Carey, Caselli, Cesaroni, Cernicharo,
  Chakrabarti, Chrysostomou, Codella, Cohen, Compiegne, Davis, de~Bernardis,
  de~Gasperis, {Di Francesco}, di~Giorgio, Elia, Faustini, Fischera, Fukui,
  Fuller, Ganga, Garcia-Lario, Giard, Giardino, Glenn, Goldsmith, Griffin,
  Hoare, Huang, Jiang, Joblin, Joncas, Juvela, Kirk, Lagache, Li, Lim, Lord,
  Lucas, Maiolo, Marengo, Marshall, Masi, Massi, Matsuura, Meny, Minier,
  Miville-Desch{\^{e}}nes, Montier, Motte, M{\"{u}}ller, Natoli, Neves, Olmi,
  Paladini, Paradis, Pestalozzi, Pezzuto, Piacentini, Pomar{\`{e}}s, Popescu,
  Reach, Richer, Ristorcelli, Roy, Royer, Russeil, Saraceno, Sauvage, Schilke,
  Schneider-Bontemps, Schuller, Schultz, Shepherd, Sibthorpe, Smith, Smith,
  Spinoglio, Stamatellos, Strafella, Stringfellow, Sturm, Taylor, Thompson,
  Tuffs, Umana, Valenziano, Vavrek, Viti, Waelkens, Ward-Thompson, White,
  Wyrowski, Yorke, Zhang, Swinyard, Bally, Barlow, Bernard, Martin, Moore,
  Noriega-Crespo, Plume, Testi, Zavagno, Abergel, Ali, Andr{\'{e}}, Baluteau,
  Benedettini, Bern{\'{e}}, Billot, Blommaert, Bontemps, Boulanger, Brand,
  Brunt, Burton, Campeggio, Carey, Caselli, Cesaroni, Cernicharo, Chakrabarti,
  Chrysostomou, Codella, Cohen, Compiegne, Davis, de~Bernardis, de~Gasperis,
  {Di Francesco}, di~Giorgio, Elia, Faustini, Fischera, Fukui, Fuller, Ganga,
  Garcia-Lario, Giard, Giardino, Glenn, Goldsmith, Griffin, Hoare, Huang,
  Jiang, Joblin, Joncas, Juvela, Kirk, Lagache, Li, Lim, Lord, Lucas, Maiolo,
  Marengo, Marshall, Masi, Massi, Matsuura, Meny, Minier,
  Miville-Desch{\^{e}}nes, Montier, Motte, M{\"{u}}ller, Natoli, Neves, Olmi,
  Paladini, Paradis, Pestalozzi, Pezzuto, Piacentini, Pomar{\`{e}}s, Popescu,
  Reach, Richer, Ristorcelli, Roy, Royer, Russeil, Saraceno, Sauvage, Schilke,
  Schneider-Bontemps, Schuller, Schultz, Shepherd, Sibthorpe, Smith, Smith,
  Spinoglio, Stamatellos, Strafella, Stringfellow, Sturm, Taylor, Thompson,
  Tuffs, Umana, Valenziano, Vavrek, Viti, Waelkens, Ward-Thompson, White,
  Wyrowski, Yorke, \& Zhang}]{Molinari2010}
Molinari, S., Bally, J., Barlow, M., {et~al.} 2010, \pasp, 122, 314

\bibitem[{Nutter {et~al.}(2008)Nutter, Kirk, Stamatellos, \&
  Ward-Thompson}]{Nutter2008}
Nutter, D., Kirk, J.~M., Stamatellos, D., \& Ward-Thompson, D. 2008, \mnras,
  384, 755

\bibitem[{Ossenkopf \& Henning(1994)}]{Ossenkopf1994}
Ossenkopf, V. \& Henning, T. 1994, \aap, 291, 943

\bibitem[{Ostriker(1964)}]{Ostriker1964}
Ostriker, J. 1964, ApJ, 140, 1056

\bibitem[{Padoan {et~al.}(2001)Padoan, Juvela, Goodman, \&
  Nordlund}]{Padoan2001}
Padoan, P., Juvela, M., Goodman, A., \& Nordlund, {\AA}. 2001, \apj, 553, 227

\bibitem[{Palmeirim {et~al.}(2013)Palmeirim, Andr{\'{e}}, Kirk, Ward-Thompson,
  Arzoumanian, K{\"{o}}nyves, Didelon, Schneider, Benedettini, Bontemps, {Di
  Francesco}, Elia, Griffin, Hennemann, Hill, Martin, Men'shchikov, Molinari,
  Motte, {Nguyen Luong}, Nutter, Peretto, Pezzuto, Roy, Rygl, Spinoglio, White,
  \& White}]{Palmeirim2013}
Palmeirim, P., Andr{\'{e}}, P., Kirk, J., {et~al.} 2013, \aap, 550, A38

\bibitem[{Panopoulou {et~al.}(2014)Panopoulou, Tassis, Goldsmith, \&
  Heyer}]{Panopoulou2014}
Panopoulou, G.~V., Tassis, K., Goldsmith, P.~F., \& Heyer, M.~H. 2014, \mnras,
  444, 2507

\bibitem[{Peretto {et~al.}(2014)Peretto, Fuller, Andr{\'{e}}, Arzoumanian,
  Rivilla, Bardeau, {Duarte Puertas}, {Guzman Fernandez}, Lenfestey, Li,
  Olguin, R{\"{o}}ck, de~Villiers, \& Williams}]{Peretto2014a}
Peretto, N., Fuller, G.~A., Andr{\'{e}}, P., {et~al.} 2014, \aap, 561, A83

\bibitem[{Peretto {et~al.}(2013)Peretto, Fuller, Duarte-Cabral, Avison,
  Hennebelle, Pineda, Andr{\'{e}}, Bontemps, Motte, Schneider, \&
  Molinari}]{Peretto2013}
Peretto, N., Fuller, G.~a., Duarte-Cabral, A., {et~al.} 2013, \aap, 555, A112

\bibitem[{Pillai {et~al.}(2006{\natexlab{a}})Pillai, Wyrowski, Carey, \&
  Menten}]{Pillai2006a}
Pillai, T., Wyrowski, F., Carey, S.~J., \& Menten, K.~M. 2006{\natexlab{a}},
  \aap, 450, 569

\bibitem[{Pillai {et~al.}(2006{\natexlab{b}})Pillai, Wyrowski, Menten, \&
  Kr{\"{u}}gel}]{Pillai2006}
Pillai, T., Wyrowski, F., Menten, K.~M., \& Kr{\"{u}}gel, E.
  2006{\natexlab{b}}, \aap, 447, 929

\bibitem[{{Planck Collaboration} {et~al.}(2014){Planck Collaboration}, Ade,
  Aghanim, Alves, Armitage-Caplan, Arnaud, Ashdown, Atrio-Barandela, Aumont,
  Aussel, Baccigalupi, Banday, Barreiro, Barrena, Bartelmann, Bartlett,
  Bartolo, Basak, Battaner, Battye, Benabed, Beno{\^{i}}t, Benoit-L{\'{e}}vy,
  Bernard, Bersanelli, Bertincourt, Bethermin, Bielewicz, Bikmaev, Blanchard,
  Bobin, Bock, B{\"{o}}hringer, Bonaldi, Bonavera, Bond, Borrill, Bouchet,
  Boulanger, Bourdin, Bowyer, Bridges, Brown, Bucher, Burenin, Burigana,
  Butler, Calabrese, Cappellini, Cardoso, Carr, Carvalho, Casale, Castex,
  Catalano, Challinor, Chamballu, Chary, Chen, Chiang, Chiang, Chon,
  Christensen, Churazov, Church, Clemens, Clements, Colombi, Colombo, Combet,
  Comis, Couchot, Coulais, Crill, Cruz, Curto, Cuttaia, {Da Silva}, Dahle,
  Danese, Davies, Davis, de~Bernardis, de~Rosa, de~Zotti, D{\'{e}}chelette,
  Delabrouille, Delouis, D{\'{e}}mocl{\`{e}}s, D{\'{e}}sert, Dick, Dickinson,
  Diego, Dolag, Dole, Donzelli, Dor{\'{e}}, Douspis, Ducout, Dunkley, Dupac,
  Efstathiou, Elsner, En{\ss}lin, Eriksen, Fabre, Falgarone, Falvella, Fantaye,
  Fergusson, Filliard, Finelli, Flores-Cacho, Foley, Forni, Fosalba, Frailis,
  Fraisse, Franceschi, Freschi, Fromenteau, Frommert, Gaier, Galeotta,
  Gallegos, Galli, Gandolfo, Ganga, Gauthier, G{\'{e}}nova-Santos, Ghosh,
  Giard, Giardino, Gilfanov, Girard, Giraud-H{\'{e}}raud, Gjerl{\o}w,
  Gonz{\'{a}}lez-Nuevo, G{\'{o}}rski, Gratton, Gregorio, Gruppuso, Gudmundsson,
  Haissinski, Hamann, Hansen, Hansen, Hanson, Harrison, Heavens, Helou, Hempel,
  Henrot-Versill{\'{e}}, Hern{\'{a}}ndez-Monteagudo, Herranz, Hildebrandt,
  Hivon, Ho, Hobson, Holmes, Hornstrup, Hou, Hovest, Huey, Huffenberger,
  Hurier, Ili{\'{c}}, Jaffe, Jaffe, Jasche, Jewell, Jones, Juvela, Kalberla,
  Kangaslahti, Keih{\"{a}}nen, Kerp, Keskitalo, Khamitov, Kiiveri, Kim, Kisner,
  Kneissl, Knoche, Knox, Kunz, Kurki-Suonio, Lacasa, Lagache,
  L{\"{a}}hteenm{\"{a}}ki, Lamarre, Langer, Lasenby, Lattanzi, Laureijs,
  Lavabre, Lawrence, Jeune, Leach, Leahy, Leonardi, Le{\'{o}}n-Tavares, Leroy,
  Lesgourgues, Lewis, Li, Liddle, Liguori, Lilje, Linden-V{\o}rnle, Lindholm,
  L{\'{o}}pez-Caniego, Lowe, Lubin, Mac{\'{i}}as-P{\'{e}}rez, MacTavish,
  Maffei, Maggio, Maino, Mandolesi, Mangilli, Marcos-Caballero, Marinucci,
  Maris, Marleau, Marshall, Martin, Mart{\'{i}}nez-Gonz{\'{a}}lez, Masi,
  Massardi, Matarrese, Matsumura, Matthai, Maurin, Mazzotta, McDonald, McEwen,
  McGehee, Mei, Meinhold, Melchiorri, Melin, Mendes, Menegoni, Mennella,
  Migliaccio, Mikkelsen, Millea, Miniscalco, Mitra, Miville-Desch{\^{e}}nes,
  Molinari, Moneti, Montier, Morgante, Morisset, Mortlock, Moss, Munshi,
  Murphy, Naselsky, Nati, Natoli, Negrello, Nesvadba, Netterfield,
  N{\o}rgaard-Nielsen, North, Noviello, Novikov, Novikov, O'Dwyer, Orieux,
  Osborne, O'Sullivan, Oxborrow, Paci, Pagano, Pajot, Paladini, Pandolfi,
  Paoletti, Partridge, Pasian, Patanchon, Paykari, Pearson, Pearson, Peel,
  Peiris, Perdereau, Perotto, Perrotta, Pettorino, Piacentini, Piat, Pierpaoli,
  Pietrobon, Plaszczynski, Platania, Pogosyan, Pointecouteau, Polenta,
  Ponthieu, Popa, Poutanen, Pratt, Pr{\'{e}}zeau, Prunet, Puget, Pullen,
  Rachen, Racine, Rahlin, R{\"{a}}th, Reach, Rebolo, Reinecke, Remazeilles,
  Renault, Renzi, Riazuelo, Ricciardi, Riller, Ringeval, Ristorcelli, Robbers,
  Rocha, Roman, Rosset, Rossetti, Roudier, Rowan-Robinson,
  Rubi{\~{n}}o-Mart{\'{i}}n, Ruiz-Granados, Rusholme, Salerno, Sandri,
  Sanselme, Santos, Savelainen, Savini, Schaefer, Schiavon, Scott, Seiffert,
  Serra, Shellard, Smith, Smoot, Souradeep, Spencer, Starck, Stolyarov,
  Stompor, Sudiwala, Sunyaev, Sureau, Sutter, Sutton, Suur-Uski, Sygnet,
  Tauber, Tavagnacco, Taylor, Terenzi, Texier, Toffolatti, Tomasi, Torre,
  Tristram, Tucci, Tuovinen, T{\"{u}}rler, Tuttlebee, Umana, Valenziano,
  Valiviita, {Van Tent}, Varis, Vibert, Viel, Vielva, Villa, Vittorio, Wade,
  Wandelt, Watson, Watson, Wehus, Welikala, Weller, White, White, Wilkinson,
  Winkel, Xia, Yvon, Zacchei, Zibin, \& Zonca}]{PlanckI2014}
{Planck Collaboration}, P., Ade, P. A.~R., Aghanim, N., {et~al.} 2014, \aap,
  571, A1

\bibitem[{Ragan {et~al.}(2014)Ragan, Henning, Tackenberg, Beuther, Johnston,
  Kainulainen, \& Linz}]{Ragan2014}
Ragan, S.~E., Henning, T., Tackenberg, J., {et~al.} 2014, \aap, 568, A73

\bibitem[{Recchi {et~al.}(2013)Recchi, Hacar, \& Palestini}]{Recchi2013}
Recchi, S., Hacar, A., \& Palestini, A. 2013, \aap, 558, A27

\bibitem[{Reid {et~al.}(2016)Reid, Dame, Menten, \& Brunthaler}]{Reid2016}
Reid, M.~J., Dame, T.~M., Menten, K.~M., \& Brunthaler, A. 2016, ApJ, 823, 77

\bibitem[{Schisano {et~al.}(2014)Schisano, Rygl, Molinari, Busquet, Elia,
  Pestalozzi, Polychroni, Billot, Carey, Paladini, Noriega-Crespo, Moore,
  Plume, Glover, \& V{\'{a}}zquez-Semadeni}]{Schisano2014}
Schisano, E., Rygl, K. L.~J., Molinari, S., {et~al.} 2014, ApJ, 791, 27

\bibitem[{Schneider {et~al.}(2010)Schneider, Csengeri, Bontemps, Motte, Simon,
  Hennebelle, Federrath, \& Klessen}]{Schneider2010}
Schneider, N., Csengeri, T., Bontemps, S., {et~al.} 2010, \aap, 520, A49

\bibitem[{Schneider \& Elmegreen(1979)}]{Schneider1979}
Schneider, S. \& Elmegreen, B.~G. 1979, ApJ, 41, 87

\bibitem[{Schuller {et~al.}(2017)Schuller, Csengeri, Urquhart, Duarte-Cabral,
  Barnes, Giannetti, Hernandez, Leurini, Mattern, Medina, Agurto, Azagra,
  Anderson, Beltr{\'{a}}n, Beuther, Bontemps, Bronfman, Dobbs, Dumke, Finger,
  Ginsburg, Gonzalez, Henning, Kauffmann, Mac-Auliffe, Menten,
  Montenegro-Montes, Moore, Muller, Parra, Perez-Beaupuits, Pettitt, Russeil,
  S{\'{a}}nchez-Monge, Schilke, Schisano, Suri, Testi, Torstensson, Venegas,
  Wang, Wienen, Wyrowski, \& Zavagno}]{Schuller2017}
Schuller, F., Csengeri, T., Urquhart, J.~S., {et~al.} 2017, \aap, 601, A124

\bibitem[{Schuller {et~al.}(2009)Schuller, Menten, Contreras, Wyrowski,
  Schilke, Bronfman, Henning, Walmsley, Beuther, Bontemps, Cesaroni, Deharveng,
  Garay, Herpin, Lefloch, Linz, Mardones, Minier, Molinari, Motte, Nyman,
  Reveret, Risacher, Russeil, Schneider, Testi, Troost, Vasyunina, Wienen,
  Zavagno, Kovacs, Kreysa, Siringo, \& Wei{\ss}}]{Schuller2009}
Schuller, F., Menten, K.~M., Contreras, Y., {et~al.} 2009, \aap, 504, 415

\bibitem[{Smith {et~al.}(2015)Smith, Glover, Klessen, Fuller, Smith, Glover,
  Klessen, \& Fuller}]{Smith2015}
Smith, R.~J., Glover, S. C.~O., Klessen, R.~S., {et~al.} 2015, \mnras, 455,
  3640

\bibitem[{Sousbie(2011)}]{Sousbie2011}
Sousbie, T. 2011, \mnras, 414, 350

\bibitem[{Stutz \& Gould(2016)}]{Stutz2016}
Stutz, A.~M. \& Gould, A. 2016, \aap, 590, A2

\bibitem[{Takahashi {et~al.}(2013)Takahashi, Ho, Teixeira, Zapata, \&
  Su}]{Takahashi2013}
Takahashi, S., Ho, P. T.~P., Teixeira, P.~S., Zapata, L.~A., \& Su, Y.-N. 2013,
  ApJ, 763, 57

\bibitem[{Teixeira {et~al.}(2016)Teixeira, Takahashi, Zapata, Ho, \&
  Ho}]{Teixeira2016}
Teixeira, P.~S., Takahashi, S., Zapata, L.~A., Ho, P. T.~P., \& Ho, P. T.~P.
  2016, \aap, 587, A47

\bibitem[{Urquhart {et~al.}(2014)Urquhart, Csengeri, Wyrowski, Schuller,
  Bontemps, Bronfman, Menten, Walmsley, Contreras, Beuther, Wienen, \&
  Linz}]{Urquhart2014}
Urquhart, J.~S., Csengeri, T., Wyrowski, F., {et~al.} 2014, \aap, 568, A41

\bibitem[{Urquhart {et~al.}(2018)Urquhart, K{\"{o}}nig, Giannetti, Leurini,
  Moore, Eden, Pillai, Thompson, Braiding, Burton, Csengeri, Dempsey, Figura,
  Froebrich, Menten, Schuller, Smith, \& Wyrowski}]{Urquhart2018}
Urquhart, J.~S., K{\"{o}}nig, C., Giannetti, A., {et~al.} 2018, \mnras, 473,
  1059

\bibitem[{Vassilev {et~al.}(2008)Vassilev, Meledin, Lapkin, Belitsky,
  Nystr{\"{o}}m, Henke, Pavolotsky, Monje, Risacher, Olberg, Strandberg,
  Sundin, Fredrixon, Ferm, Desmaris, Dochev, Pantaleev, Bergman, \&
  Olofsson}]{Vassilev2008}
Vassilev, V., Meledin, D., Lapkin, I., {et~al.} 2008, \aap, 490, 1157

\bibitem[{Wang {et~al.}(2016)Wang, Testi, Burkert, Walmsley, Beuther, \&
  Henning}]{Wang2016}
Wang, K., Testi, L., Burkert, A., {et~al.} 2016, \apjs, 226, 9

\bibitem[{Wang {et~al.}(2015)Wang, Testi, Ginsburg, Walmsley, Molinari, \&
  Schisano}]{Wang2015}
Wang, K., Testi, L., Ginsburg, A., {et~al.} 2015, \mnras, 450, 4043

\bibitem[{Wang {et~al.}(2014)Wang, Zhang, Testi, Tak, Wu, Zhang, Pillai,
  Wyrowski, Carey, Ragan, \& Henning}]{Wang2014}
Wang, K., Zhang, Q., Testi, L., {et~al.} 2014, \mnras, 439, 3275

\bibitem[{Wang {et~al.}(2012)Wang, Zhang, Wu, Li, \& Zhang}]{Wang2012}
Wang, K., Zhang, Q., Wu, Y., Li, H.~B., \& Zhang, H. 2012, \apj, 745

\bibitem[{Zucker {et~al.}(2015)Zucker, Battersby, \& Goodman}]{Zucker2015}
Zucker, C., Battersby, C., \& Goodman, A. 2015, ApJ, 815, 23

\bibitem[{{Zucker} {et~al.}(2017){Zucker}, {Battersby}, \&
  {Goodman}}]{Zucker2017}
{Zucker}, C., {Battersby}, C., \& {Goodman}, A. 2017, ArXiv e-prints
  [\eprint[arXiv]{1712.09655}]

\end{thebibliography}

\bibliographystyle{aa}

\longtab[1]{
\begin{landscape}
\begin{longtable}{lcclcccccccccc}
\caption{tbd}
\label{final-cat1}\\
\hline\hline
Filament ID & $\ell$ & $b$ & Status & $N_\text{c}$ & $v_{lsr}$\small{($^{13}$CO)} & $v_{lsr}$\small{(C$^{18}$O)} & $\sigma(v_{\rm ^{13}CO}) $ & $\sigma(v_{\rm C^{18}O})$ & $\sigma_\text{v}$ & $\sigma_\text{v}$\small{($^{13}$CO)} & $\sigma_\text{v}$\small{(C$^{18}$O)} & $\sigma_\text{v,t}$\small{($^{13}$CO)} & $\sigma_\text{v,t}$\small{(C$^{18}$O)} \\
 & $\degree$ & $\degree$ &  &  & $\rm km\,s^{-1}$ &  $\rm km\,s^{-1}$ &  $\rm km\,s^{-1}$ &  $\rm km\,s^{-1}$ &  $\rm km\,s^{-1}$ &  $\rm km\,s^{-1}$ &  $\rm km\,s^{-1}$ &  $\rm km\,s^{-1}$ &  $\rm km\,s^{-1}$ \\
\hline
\endfirsthead
\caption{continued.}\\
\hline\hline
Filament ID & $\ell$ & $b$ & Status & $N_\text{c}$ & $v_{lsr}$\small{($^{13}$CO)} & $v_{lsr}$\small{(C$^{18}$O)} & $\sigma(v_{\rm ^{13}CO}) $ & $\sigma(v_{\rm C^{18}O})$ & $\sigma_\text{v}$ & $\sigma_\text{v}$\small{($^{13}$CO)} & $\sigma_\text{v}$\small{(C$^{18}$O)} & $\sigma_\text{v,t}$\small{($^{13}$CO)} & $\sigma_\text{v,t}$\small{(C$^{18}$O)} \\
 & $\degree$ & $\degree$ &  &  & $\rm km\,s^{-1}$ &  $\rm km\,s^{-1}$ &  $\rm km\,s^{-1}$ &  $\rm km\,s^{-1}$ &  $\rm km\,s^{-1}$ &  $\rm km\,s^{-1}$ &  $\rm km\,s^{-1}$ &  $\rm km\,s^{-1}$ &  $\rm km\,s^{-1}$ \\
\hline
\endhead
\hline
\endfoot
G000.615-00.448\_0 & 0.61 & -0.45 & full. cor. & 1 & 15.69 & 15.75 & 0.56 & 7.18 & 0.84 & 0.95 & 0.5 & 1.19 & 1.16 \\
G000.675+00.310\_0 & 0.67 & 0.31 & part. cor. & 1 & 136.44 & - & 1.33 & 0.0 & 1.72 & 1.73 & - & 3.79 & - \\
G000.914+00.308\_0 & 0.91 & 0.31 & part. cor. & 1 & 174.07 & 173.73 & 1.99 & 0.0 & 1.94 & 1.95 & - & 4.46 & 4.88 \\
G001.717+00.359\_0 & 1.72 & 0.36 & part. cor. & 3 & 3.07 & 2.34 & 0.36 & 0.53 & 0.79 & 0.92 & 0.46 & 1.24 & 0.81 \\
G001.717+00.359\_1 & 1.72 & 0.36 & part. cor. & 3 & 8.93 & - & 0.44 & 1.19 & 0.97 & 1.07 & 0.61 & 1.76 & - \\
G001.717+00.359\_2 & 1.72 & 0.36 & part. cor. & 3 & -17.3 & - & 3.17 & 0.0 & 1.73 & 1.74 & - & 4.65 & - \\
G002.128+00.295\_0 & 2.13 & 0.3 & full. cor. & 1 & 3.86 & 2.85 & 0.36 & 1.66 & 1.13 & 1.23 & 0.71 & 1.98 & - \\
G002.210-00.049\_0 & 2.21 & -0.05 & part. cor. & 2 & 11.27 & 8.39 & 2.05 & 0.0 & 2.77 & 2.77 & - & 5.62 & 5.95 \\
\end{longtable}
\end{landscape}
}

\longtab[1]{
\begin{landscape}
\begin{longtable}{lccccccccccccc}
\caption{tbd}
\label{final-cat2}\\
\hline\hline
Filament ID & $d$ & $l$ & $l(d)$ & $M(\rm ATG)$ & $M(\rm ATG+P)$ & $M(\rm dust)$ & $M(\rm gas)$ & $m_\text{crit,nt}$ & $m_\text{obs}$ & det. $^{13}$CO & det. C$^{18}$O & edge flag & $d$ flag \\
 & kpc & $\degree$ & pc & $\rm M_\odot$ & $\rm M_\odot$ & $\rm M_\odot$ & $\rm M_\odot$ & $\rm M_\odot \, pc^{-1}$ & $\rm M_\odot \, pc^{-1}$ &  &  &  &  \\
\hline
\endfirsthead
\caption{continued.}\\
\hline\hline
Filament ID & $d$ & $l$ & $l(d)$ & $M(\rm ATG)$ & $M(\rm ATG+P)$ & $M(\rm dust)$ & $M(\rm gas)$ & $m_\text{crit,nt}$ & $m_\text{obs}$ & det. $^{13}$CO & det. C$^{18}$O & edge flag & $d$ flag \\
 & kpc & $\degree$ & pc & $\rm M_\odot$ & $\rm M_\odot$ & $\rm M_\odot$ & $\rm M_\odot$ & $\rm M_\odot \, pc^{-1}$ & $\rm M_\odot \, pc^{-1}$ &  &  &  &  \\
\hline
\endhead
\hline
\endfoot
G000.615-00.448\_0 & - & - & 0.11 & - & - & - & - & - & 351 & 1.0 & 0.28 & 0 & 0 \\
G000.675+00.310\_0 & - & - & 0.03 & - & - & - & - & - & 1397 & 0.18 & 0.0 & 0 & 0 \\
G000.914+00.308\_0 & - & - & 0.04 & - & - & - & - & - & 1769 & 0.25 & 0.0 & 0 & 0 \\
G001.717+00.359\_0 & - & - & 0.03 & - & - & - & - & - & 312 & 0.12 & 0.03 & 0 & 0 \\
G001.717+00.359\_1 & - & - & 0.06 & - & - & - & - & - & 459 & 0.18 & 0.02 & 0 & 0 \\
G001.717+00.359\_2 & - & - & 0.06 & - & - & - & - & - & 1410 & 0.22 & 0.0 & 0 & 0 \\
G002.128+00.295\_0 & - & - & 0.06 & - & - & - & - & - & 613 & 1.0 & 0.19 & 0 & 0 \\
G002.210-00.049\_0 & - & - & 0.04 & - & - & - & - & - & 3576 & 0.5 & 0.0 & 0 & 0 \\
\end{longtable}
\end{landscape}
}

\begin{appendix}

\section{gas dust correlation examples}
\label{cor_examples}

\begin{figure}
\centering
\includegraphics[width=0.5\textwidth, clip=true, trim= 0cm 0cm 0cm 0cm]{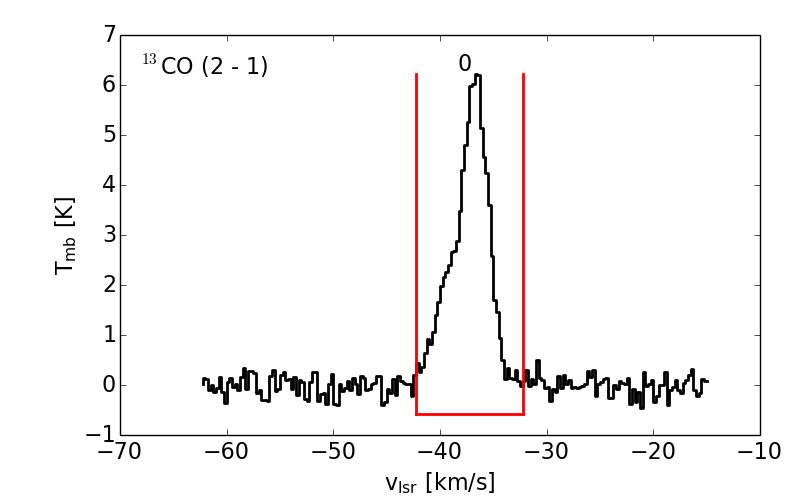}
\includegraphics[width=0.5\textwidth, clip=true, trim= 0cm 0cm 0cm 0cm]{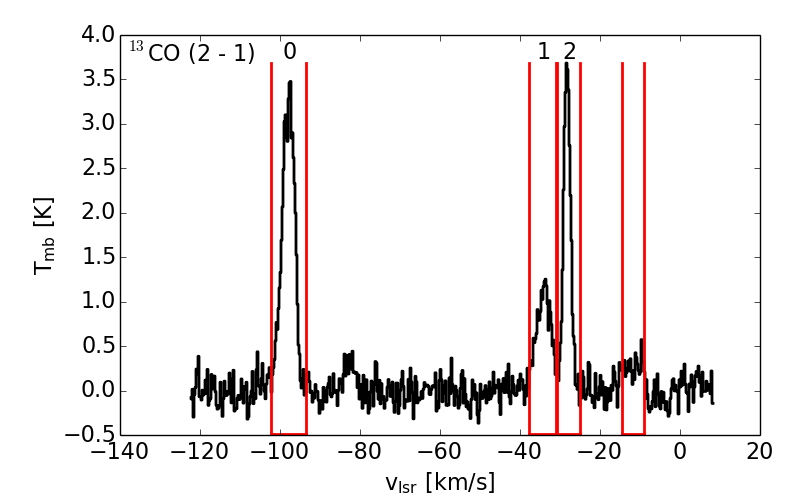}
\caption{Average \cco  spectrum over the skeleton of filament candidate ``G339.116-00.405'' (top) and ``G346.293+00.109'' (bottom). The red lines mark the identified emission intervals named by letters.}
\label{spectrum_app}
\end{figure}

\begin{figure*}
\centering
\begin{minipage}{0.32\textwidth}
\includegraphics[width=\textwidth, clip=true, trim= 0.5cm 1cm 0.0cm 0.0cm]{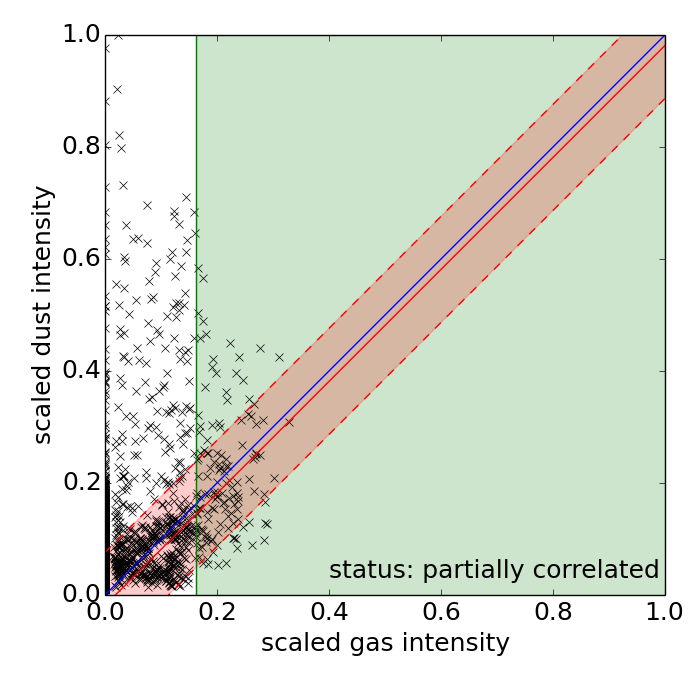}
\end{minipage}
\begin{minipage}{0.32\textwidth}
\includegraphics[width=\textwidth, clip=true, trim= 0.5cm 1cm 0.0cm 0.0cm]{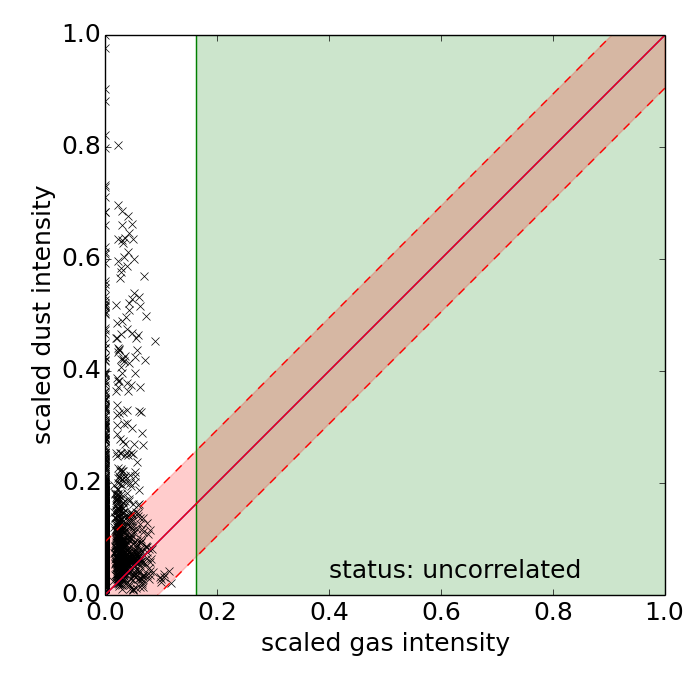}
\end{minipage}
\begin{minipage}{0.32\textwidth}
\includegraphics[width=\textwidth, clip=true, trim= 0.5cm 1cm 0.0cm 0.0cm]{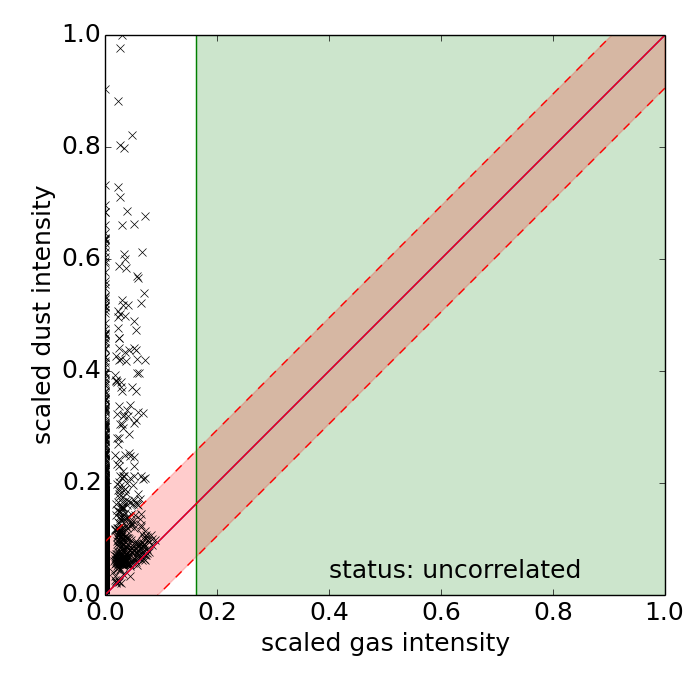}
\end{minipage}
\caption{Gas-dust correlation plots of the velocity components ``b'',``c'', and ``d'' of the filament candidate G333.297+00.073 (see Figs. \ref{spectrum}, \ref{pv-plot}, and \ref{integrated}). The blue line gives the one-to-one correlation. The green area indicates values above the $\sigma_\text{gas}$ limit. The red line shows the fitting result, and the area within the dashed red lines marks the $\pm \sigma_\text{cor}$ surrounding. $p_\text{cor, gas}$ is estimated from the overlap of these areas.}
\label{cor_rest}
\end{figure*}

\begin{figure*}
\centering
\begin{minipage}{0.59\textwidth}
\includegraphics[width=\textwidth, clip=true, trim= 0.cm 2.cm 0.0cm 3.0cm]{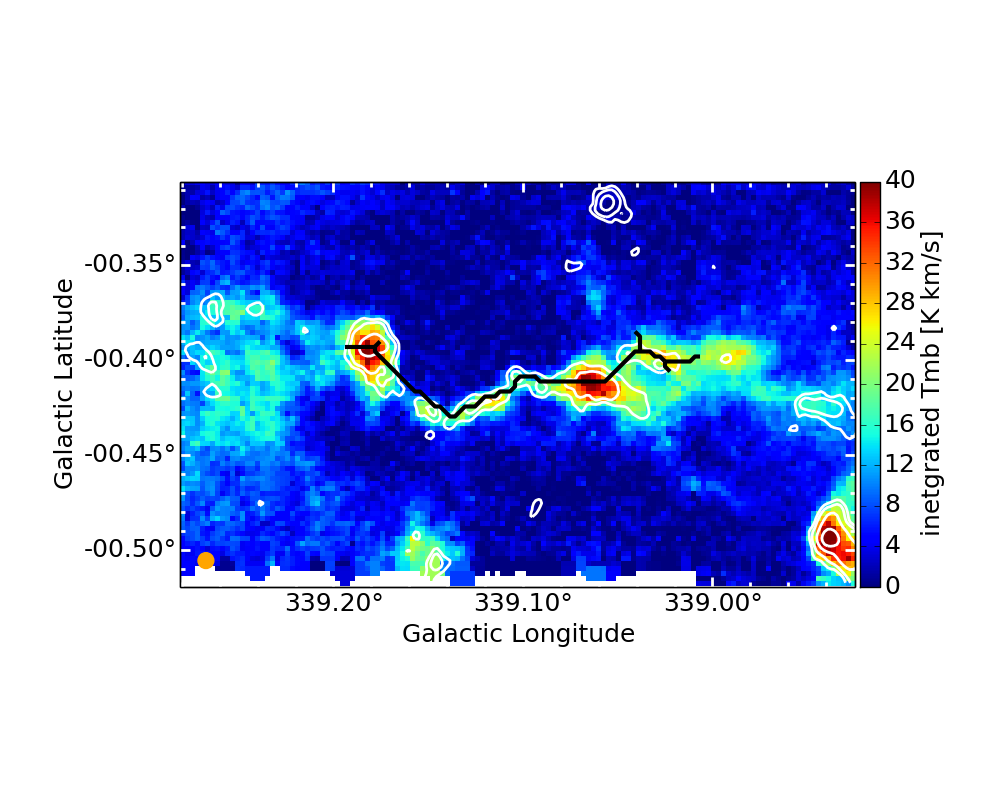}
\end{minipage}
\begin{minipage}{0.37\textwidth}
\includegraphics[width=\textwidth, clip=true, trim= 0.5cm 1cm 0.0cm 0.0cm]{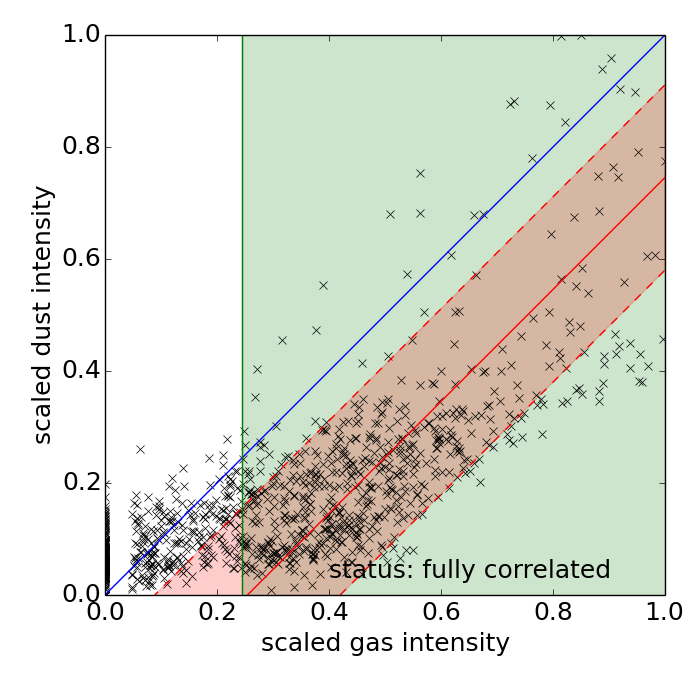}
\end{minipage}
\caption{Integrated \cco intensity map (left) and gas-dust correlation plot (right, description as in Fig. \ref{cor_rest}) of the single velocity component filament candidate ``G339.116-00.405''. Because of the good correlation between the only identified velocity component with the ATLASGAL emission, it was categorized as fully correlated filament and can be found as ``G339.116-00.405\_0'' in the final catalogue.}
\label{single_example}
\end{figure*}

\begin{figure*}
\centering
\begin{minipage}{0.59\textwidth}
\includegraphics[width=\textwidth, clip=true, trim= 0.cm 2.cm 0.0cm 3.0cm]{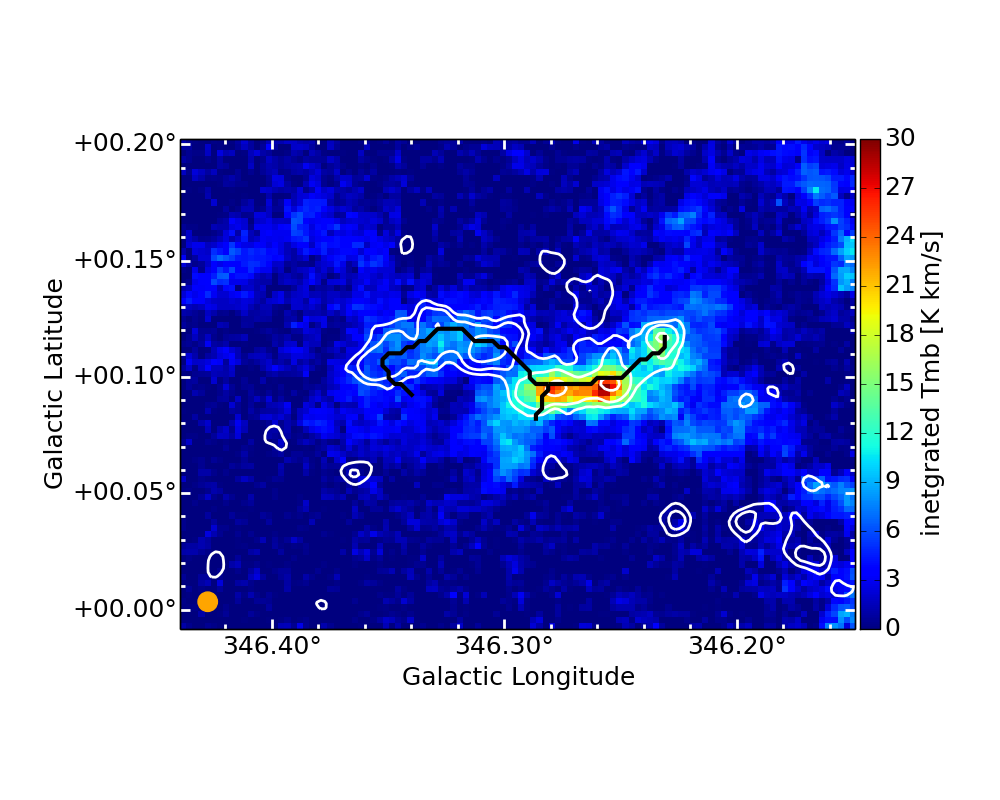}
\end{minipage}
\begin{minipage}{0.37\textwidth}
\includegraphics[width=\textwidth, clip=true, trim= 0.5cm 1cmcm 0.0cm 0.0cm]{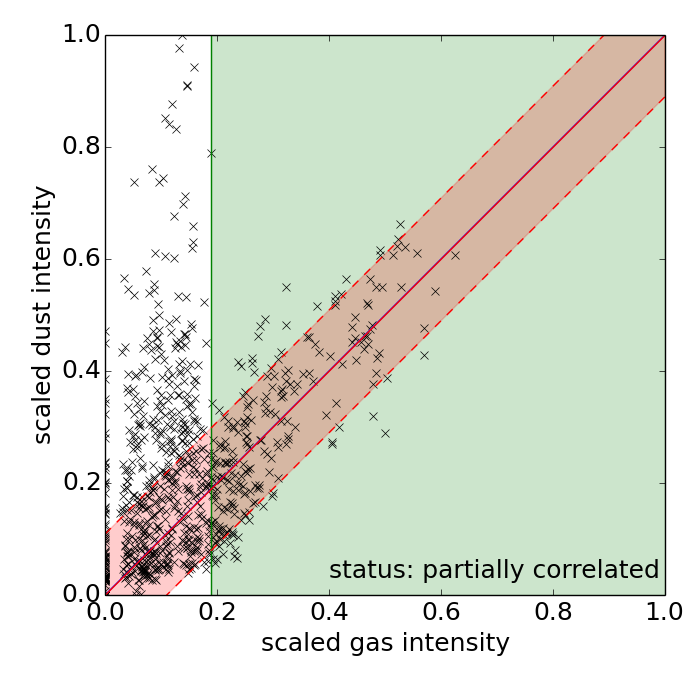}
\end{minipage}
\caption{Integrated \cco intensity map left) and gas-dust correlation plot (right, description as in Fig. \ref{cor_rest})) of the first velocity component of the filament candidate ``G346.293+00.109''. The partially correlated filament is listed as ``G346.293+00.109\_0''. The other three velocity components of the candidate are shown in Fig. \ref{example2}. }
\label{example1}
\end{figure*}

\begin{figure*}
\centering
\begin{minipage}{0.59\textwidth}
\includegraphics[width=\textwidth, clip=true, trim= 0.cm 2.cm 0.0cm 3.0cm]{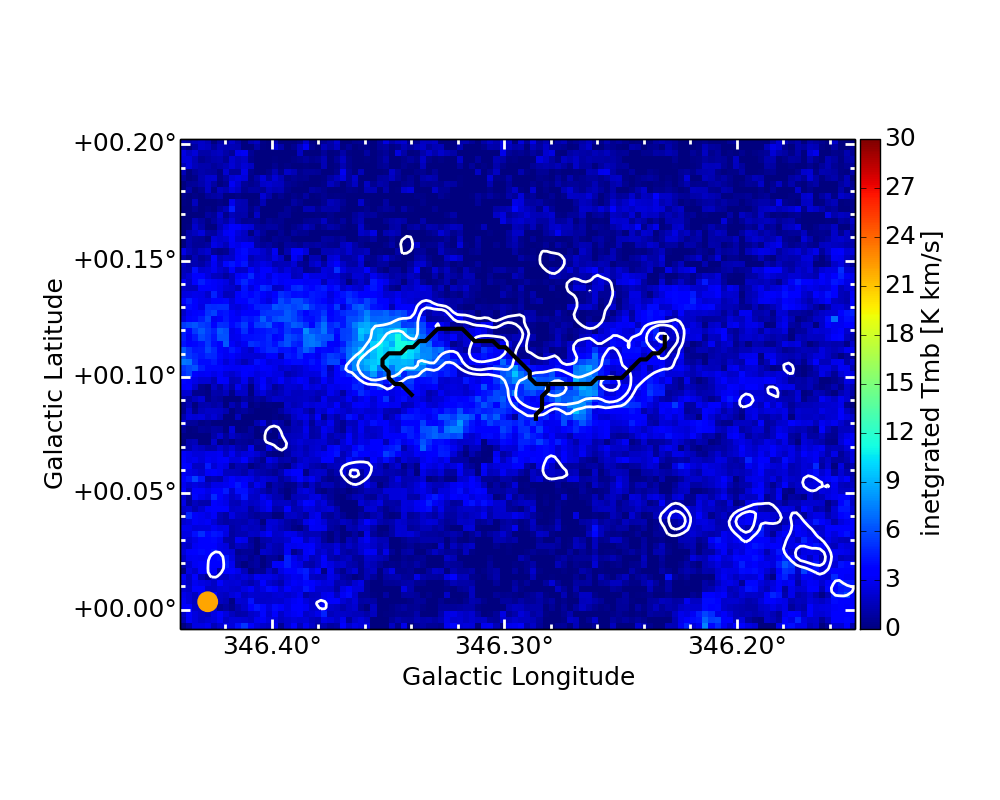}
\end{minipage}
\begin{minipage}{0.37\textwidth}
\includegraphics[width=\textwidth, clip=true, trim= 0.5cm 1cmcm 0.0cm 0.0cm]{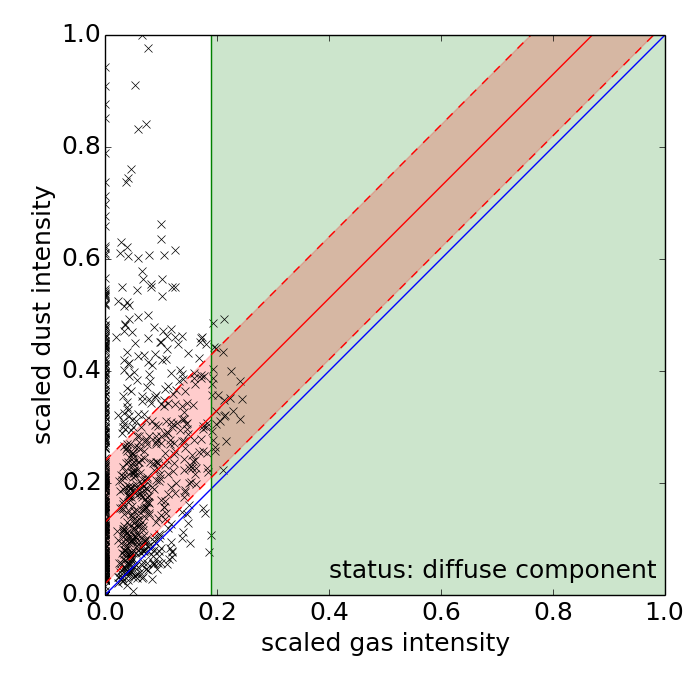}
\end{minipage}

\begin{minipage}{0.59\textwidth}
\includegraphics[width=\textwidth, clip=true, trim= 0.cm 2.cm 0.0cm 3.0cm]{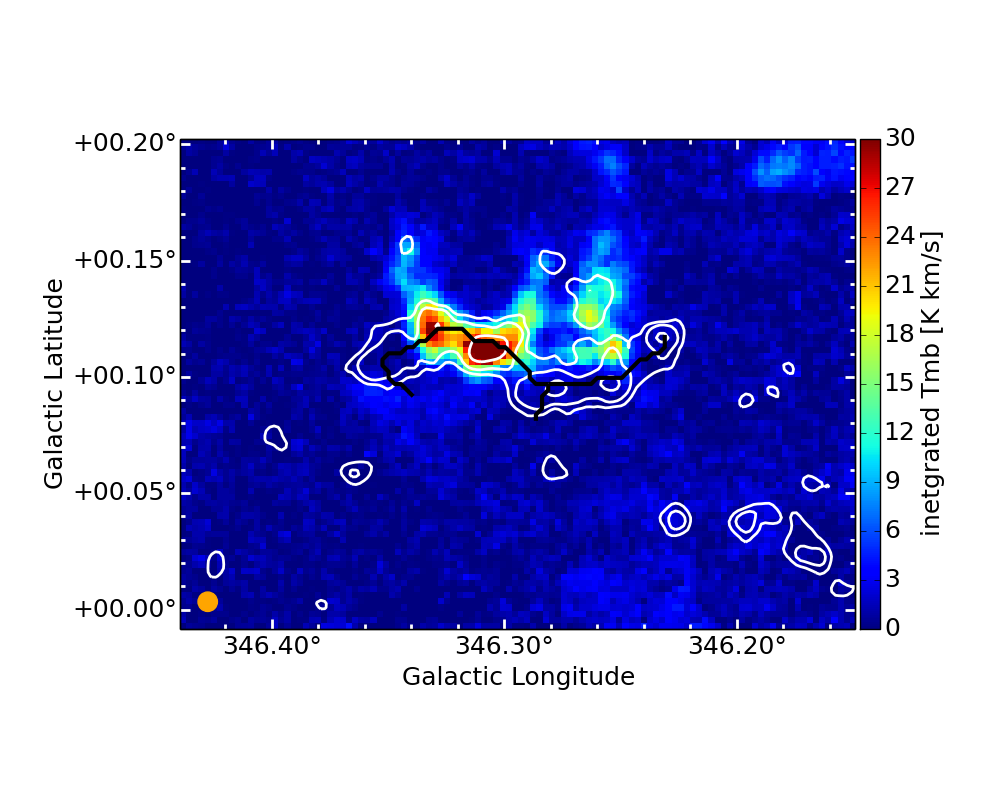}
\end{minipage}
\begin{minipage}{0.37\textwidth}
\includegraphics[width=\textwidth, clip=true, trim= 0.5cm 1cmcm 0.0cm 0.0cm]{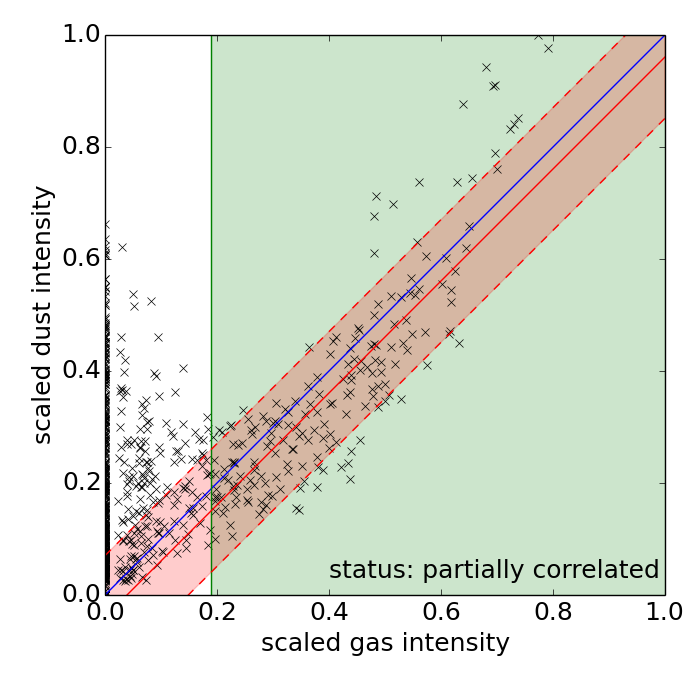}
\end{minipage}

\begin{minipage}{0.59\textwidth}
\includegraphics[width=\textwidth, clip=true, trim= 0.cm 2.cm 0.0cm 3.0cm]{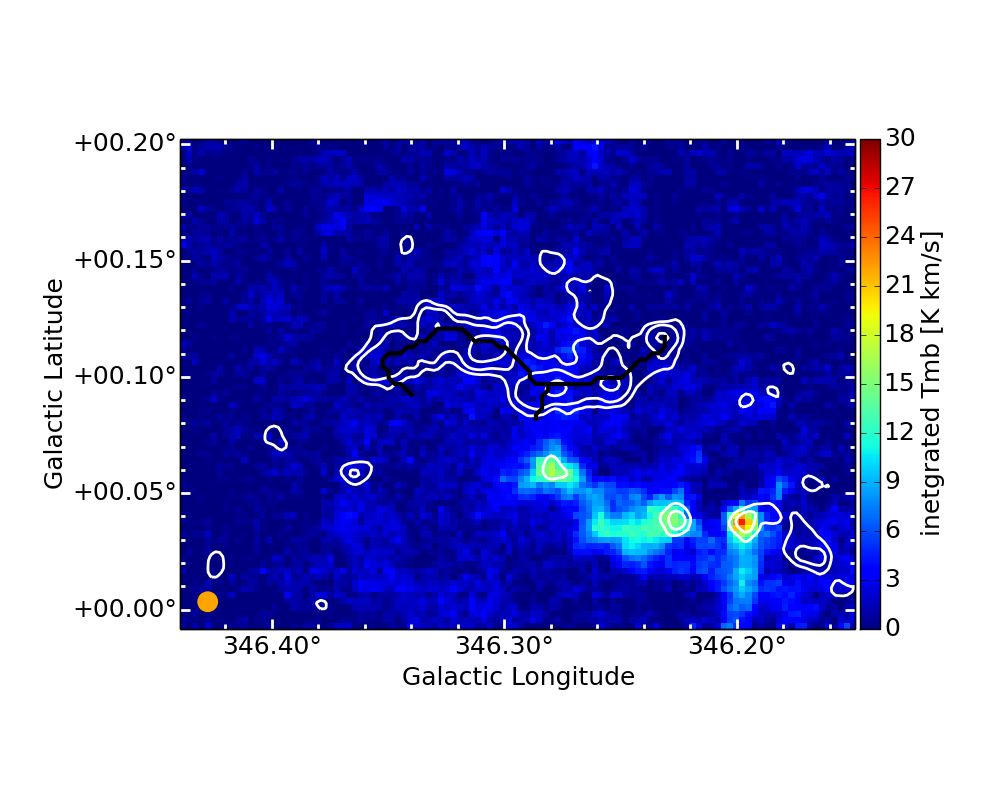}
\end{minipage}
\begin{minipage}{0.37\textwidth}
\includegraphics[width=\textwidth, clip=true, trim= 0.5cm 1cm 0.0cm 0.0cm]{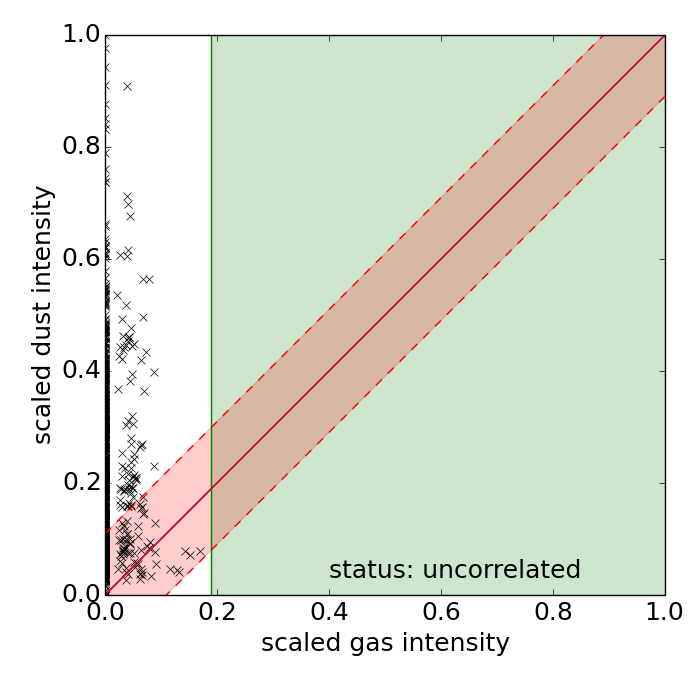}
\end{minipage}
\caption{Integrated \cco intensity maps (left) and gas-dust correlation plots (right, description as in Fig. \ref{cor_rest})) of the second (top), third (middle), and fourth (bottom) velocity component of the filament candidate ``G346.293+00.109''. The second, diffuse component is listed as ``G346.293+00.109\_1'' and the third, partially correlated component is listed as ``G346.293+00.109\_2''. The uncorrelated component is not listed in the catalogue. The first velocity components of the candidate is shown in Fig. \ref{example1}. }
\label{example2}
\end{figure*}

In section \ref{correlation} we are describing the method of comparing the ATLASGAL dust intensities with the integrated \cco intensity maps to identify the spatially correlated velocity components. Based on this correlation we assign one of four categories to the structure, which are: uncorrelated, completely correlated, partially correlated, and diffuse component. To visualize this categorization we show one example in the section \ref{correlation} (Figs. \ref{integrated}, \ref{gas-dust-correlation}). To give a complete picture we present here the correlation plots of the other components of the filament candidate ``G333.297+00.073'' (Fig. \ref{cor_rest}, for intensity maps see Fig. \ref{integrated}), and \cco spectra (Fig. \ref{spectrum_app}), integrated intensity maps and the corresponding correlation plots for the filament candidates ``G339.116-00.405'' (Fig. \ref{single_example}) and ``G346.293+00.109'' (Figs. \ref{example1} and \ref{example2}).

\section{Filament profiles}
\label{other_profiles}

\begin{figure*}
\centering
\begin{minipage}{0.505\textwidth}
\includegraphics[width=\textwidth, clip=true, trim= 1.cm 1.5cm 0.5cm 0.3cm]{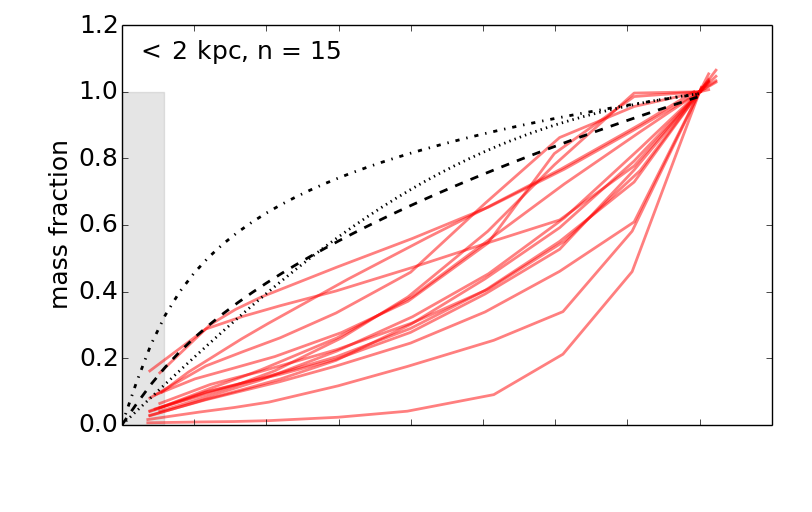}
\end{minipage}
\begin{minipage}{0.48\textwidth}
\includegraphics[width=\textwidth, clip=true, trim= 2.cm 1.5cm 0.5cm 0.3cm]{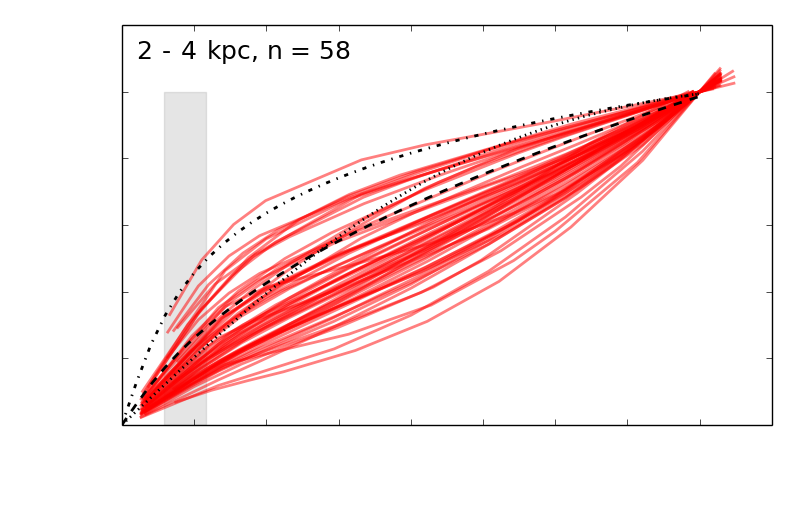}
\end{minipage}

\begin{minipage}{0.505\textwidth}
\includegraphics[width=\textwidth, clip=true, trim= 1.cm 0.5cm 0.5cm 0.3cm]{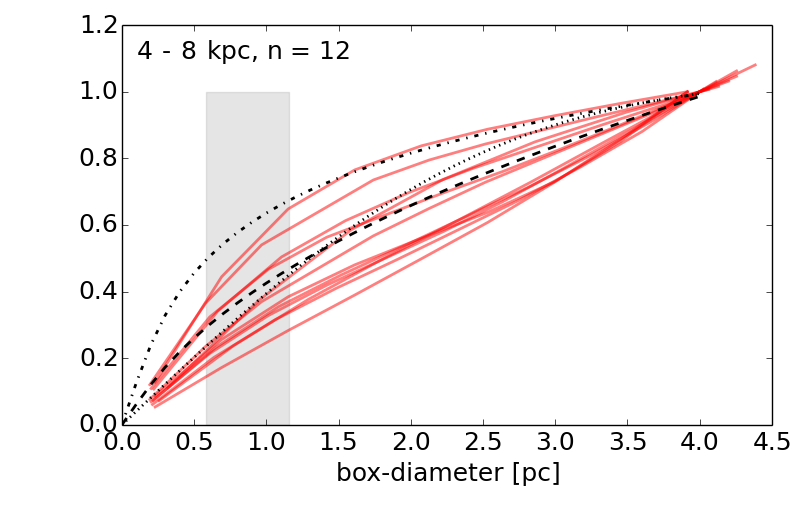}
\end{minipage}
\begin{minipage}{0.48\textwidth}
\includegraphics[width=\textwidth, clip=true, trim= 2.cm 0.5cm 0.5cm 0.3cm]{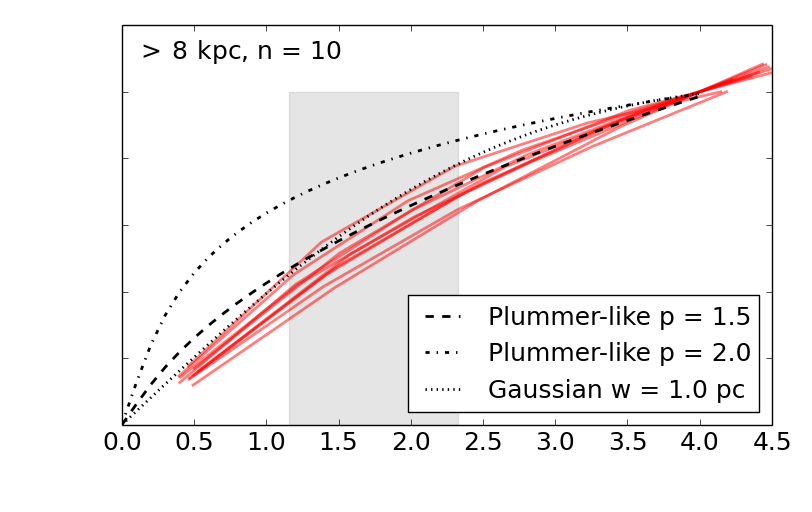}
\end{minipage}

\caption{Fraction of the filament mass derived from corrected ATLASGAL+PLANCK dust continuum emission dependent on the box-diameter of the mask separated with distances. \emph{Top left:} $d < 2~\text{kpc}$, \emph{Top right:} $2~\text{kpc} < d < 4~\text{kpc}$, \emph{Bottom left:} $4~\text{kpc} < d < 8~\text{kpc}$, \emph{Bottom right:} $d > 8~\text{kpc}$. The gray lines indicate the physical beam size at these distances. The black lines show the theoretical profiles, which describe a Plummer-like $p=1.5$ (dashed) or $p=2.0$ (dash-dotted), and a Gaussian with $w = 1.0$ (dotted).}
\label{mass-width-dust}
\end{figure*}

\begin{figure*}
\centering
\begin{minipage}{0.505\textwidth}
\includegraphics[width=\textwidth, clip=true, trim= 1.cm 1.5cm 0.5cm 0.3cm]{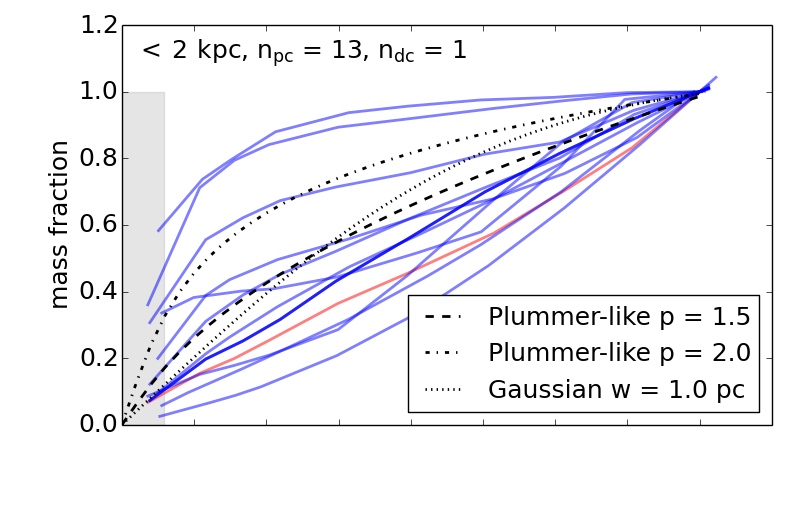}
\end{minipage}
\begin{minipage}{0.48\textwidth}
\includegraphics[width=\textwidth, clip=true, trim= 2.cm 1.5cm 0.5cm 0.3cm]{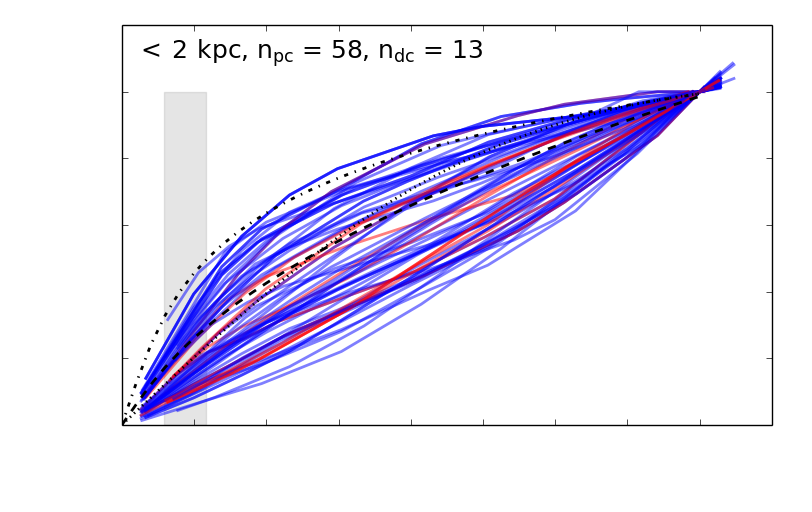}
\end{minipage}

\begin{minipage}{0.505\textwidth}
\includegraphics[width=\textwidth, clip=true, trim= 1.cm 0.5cm 0.5cm 0.3cm]{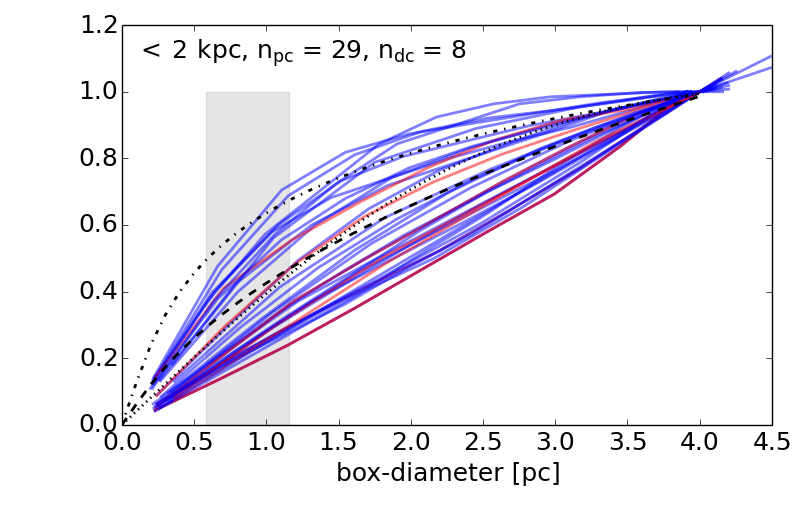}
\end{minipage}
\begin{minipage}{0.48\textwidth}
\includegraphics[width=\textwidth, clip=true, trim= 2.cm 0.5cm 0.5cm 0.3cm]{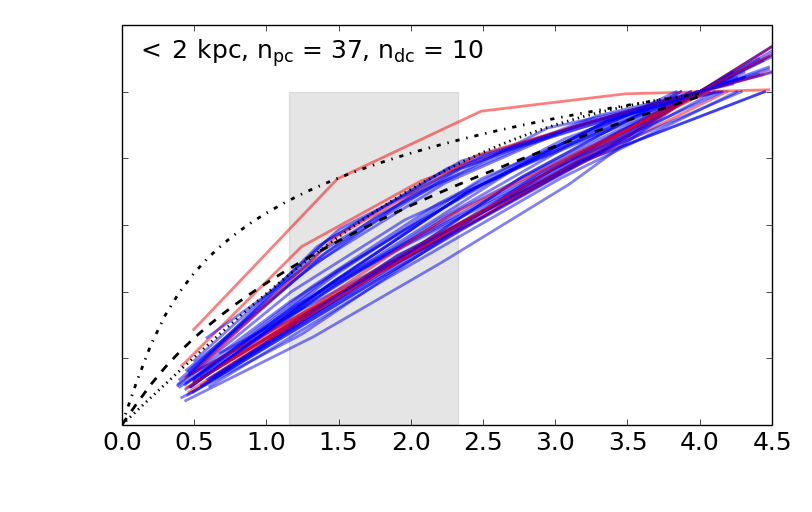}
\end{minipage}

\caption{Fraction of the filament mass for partially correlated (blue) and diffuse component (red) filaments derived from \cco emission dependent on the box-diameter of the mask separated with distances. \emph{Top left:} $d < 2~\text{kpc}$, \emph{Top right:} $2~\text{kpc} < d < 4~\text{kpc}$, \emph{Bottom left:} $4~\text{kpc} < d < 8~\text{kpc}$, \emph{Bottom right:} $d > 8~\text{kpc}$. The vertical gray lines indicate the physical beam size at these distances. The black lines show the theoretical profiles, which describe a Plummer-like $p=1.5$ (dashed) or $p=2.0$ (dash-dotted), and a Gaussian with $w = 1.0$ (dotted).}
\label{mass-width-lescor}
\end{figure*}

In Section \ref{Filament-profile} we showed how the gas mass of correlated filaments is increasing with increasing radius. To be complete, we also show the mass curves of the correlated filaments based on the corrected ATLASGAL+PLANCK data (Fig. \ref{mass-width-dust}), and the mass curves of the partially correlated (blue, Fig. \ref{mass-width-lescor}) and diffuse component filaments (red, Fig. \ref{mass-width-lescor}) based on the integrated \cco observations.

The mass curves of of the correlated filaments are in mostly in agreement with \cco observations. However, for the most nearby filaments ($\rm < 2~kpc$), and especially in the continuum data, we find profiles indicating a slope of $p < 0$. This can be explained by line-of-sight confusion within the large boxes around the filament skeleton. As the dust continuum data traces all emission along the line-of-site it is possible that strong emission, which is not related to the filament but located nearby, is taken into account for the mass estimate for larger radii. Therefore, the masses will be overestimated. This effect is more likely for nearby filaments, as larger angular sizes are taken into account for the same physical size. The gas mass curves of the partially correlated and diffuse component filaments show on average similar results as the correlated filaments, but with a larger scatter, as the skeleton no longer necessarily represents the shape of the structure.

\end{appendix}

\end{document}